\documentstyle{book}               

\newfont{\twolineletters}{msbm10}
\catcode`@=11
\@addtoreset{equation}{section}
\def\theequation{\thesection.\arabic{equation}}

\def\lhead{\@ifnextchar[{\@xlhead}{\@ylhead}}
\def\@xlhead[#1]#2{\gdef\@elhead{#1}\gdef\@olhead{#2}}
\def\@ylhead#1{\gdef\@elhead{#1}\gdef\@olhead{#1}}

\def\chead{\@ifnextchar[{\@xchead}{\@ychead}}
\def\@xchead[#1]#2{\gdef\@echead{#1}\gdef\@ochead{#2}}
\def\@ychead#1{\gdef\@echead{#1}\gdef\@ochead{#1}}

\def\rhead{\@ifnextchar[{\@xrhead}{\@yrhead}}
\def\@xrhead[#1]#2{\gdef\@erhead{#1}\gdef\@orhead{#2}}
\def\@yrhead#1{\gdef\@erhead{#1}\gdef\@orhead{#1}}

\def\lfoot{\@ifnextchar[{\@xlfoot}{\@ylfoot}}
\def\@xlfoot[#1]#2{\gdef\@elfoot{#1}\gdef\@olfoot{#2}}
\def\@ylfoot#1{\gdef\@elfoot{#1}\gdef\@olfoot{#1}}

\def\cfoot{\@ifnextchar[{\@xcfoot}{\@ycfoot}}
\def\@xcfoot[#1]#2{\gdef\@ecfoot{#1}\gdef\@ocfoot{#2}}
\def\@ycfoot#1{\gdef\@ecfoot{#1}\gdef\@ocfoot{#1}}

\def\rfoot{\@ifnextchar[{\@xrfoot}{\@yrfoot}}
\def\@xrfoot[#1]#2{\gdef\@erfoot{#1}\gdef\@orfoot{#2}}
\def\@yrfoot#1{\gdef\@erfoot{#1}\gdef\@orfoot{#1}}

\newdimen\headrulewidth
\newdimen\footrulewidth
\newdimen\plainheadrulewidth
\newdimen\plainfootrulewidth
\newdimen\headwidth
\newif\if@fancyplain \@fancyplainfalse
\def\fancyplain#1#2{\if@fancyplain#1\else#2\fi}

\footrulewidth\z@
\plainheadrulewidth\z@
\plainfootrulewidth\z@

\def\@fancyhead#1#2#3#4#5{#1\hbox to\headwidth{\vbox{\hbox
{\rlap{\parbox[b]{\headwidth}{\raggedright#2\strut}}\hfill
\parbox[b]{\headwidth}{\centering#3\strut}\hfill
\llap{\parbox[b]{\headwidth}{\raggedleft#4\strut}}}\headrule}}#5}

\def\@fancyfoot#1#2#3#4#5{#1\hbox to\headwidth{\vbox{\footrule
\hbox{\rlap{\parbox[t]{\headwidth}{\raggedright#2\strut}}\hfill
\parbox[t]{\headwidth}{\centering#3\strut}\hfill
\llap{\parbox[t]{\headwidth}{\raggedleft#4\strut}}}}}#5}

\def\headrule{{\if@fancyplain\headrulewidth\plainheadrulewidth\fi
\hrule\@height\headrulewidth\@width\headwidth \vskip-\headrulewidth}}

\def\footrule{{\if@fancyplain\footrulewidth\plainfootrulewidth\fi
\vskip-0.3\normalbaselineskip\vskip-\footrulewidth
\hrule\@width\headwidth\@height\footrulewidth\vskip0.3\normalbaselineskip}}

\def\ps@fancy{
\let\@mkboth\markboth
\@ifundefined{chapter}{\def\sectionmark##1{\markboth
{\ifnum \c@secnumdepth>\z@
 \thesection\hskip 1em\relax \fi ##1}{}}
\def\subsectionmark##1{\markright {\ifnum \c@secnumdepth >\@ne
 \thesubsection\hskip 1em\relax \fi ##1}}}
{\def\chaptermark##1{\markboth {\ifnum \c@secnumdepth>\m@ne
 \@chapapp\ \thechapter. \ \fi ##1}{}}
\def\sectionmark##1{\markright{\ifnum \c@secnumdepth >\z@
 \thesection. \ \fi ##1}}}
\def\@oddhead{\@fancyhead\relax\@olhead\@ochead\@orhead\hss}
\def\@oddfoot{\@fancyfoot\relax\@olfoot\@ocfoot\@orfoot\hss}
\def\@evenhead{\@fancyhead\hss\@elhead\@echead\@erhead\relax}
\def\@evenfoot{\@fancyfoot\hss\@elfoot\@ecfoot\@erfoot\relax}
\headwidth\textwidth}
\def\ps@fancyplain{\ps@fancy \let\ps@plain\ps@plain@fancy}
\def\ps@plain@fancy{\@fancyplaintrue\ps@fancy}

\catcode`\@=12

\begin{document}

\pagestyle{fancy}

\onecolumn      

\thispagestyle{empty}
\begin{flushright}
          hepth@xxx/yymmxxx\\
          December 1995\\
\end{flushright}
\begin{center}
\vfill \vfill \vfill \vfill
\begin{minipage}{12.5cm}
\begin{center}
{\huge\bf Covariant Quantisation \\ in the \\
Antifield Formalism$^{1}$\\ }
\end{center}
\end{minipage}
\vfill \vfill \vfill
\begin{minipage}{12.5cm}
\begin{center}
{\large\bf Stefan Vandoren$^{2}$ \\ }
\end{center}
\end{minipage}
\vfill
\begin{minipage}{12.5cm}
\begin{center}
{\large Instituut voor Theoretische Fysica\\
   K.U.Leuven, Celestijnenlaan 200D\\
    B-3001 Leuven, Belgium \\}
\end{center}
\end{minipage}
\vfill
\begin{center}
{\bf Abstract}
\end{center}
\begin{quote}
\small
In this thesis we give an overview of the antifield formalism and show how it
must be used to quantise arbitrary gauge theories. The formalism is further
developed and illustrated in several examples, including Yang-Mills theory,
chiral $W_3$ and $W_{2,5/2}$ gravity, strings in curved backgrounds and
topological field theories. All these models are characterised by their gauge
algebra, which can be open, reducible, or even infinitly reducible. We show
in detail how to perform the gauge fixing and how to compute the anomalies
using Pauli-Villars regularisation and the heat kernel method. Finally, we
discuss the geometrical structure of the antifield formalism.
\vspace{2mm} \vfill \hrule width 3.cm
{\footnotesize
\noindent $^{1}$ PhD thesis, promotor: A. Van Proeyen \\
\noindent $^{2}$ E--mail : stefan.vandoren@fys.kuleuven.ac.be}
\normalsize
\end{quote}
\end{center}
\newpage


\parskip 6pt plus 1pt minus 1pt
\lhead[\fancyplain{}{}]{\fancyplain{}{}}
\rhead[\fancyplain{}{}]{\fancyplain{}{}}
\chead{\fancyplain{}{\bf Contents}}
\cfoot{\fancyplain{}{}}
\tableofcontents
\parskip 5pt plus 1pt minus 1pt
\newpage

\lhead[\fancyplain{}{\bf\thepage}]{\fancyplain{}{}}
\chead[\fancyplain{}{\bf\leftmark}]{\fancyplain{}{\bf\rightmark}}
\rhead[\fancyplain{}{}]{\fancyplain{}{\bf\thepage}}
\cfoot{\fancyplain{\bf\thepage}{}}

\def\W52{$W_{2,5/2}$}
\def\ie{{\sl i.e.\ }}
\def\eg{{\sl e.g.\ }}
\def\rhs{{\sl rhs}}
\def\lhs{{\sl lhs}}
\def\emt{en\-er\-gy--mo\-men\-tum tensor}
\def\emo{en\-er\-gy--mo\-men\-tum operator}
\def\WA{$W$--al\-ge\-bra}
\def\WS{$W$--string}
\def\KA{Ka\v{c}--Moody al\-ge\-bra}
\def\be{\begin{equation}}
\def\ee{\end{equation}}
\def\bea{\begin{eqnarray}}
\def\beastar{\begin{eqnarray*}}
\def\eea{\end{eqnarray}}
\def\eeastar{\end{eqnarray*}}
\def\nonu{\nonumber \\}
\def\leqn#1{\lefteqn{#1}\nonu}

\newcommand{\dd}[1]{{\delta \over \delta #1}}
\newcommand{\ddt}[2]{{\delta #1\over \delta #2}}
\newcommand{\del}{\partial}
\def\db{\bar{\partial}}
\def\dz{\partial}
\newcommand{\bp}{{\bar \psi }}
\newcommand{\bd}{{\bar \partial }}
\newcommand{\nablab}{\overline{\nabla}}
\def\SBV{S_BV}

\def\ft#1#2{{\textstyle{{#1}\over{#2}}}}


\newcommand{\cA}{{\cal A}}
\newcommand{\cB}{{\cal B}}
\newcommand{\cC}{{\cal C}}
\newcommand{\cD}{{\cal D}}
\newcommand{\cE}{{\cal E}}
\newcommand{\cF}{{\cal F}}
\newcommand{\cG}{{\cal G}}
\newcommand{\cH}{{\cal H}}
\newcommand{\cI}{{\cal I}}
\newcommand{\cJ}{{\cal J}}
\newcommand{\cK}{{\cal K}}
\newcommand{\cL}{{\cal L}}
\newcommand{\cM}{{\cal M}}
\newcommand{\cN}{{\cal N}}
\newcommand{\cO}{{\cal O}}
\newcommand{\cP}{{\cal P}}
\newcommand{\cQ}{{\cal Q}}
\newcommand{\cR}{{\cal R}}
\newcommand{\cS}{{\cal S}}
\newcommand{\cT}{{\cal T}}
\newcommand{\cU}{{\cal U}}
\newcommand{\cV}{{\cal V}}
\newcommand{\cW}{{\cal W}}
\newcommand{\cX}{{\cal X}}
\newcommand{\cY}{{\cal Y}}
\newcommand{\cZ}{{\cal Z}}


\newcommand{\links}{\hspace{-10mm}}


\newsavebox{\uuunit}
\sbox{\uuunit}
    {\setlength{\unitlength}{0.825em}
     \begin{picture}(0.6,0.7)
        \thinlines
        \put(0,0){\line(1,0){0.5}}
        \put(0.15,0){\line(0,1){0.7}}
        \put(0.35,0){\line(0,1){0.8}}
       \multiput(0.3,0.8)(-0.04,-0.02){12}{\rule{0.5pt}{0.5pt}}
     \end {picture}}
\newcommand {\unity}{\mathord{\!\usebox{\uuunit}}}


\def\lefthook{{\vrule height5pt width0.4pt depth0pt}}
\def\righthook{{\vrule height5pt width0.4pt depth0pt}}
\def\leftrighthookfill{$\mathsurround=0pt \mathord\lefthook
     \hrulefill\mathord\righthook$}
\def\underhook#1{\vtop{\ialign{##\crcr$\hfil\displaystyle{#1}\hfil$\crcr
      \noalign{\kern-1pt\nointerlineskip\vskip2pt}
      \leftrighthookfill\crcr}}}


\newcommand{\zb}{\bar{z}}
\newcommand{\wb}{\bar{w}}
\newcommand{\bb}{\overline{b}}
\newcommand{\cb}{\overline{c}}
\newcommand{\betb}{\bar{\b}}
\newcommand{\gb}{\bar{g}}
\newcommand{\eb}{\bar{\e}}
\newcommand{\mb}{\bar{\m}}
\newcommand{\pba}{\bar{p}}
\newcommand{\qb}{\bar{q}}

\newcommand{\bX}{\bar{X}}
\newcommand{\bH}{\bar{H}}
\newcommand{\bx}{\bar{\xi }}
\newcommand{\bF}{\bar{F}}

\newcommand{\delb}{\bar{\partial}}
\newcommand{\edb}{\frac{1}{\bar\partial}}         
\newcommand{\ddb}{\frac{\partial}{\bar\partial}}  %
\newcommand{\nab}{\nabla}
\newcommand{\nabb}{\overline{\nabla}}
\newcommand{\ope}[2]{\frac{#1}{(z-w)^#2}}
\newcommand{\opee}[1]{\frac{#1}{z-w}}
\newcommand{\no}{:}
\newcommand{\zw}{(z-w)}
\newcommand{\maat}[1]{\frac{#1}{2\pi i}}
\newcommand{\pinv}{\frac{1}{\p}}

\newcommand{\invac}{\left| 0 \right> }
\newcommand{\outvac}{\left< 0 \right| }
\newcommand{\Lb}{\overline{L}}
\newcommand{\Tb}{\overline{T}}
\newcommand{\Jb}{\overline{J}}
\newcommand{\KM}{Ka\v{c}-Moody}
\newcommand{\ktwo}{\frac{\kappa }{2}}

\newcommand{\wh}[1]{\widehat{#1}}
\def\th{\tilde h }
\def\tg{\tilde g }
\def\tX{\tilde X }
\def\tJ{\tilde J }
\def\cJ{\check{J}}
\def\Ahb{\wh{\!\bar A}}
\def\Aha{\wh{\! A}}
\def\qq{\qquad}

\newcommand{\dr}{\raise.3ex\hbox{$\stackrel{\leftarrow}{\partial }$}{}}
\newcommand{\dl}{\raise.3ex\hbox{$\stackrel{\rightarrow}{\partial}$}{}}
\newcommand{\delr}{\raise.3ex\hbox{$\stackrel{\leftarrow}{\partial}$}
{}}
\newcommand{\dell}{\raise.3ex\hbox{$\stackrel{\rightarrow}{\partial}$}
{}}
\newcommand{\rd}{\raise.3ex\hbox{$\stackrel{\leftarrow}{\delta}$}
{}}
\newcommand{\ld}{\raise.3ex\hbox{$\stackrel{\rightarrow}{\delta}$}
{}}
\newcommand{\eqn}[1]{(\ref{#1})}

\newcommand{\ind}[1]{\mbox{\footnotesize #1}}
\newcommand{\inds}[1]{\mbox{\scriptsize #1}}
\newcommand{\DS}{\mbox{\scriptsize DS }}
\newcommand{\dis}{\displaystyle}
\newcommand{\cl}{{\rm cl}}
\newcommand{\eff}{{\rm eff}}
\newcommand{\indu}{{\rm ind}}
\newcommand{\naar}{\mapsto}
\newcommand{\tr}{\mbox{tr}}
\newcommand{\half}{\frac{1}{2}}
\newcommand{\shalf}[1]{\mbox{$\scriptstyle\frac{\mbox{\footnotesize $#1$}}
{\mbox{\footnotesize2}}$}}
\newcommand{\sfrac}[2]{\mbox{$\scriptstyle\frac{\mbox{$\scriptstyle #1$}}
{\mbox{$\scriptstyle #2$}}$}}

\newcommand{\cmp}[3]{Commun. Math. Phys. {\bf #1} (19{#2}) #3}
\newcommand{\prl}[3]{Phys. Rev. Lett. {\bf #1} (19{#2}) #3}
\newcommand{\jmp}[3]{J. Math. Phys. {\bf #1} (19{#2}) #3}
\newcommand{\npb}[3]{Nucl. Phys. {\bf B#1} (19{#2}) #3}
\newcommand{\mpl}[3]{Mod. Phys. Lett. {\bf A#1} (19{#2}) #3}

\newcommand{\pl}{(\partial X^\mu )}
\newcommand{\plb}{(\partial X^\nu )}
\newcommand{\plg}{(\partial X^\rho )}
\newcommand{\plr}{(\partial X^\sigma )}
\newcommand{\pls}{(\partial X^\tau )}
\newcommand{\bpl}{(\bar{\partial }X^\mu )}
\newcommand{\ds}{d_{\mu \nu \rho }}
\newcommand{\na}{\nabla }
\newcommand{\dkt}{\delta _{KT}}
\newcommand{\QED}{{\hspace*{\fill}\rule{2mm}{2mm}\linebreak}}
\newtheorem{lemma}{Lemma}
\renewcommand{\thelemma}{\thesection.\arabic{lemma}}
\newtheorem{theorem}{Theorem}
\renewcommand{\thetheorem}{\thesection.\arabic{theorem}}
\newtheorem{defn}{Definition}
\renewcommand{\thedefn}{\thesection.\arabic{defn}}
\renewcommand{\theequation}{\thesection.\arabic{equation}}

\newcommand{\gras}[1]{\epsilon_{#1}}
\newcommand{\gh}[1]{\mbox{gh} \left( #1 \right)}
\newcommand{\sdet}{\mbox{sdet}}
\newcommand{\str}{\mbox{str}}
\newcommand{\ihbar}{\frac{i}{\hbar}}
%
\newcommand{\sqrg}{\sqrt{g}}
\newcommand{\sqrabsg}{\sqrt{\vert g \vert}}
\newcommand{\ddr}{\raise.3ex\hbox{$\stackrel{\leftarrow}{d}$}}
\newcommand{\ddl}{\raise.3ex\hbox{$\stackrel{\rightarrow}{d}$}}

\newcommand{\beq}{\begin{equation}}
\newcommand{\eeq}{\end{equation}}

\newcommand{\for}{{\textstyle\frac{1}{4}}}

\newcommand{\dirac}[1]{/ \!\!\!#1}
\newcommand{\vgl}[1]{eq.(\ref{#1})}
\newcommand{\gv}{\gamma^5}
\newcommand{\gu}[1]{\gamma^{#1}}
\newcommand{\gd}[1]{\gamma_{#1}}

\newcommand{\wt}{\widetilde}
\def\sepand{\rule{14cm}{0pt}\and}
\def\gtwid{\raise.3ex\hbox{$>$\kern-.75em\lower1ex\hbox{$\sim$}}}
\def\ltwid{\raise.3ex\hbox{$<$\kern-.75em\lower1ex\hbox{$\sim$}}}

%
%
\newcommand{\CD}[2]{\nabla^{(#1)}_{#2}}

\newcommand{\sqrabsh}{\sqrt{\vert h \vert}}

\chapter{Introduction}
\section{The general context}
The aim of physics is to understand and predict the phenomena that occur
in nature. Perhaps the most surprising fact is that nature behaves
differently at different
scales. For example, when two particles with velocities $v_1$ and $v_2$
(small compared to the speed of light)
pass each other, their relative velocity is simply the sum of the two~:
$v_{rel}=v_1+v_2$.
At higher velocities, approaching the
speed of light $c$, this no longer holds and the relative
velocity is smaller then the sum, i.e.
$v_{rel}=(v_1+v_2)/(1+v_1v_2/c^2)$.
In other words, at high velocities, Newton's theory of classical mechanics
\cite{Newton} breaks down and one has to replace it by Einstein's special
relativity \cite{SpRel}. The same thing happens when going to microscopic
scales. Take for example the atom. According to the theory of
electromagnetism, the electrons moving in orbits
around the nucleus will lose energy by radiation. The radius of the orbit
will therefore decrease and eventually the atom will collapse.
This shows that we need to look for a new physical
principle that explains the phenomena at microscopic scales.
To understand these short distance effects, we have to replace classical
mechanics by quantum mechanics \cite{QM}.

Scientists believe that nature can be described in terms of physical
laws, which can be translated into mathematical equations. To understand
and predict nature from
first principles, we need to know all the different elementary particles
and the forces that act between them.
Four forces are known~: gravity, electromagnetism, the weak and the strong
(nuclear)
force. The classical descriptions of these forces were
developed by Newton, Maxwell, Fermi and Yukawa
\cite{Newton,Maxwell,Fermi,Yukawa}.
However, as mentioned above, one has to extend these theories to relativistic
and quantum mechanical scales. E.g. Newton's theory of gravity becomes
Einstein's general relativity \cite{GenRel} when going to relativistic
scales. Maxwell's
theory of electromagnetism (coupled to matter) becomes quantum
electrodynamics \cite{QED} which can be applied for both
relativistic and quantum mechanical systems.

Electromagnetic, weak and strong interactions can be
unified in a single theory
called the Standard Model \cite{SM}. It is the theory of quarks and leptons
and
their interactions. When we talk about "theories" in this thesis, we always
mean {\it field theories}~: all particles are represented
by fields $\phi ^i$, and the classical dynamics and interactions are
determined by specifying an action functional depending on these fields~:
\begin{equation}
S(\phi ^i)=\int d^4x \cL(\phi ^i,\partial _\mu \phi ^i,x^\mu )\ ,
\end{equation}
where $\cL$ is called the Lagrangian. To make this theory applicable at
relativistic energies we have to write the action in a
Lorentz covariant way, according to the principles of special
relativity. Therefore, the action functional is an integral over space-time
with coordinates $x^\mu $.
The construction of the action
functional for the Standard Model is based on the principle of local gauge
invariance. This means that the action possesses a number $a=1,...,n$ of
local symmetries
\begin{equation}
\delta _{\epsilon ^a}\phi ^i=R^i_a\epsilon ^a \Rightarrow \delta
_{\epsilon ^a}S=0 \ .\label{Rmatrix}
\end{equation}
Local means that the parameters depend on the chosen point in space-time,
i.e. $\epsilon ^a=\epsilon ^a(x)$. $R^i_a$ are called the generators of the
symmetries. In the case of the Standard Model, these generators form the
Lie algebra based on the Lie group $SU(3)\times SU(2) \times U(1)$. The
three factors of this Lie group correspond to the strong, weak and
electromagnetic forces. It is surprising that requiring local gauge
invariance generates forces between particles.

The same principle can be used to construct Einstein's theory of general
relativity. This theory describes the interaction of
matter
and gravity, using the principle of general coordinate invariance. By the
latter we mean that the action $S(\phi ,g_{\mu \nu })$ is invariant
under the transformations $\delta _{\epsilon ^\mu
}g_{\mu \nu }=
\epsilon ^\gamma\partial_
\gamma g_{\mu \nu}+\partial_\mu
\epsilon ^\gamma g_{\gamma \nu }+\partial_\nu  \epsilon ^\gamma g_{\mu
\gamma}$ for the metric and
$\delta
_{\epsilon ^\mu }\phi ^i=\epsilon ^\mu \partial _\mu \phi ^i+...$ for the
matter fields.
The ellipsis denote
terms depending on the chosen type of matter fields, e.g. scalar
fields, spin 1/2 fermions, tensor fields,... . In this way Einstein's
theory of gravity is a gauge theory, the generators forming the group of
general coordinate transformations. The difference with the Standard Model
is that the transformation rules on the fields are now induced by a local
transformation on the space-time coordinates, i.e. $x^\mu \rightarrow
x^\mu +\epsilon ^\mu (x)$.

The next step is to make these theories applicable at short distances,
i.e.
when quantum effects become important. Therefore, we have to study {\it
relativistic quantum field theory}. This can be done using path
integral techniques. One can compute transition amplitudes and correlation
functions from the generating functional
\begin{equation}
Z(J)=\int \cD\phi \exp^{\frac{i}{\hbar}(S(\phi )+J\phi )}\ .
\end{equation}
Mathematically speaking, we are now entering the subject of functional
integration.
An unsolved problem here is now to define a measure on the configuration
space, i.e. what we have written as $\cD \phi $. Surprisingly, physicists
are able, to a certain extent, to circumvent this problem and to make
very precise predictions using path integral techniques, in
(almost perfect) agreement with experiments. The generating functional
can be computed by setting up a perturbation theory (a loop expansion) in
terms of Feynman diagrams. Certainly for the Standard Model, this
diagrammatical expansion has been very succesful.

Although gauge invariance plays an important role in describing the
interactions between elementary particles, the path integral measure again
poses a problem.
It is clear that all field configurations differing
by a local gauge transformation, have the same action. Because
of this, forgetting for
the moment the source term $J\phi $, the integration will lead to diverging
results. This cannot lead to a physical theory with sensible predictions.
To cure this problem, we have to change something by hand. Out of each
class of gauge equivalent field configurations, we will pick only one.
This procedure is called gauge fixing and is the first step in
the quantisation  of a gauge theory. The way to implement the gauge fixing
procedure in the path integral was first demonstrated in quantum
electrodynamics by Faddeev and Popov \cite{FP}. Later on, this technique
was generalised to the Standard Model and General Relativity,
using a method developed by by Becchi, Rouet and Stora, and also
independent of them, by Tyutin, see \cite{BRST}. We will comment on these
methods in the next section.

Nowadays, we have more complicated
gauge theories, which we will discuss below. The gauge fixing for these
models is more involved, and the BRST method is not applicable anymore. In
this thesis, we will present a new formalism, called {\it antifield
formalism} and developed by Batalin and Vilkovisky \cite{BV1,BV2}, that
enables us to do a proper gauge fixing for {\it all} gauge theories.
As we will explain in the next
section, it encompasses the different quantisation schemes mentioned
above.

The next problem when evaluating the path integral are the ultraviolet
divergencies suffered by physical amplitudes and correlation functions.
These are due to the short distance behaviour of the
interactions between two particles. For a theory to make
sense at short distances, one has to follow a regularisation and
renormalisation prescription. The Standard Model is an example where this
prescription leads to a consistent and well defined theory, where
predictions can be made. Such a theory is called renormalisable.

Unfortunately, General Relativity is not renormalisable. We seem to
have no good theory for quantum gravity. For some time, people thought that
supergravity \cite{sugraFrPvNFer}, a combination of gravity and
supersymmetry, could save
the day. A lot of the divergences in correlation functions then disappear
because the infinities of the bosonic sector cancel against the
infinities of the fermionic sector. However, not all divergences can be
removed in this way, and supergravity turns out to be non--renormalisable.
But still,
theoreticians believe that in search for a renormalisable
theory including gravity, supersymmetry
will be of crucial importance. One may also look for supersymmetric
extensions of the Standard Model, based on grand unification groups like
$SU(5)$ or $SO(10)$. In these models, supersymmetry is essential because
it solves the so called the hierarchy problem. This can be understood as
follows. In grand unified theories there are two scales. First there is the
the grand unification scale ($10^{15}-10^{17}$ GeV) where the $SU(5)$
symmetry is spontaneously broken, due to the Higgs effect, to $SU(3)\times
SU(2) \times U(1)$. And then, there is the electroweak scale at energies
around $250$ GeV. The particles of the Standard Model have masses of only
a few GeV. In grand unified theories, these masses will receive
radiative corrections lifting them to the grand unification scale. This
problem is called the hierarchy problem. When combining grand unified
theories with supersymmetry, these radiative corrections are absent due to
a boson-fermion cancellation.

On the other hand, supersymmetry is not observed
in experiments, so it must be broken at a certain scale,
which is believed to be around 1 TeV. The LHC
accelerator at CERN will reach this energy and will
hopefully find, besides the Higgs boson, the first
supersymmetric particle in nature. It would be great if we
could start a new millenium with such a discovery.

There is however new hope for a candidate that includes a quantum theory of
gravity, namely superstring theory \cite{string}. It replaces the concept
of a particle as a pointlike object by a new physical principle~: that of
a particle
as an excitation of a string. Again, we see that nature, if string theory
describes it, behaves differently
at different scales. The theory contains only one parameter,
the string tension. The spectrum of the theory contains a massless
spin two particle, which can be interpreted as the graviton. For
this interpretation, the string tension must correspond to energies around
the Planck scale, which is $10^{19}$ GeV. Superstring theory
is strongly believed to be free from divergences and, as a bonus, unifies
the (supersymmetric) Standard Model with General Relativity.

String theory is another example of a (2 dimensional) gauge theory. The
gauge group is the group of conformal transformations and it is infinite
dimensional. As a consequence of gauge invariance, some degrees of freedom
are absent, i.e. they can be gauged away. This is true for all gauge
theories. The simplest example is the photon~: it has only two physical
degrees of freedom, but is described by four gauge fields $A_\mu $.
The fact that we describe the photon by 4 fields instead of two is
dictated by Lorentz covariance. As stated earlier, we want to keep
this covariance manifest in our description, both classically and
quantum mechanically.

Using
the gauge symmetry, one can count the number of classical physical degrees
of freedom. When going to the quantum theory, i.e the path integral, the
question arises whether these degrees of freedom are still physical. This
is only
guaranteed if the gauge symmetry survives the
quantum theory. It can indeed happen that the
regularisation procedure
does not respect the symmetries of the classical theory. One says that the
theory suffers from anomalies.
For example, superstring theory has an
anomaly, except when the theory is formulated in a 10 dimensional
space--time.
To make contact with the real world, one must compactify 6 dimensions
to a scale which cannot be observed. This can be done in various ways, and
research is still going on to find the correct string vacuum
that describes nature as observed at low energies.

The antifield formalism which we will present here can be formulated at
the quantum level. We will show in this thesis, in various examples,
how it can be used in a regularised path integral and how one can compute
anomalies within this framework.

\section{Unifying different quantisation schemes}
As mentioned in the previous section, two things have to be done
before one can evaluate the path integral. The first is to fix the gauge,
the second is chosing a regularisation (and renormalisation)
scheme. Let us first concentrate on the gauge fixing procedure.

Gauge theories describing physics at higher energies turn out to be
more complicated. It is
then not surprising that the gauge fixing procedure becomes more
involved, as we will now explain. We start with the simplest
gauge theory, namely Maxwell's theory of electromagnetism, which is based
on a $U(1)$ gauge group. A naive way of gauge fixing is
simply adding a term to the action to break the gauge invariance
explicitly. There are of course severe restrictions on the terms that
can be added.
In fact, one must require that the physical quantities are
independent of the chosen terms. It is remarkable that this way of gauge
fixing works in electromagnetism. Examples are the Feynman
gauge or the Landau gauge.

Before turning to other theories, let us mention that this naive gauge
fixing
procedure was later explained by Faddeev and Popov \cite{FP}. They showed
how one
can restrict the measure to integrate only over gauge inequivalent field
configurations. It is based on inserting extra delta functions in the
measure containing an appropriate gauge fixing function.
To do this properly, they
introduced new fields called ghosts and antighosts. Roughly speaking,
these fields eat up the unphysical degrees of freedom. For example, take
again the photon field $A_\mu $, where there are only 2 physical
degrees  of freedom. When the ghost and antighost are included, two
components of the gauge field drop out of the physical
spectrum of the theory.

We want to stress that the gauge fixing procedure should preserve the
relativistic properties (i.e. Lorentz covariance) of the theory. One could
for instance decide from the beginning to eleminate two of the
four components of the photon field $A_\mu $. Doing this, one has no gauge
symmetries anymore but one loses the manifest
relativistic covariance.
More generally, we want the gauge fixing procedure to respect all the
rigid symmetries of the theory. These rigid symmetries can be very useful
for computations in the quantum theory, e.g. when proving
renormalisability or unitarity. One of the main advantages of the
quantisation method used in this thesis is that all rigid symmetries, and
thus relativistic covariance, are kept manifest.

For theories like the Standard Model or General Relativity,
the naive or Faddeev--Popov procedure no longer works ~\footnote{Although
very recently, a formulation for
quantising
Yang--Mills theories without ghosts was given in \cite{BDW}.}. Instead, one
uses the BRST method \cite{BRST}, which can be
applied to theories where the generators of the gauge symmetry form a Lie
algebra. Maxwell theory is a special case in the sense that
the Lie
algebra is commutative. Using ghosts and antighosts, BRS and T
showed that one can replace the local gauge symmetry by the so called
(rigid) BRST symmetry. This symmetry looks the same as the gauge symmetry,
but the local parameter is replaced by a ghost field multiplied with a
constant
anticommuting parameter. One then constructs an action, depending on the
original fields, ghosts and antighosts, that is BRST invariant and has no
further local gauge symmetries. With these BRST
transformation rules one can construct a nilpotent BRST operator.
Physical states are then defined as the elements of the BRST cohomology.

Certain supergravity theories introduce a new complication in the gauge
fixing procedure. In
these theories, the generators of the gauge transformations only form a Lie
algebra when using the
field equations of the action. The gauge algebra is then said to be open.
The usual BRST method is no longer
applicable. However, as was shown in \cite{sugraLagr}
for supergravity
and later on by de Wit and van Holten \cite{deWitvH} for arbitrary open
algebras, one can
generalise the BRST formalism to include also these cases. A disadvantage
of their method is that the extended BRST operator is only
nilpotent on--shell, i.e. modulo field equations. So one can only define
a weak (i.e. on--shell) cohomology.

We still have one scale left, that is the Planck scale where superstring
theory is expected to play an important role. There are two formulations of
superstring theory, namely the RNS (Ramond-Neveu-Schwarz) \cite{RNS} and
the GS
(Green-Schwarz) \cite{GS} formulations. The RNS superstring can easily be
gauge
fixed (using BRST), but in this formalism space-time supersymmetry is not
manifest. On the other hand, the GS superstring preserves
space-time supersymmetry manifestly, but its covariant quantisation is very
difficult. This is because the gauge generators are not linearly
independent. The theory is then said to be reducible. Reducible theories
were already known from quantising theories with an antisymmetric tensor
\cite{Siegelgfg}. They have the property that the matrix $R^i_a$ (see
\eqn{Rmatrix}) has null vectors, called zero modes.
To do a proper gauge fixing, one has to introduce for each zero mode
an extra field, called ghost for ghosts. In the case of the GS
superstring, these zero modes themselves have further zero modes, and this
repeats itself ad infinitum, i.e. the theory is infinitly reducible.

The situation presented above is unsatisfactory. Each time the gauge theory
becomes a little bit more complicated, we have to change our methods to do
the gauge fixing. What we want is one formalism to cover
all gauge theories. It should be applicable to ordinary Maxwell theory as
well as to the Green--Schwarz superstring. Such a formalism exists. It was
developed in the beginning of the eighties by Batalin and Vilkovisky and
is called {\it field--antifield}, {\it BV}, or simply {\it antifield}
formalism \cite{BV1,BV2}. It unifies
all previous (Lagrangian) quantisation methods into a single formalism.
On top of that, the
formalism has an underlying intrinsic and elegant geometrical structure.
In this thesis, we will discuss in detail what the BV formalism is
and how it can be used.

As we have already said, after the gauge fixing, one must choose a
regularisation scheme. There are many choices~: dimensional regularisation,
point splitting, lattice regularisation, Pauli-Villars regularisation (in
combination with higher derivative terms), non-local regularisation, the
BPHZ method, ... . Each method has
its own advantages and disadvantages. We will choose Pauli-Villars (PV)
regularisation \cite{PV}, as it allows for a clear interpretation of all
manipulations on the path integral. Maybe more important, this scheme can
be implemented in the BV formalism in a very transparent way \cite{anombv}.
Using
the antifield formalism in combination with PV regularisation, one can
compute the anomalies of a theory very efficiently. Besides that,
there exists a nilpotent - without using field
equations - operator which generates the quantum symmetries (in the
case there
are no anomalies). Physical states can then be defined as
elements in the cohomology of this operator. All this will be explained in
general and applied to several examples.

\section{Outline}
An essential ingredient of the BV formalism is the doubling of the
complete set of fields.
To each field one associates an antifield with opposite statistics. In the
next chapter,
we will explain what these antifields can be used for. We give a short
overview of what was known before the formalism was actually developed.
Then, we introduce the basic ingredients of the BV formalism and discuss
the
geometrical idea behind it. Basically, this chapter serves as a warm--up
and as a preparation for a systematic treatment of the antifield formalism.

Chapter 3 deals with the general theory. The central objects are the
antibrackets and the extended action, a functional of fields and
antifields that satisfies the
classical master equation. The solution of this equation is guaranteed by
the acyclicity and nilpotency of the Koszul-Tate operator.
We give several examples to
illustrate the general idea behind the construction. Given the extended
action, one can perform canonical transformations between fields and
antifields to do the gauge fixing. Physical
states are defined as the elements of the antibracket cohomology.

As we have already said in the previous section, gauge theories can be
reducible or even infinitly reducible. In the latter case, the extended
action will contain an infinite number of ghosts and antifields. One has to
be very careful when quantising such systems. We therefore continue in
chapter 4 the
investigation of infinitly reducible systems by giving two new examples. We
show how to deal with zero modes that vanish on--shell. The gauge fixing
can be done properly within the BV formalism. Moreover, the second
example
provides a new type of gauge theory not yet discussed in the literature.

Then we consider the quantum theory and the path intergal, in chapter 5. We
present the technique of
Pauli-Villars regularisation and illustrate it with new examples. It is
shown how anomalies arise as a non--invariance of the PV mass term. If
the classical theory has more than one gauge symmetry, one can choose
which symmetry becomes anomalous and one
can move the anomaly from one symmetry to another. This can be done by
adding counterterms, that can be computed using
the interpolating formula. We also show that this interpolation between
anomalies does not work for rigid symmetries.

In chapter 6 we show how to study the path integral in the BV
formalism. For a theory to be free of anomalies, the quantum master
equation must be satisfied. This is an infinite tower of equations which
have to be solved at each order in $\hbar$. For $\hbar=0$, this is the
above mentioned classical master equation. At one loop, we solve this
equation for the example of Yang-Mills theory, again using Pauli-Villars
regularisation. When no solution can be found, the theory suffers from
anomalies. This is the case for chiral $W_3$ gravity. We compute the one
loop anomaly in this model and it turns out that it is antifield dependent.
A mechanism to cancel the one loop anomaly is to introduce background
charges. We show how to implement this idea in the language of the antifield
formalism and remark about higher loop anomalies. Finally, we prove that
our method always gives consistent anomalies.

Another type of field theories are the topological field theories, which
are important for the study of non--pertubative phenomena, like instantons
and solitons.
These theories are characterised by the fact that the path integral is
independent
of the chosen background metric. The BV formalism turns out to be a very
useful scheme to describe topological field theories, as we show in chapter
7. All manipulations from the BRST approach are more transparent in the
BV approach. As examples, we treat topological Yang-Mills theory and
topological Landau-Ginzburg models and show that these theories are indeed
metric independent at the classical level. At higher order in $\hbar $,
extra conditions have to be satisfied.

In the last chapter we discuss the geometry behind the BV formalism.
In the Hamiltonian formalism, one can define momenta and
Poissonbrackets. In a more geometrical language, Poisson brackets are
defined in terms of a symplectic 2 form.
In the Lagrangian framework, the geometry is
based on the existence of an odd symplectic 2 form\footnote{Let us remark
that also in Hamiltonian systems
one can define antifields and antibrackets, see the first reference
of \cite{HennKT}. We prefer however the Lagrangian formalism, since it is
manifestly Lorentz covariant.}. This means that bosons are
conjugated to fermions via the antibracket defined by the odd symplectic
structure.
Instead of going to Darboux coordinates, we will build up the formalism in
a manifestly covariant way.
As an example we consider the case when the manifold is K\"ahler and show
how to solve the master equation.

My contribution to the development of the BV method is
twofold. Firstly, I contributed to the further development of the formalism
itself. At the classical level, one can find new results in sections 3.2
and 3.3, and also in chapter 4. At the quantum level, I showed how
anomalies can easiliy be computed in this framework, see sections 5.1 and
6.1. New results about the geometrical aspects of the antifield formalism
are found in sections 8.5, 8.6 and 8.7.
Secondly, I could use this formalism to investigate
several theories. Especially, I want to mention the computation of
the anomalies in chiral $W_3$ gravity, section 6.3. This provided a new
test of the BV formalism in a complicated model. I also showed how
topological field theories can be constructed using BV theory, see chapter
7.

The aim of all this is
of course to get a clearer insight in gauge theories and to understand and
predict new phenomena in the physics of elementary particles.

The results I obtained, together with several
collaborators, during my PhD can also be found in
\cite{UHWR,Shogo,anomw3,regYM,FSbv,W52,tlg,tsm}.

\chapter{A taste of antifields}
\section{Hamiltonian mechanics}

Let us start with the familiar Hamiltonian formalism. We
consider a classical system with a finite number of degrees of
freedom, say $q^i, i=1,...,N$ and a Lagrangian $L(q,\dot{q})$. In the
Hamiltonian
formalism one associates a momenta conjugated to each coordinate via the
formula
\begin{equation}
p_i=\frac{\partial L}{\partial \dot{q}^i}\ .
\end{equation}
The Hamilatonian $H(p,q)$ is then defined as the Legendre
transformation of the Lagrangian~:
\begin{equation}
H(p,q)=p\dot{q}-L(q,\dot{q})\ .\label{Ham}
\end{equation}
The Hamiltonian is the generator of time
translations. This is best expressed in terms of Poisson brackets
\begin{equation}
\{q^i,p_j\}=\delta ^i_j \ .\label{PoiBr}
\end{equation}
Time evolution of an operator (a function on phase space) $F(p,q)$
is determined via the formula
\begin{equation}
\frac{dF}{dt}=\{F,H\}\ ,
\end{equation}
where we assumed that $F$ contains no explicit time dependence. An operator
is therefore time invariant (or conserved) if it commutes with the
Hamiltonian, i.e.
when its Poisson bracket with $H$ vanishes. From the general definition of
Poisson brackets we also have that $\{F,G\}=-\{G,F\}$. In
particular this means that $\{F,F\}=0$, for any function
$F$, e.g. the Hamiltonian itself.

We will now set up an analogous construction in the Lagrangian formalism.
Motivated by the interest in studying gauge theories, we will replace the
concept of time invariance or evolution under time translations in the
Hamiltonian formalism
by gauge invariance or gauge transformations in the (covariant) Lagrangian
formalism. As the Hamiltonian was generator of time translations, the
question is, what is  the generator of gauge transformations in the
Lagrangian framework ? What are the analogues of the momenta and the
Poisson brackets ?
To answer these questions we consider a field theory with classical fields
$\phi ^i$, which we assume to be bosonic, just like the coordinates $q^i$.
The
classical action $S^0(\phi )$ has $n$ gauge symmetries characterised
by the transformation rules
\begin{equation}
\delta \phi ^i=R^i_a\epsilon ^a\ ,
\end{equation}
where $a=1,...,n$. We will now define an antifield $\phi ^*_i$ conjugated
to each field. The crucial difference with the momenta is that we
will choose the statistics (i.e. the grassmann parity) of the antifield to
be opposite to the statistics of its conjugated field, i.e. $\phi ^{i*}$
is a fermion. The reason of this
will become clear later. Remember that we have already argued in
the introduction that
after the gauge fixing the $n$ local gauge symmetries get replaced by one
rigid fermionic symmetry. The parameters $\epsilon ^a$ are then
replaced by the so called ghosts $c^a$, with statistics opposite to
$\epsilon ^a$. This rigid fermionic (BRST) invariance
is sometimes also called gauge invariance. It will always be clear from the
context if we mean the local (bosonic) or rigid (fermionic) symmetry.

With these antifields, one can
define antibrackets, analogous to \eqn{PoiBr}
\begin{equation}
(\phi ^i,\phi ^*_j)=\delta ^i_j\ .      \label{AntiBr}
\end{equation}
From this it follows that antibrackets change the statistics too. The phase
space of the Hamiltonian formalism is now replaced by the space of fields
and antifields. Functions on the phase space are now functions of fields
and antifields $F(\Phi ,\Phi ^*)$, where
$\Phi $ stands collectively for the classical fields and the ghosts $c^a$.
Also to the latter
we associate a corresponding conjugated antifield $c^*_a$ of opposite
statistics.
The antibrackets of two functions $F$ and $G$ is then defined
analogous to Poisson brackets, but with the momenta replaced by the
antifields. We will explain this in more detail in section 2.7 .
Due to the change in statistics, we will see that
two bosonic
functions (i.e. functions with zero grassmann parity) commute in the
antibracket, i.e.
$(F,G)=(G,F)$. In particular this means that $(F,F)$ will in general be
different from zero.

The analogue of the Hamiltonian will now
be an "extended" action $S(\Phi ,\Phi ^*)$, defined on the space of fields
and antifields. Time evolution was defined by taking the Poisson bracket
with the Hamiltonian. Gauge evolution will be defined by taking the
antibracket with the extended action, i.e.
\begin{equation}
\delta F=(F,S)\ .\label{gtr}
\end{equation}
We still don't know how this action $S$ is defined, i.e. what is the
analogue of the Legendre transformation \eqn{Ham} ? To determine
$S(\Phi ,\Phi ^*)$, we will
require that it is gauge invariant. This is translated
in terms of the antibracket as \footnote{We are working with the DeWitt
convention. Each index represents an internal index as well as a space
time point. When an index is repeated, there is a summation over the
internal index and an integration over space time. For more explanation
and an example, see section 6.}
\begin{equation}
(S,S)=2\frac{\dr S }{\partial \Phi ^A}\frac{\dl S }{\partial \Phi
^*_A}=0\ .\label{SkommaS=0}
\end{equation}
This is indeed a requirement since, in general $(F,F)\neq 0$.
It is the analogue of $\{H,H\}=0$, but for the Hamiltonian this is an
identity because of the properties
of the Poisson brackets. For the extended action, it is a condition that
determines $S(\Phi ,\Phi ^*)$. \eqn{SkommaS=0} is called the {\it
classical master equation}.
Due to the Jacobi identity for the antibracket, see section 7, the
transformation \eqn{gtr} is nilpotent when the classical master equation
is satisfied. The transformation rules
on the classical fields, i.e. $\delta \phi ^i=R^i_ac^a$, can be written as
$\delta \phi ^i=(\phi ^i,S)$. An action that generates this rule is
\begin{eqnarray*}
S(\Phi ,\Phi ^*)=S^0(\phi )+\phi ^*_iR^i_ac^a\ .
\end{eqnarray*}
We see that the antifield $\phi ^*_i$ acts as a source term for the gauge
symmetry. Notice that it
satisfies $(S,S)=0$ if the gauge generators form an ordinary Lie
algebra.
If we have an open gauge algebra, see section 2.5, further terms are needed
to make $S$ gauge invariant.
To determine these terms we will use the techniques of chapter 3.

\noindent
From this section we remember~:
\begin{itemize}
\item Antifields and antibrackets have some analogy with momenta and
Poisson brackets in the Hamiltonian formalism. The important difference is
that antifields have opposite statistics to their fields.
\item The Hamiltonian
is replaced by the extended action
$S(\Phi ,\Phi ^*)$ which is required to be gauge invariant in the sense of
\eqn{SkommaS=0}.
\item Antifields $\phi ^*_i$ are sources for the
gauge transformations of their conjugated fields $\phi ^i$.
\end{itemize}

\section{Maxwell theory}
As an illustration, we consider classical electromagnetism. The photon is
described by gauge fields $A_\mu $ and the classical action is
\begin{equation}
S=\frac{1}{4}F_{\mu \nu }F^{\mu \nu }\ ,
\end{equation}
where $F_{\mu \nu }=\partial _\mu A_\nu -\partial _\nu A_\mu $ and the
gauge transformation is $\delta A_\mu =\partial _\mu \epsilon $. The
action is of course understood to be integrated over space time. The
integration symbol will, as a convention from now on, never be explicitly
written.
To construct the extended action we introduce an antifield $A^*_\mu $ and a
ghost $c$ corresponding to the local gauge symmetry. The solution of the
classical master equation $(S,S)=0$ is now
\begin{equation}
S=\frac{1}{4}F_{\mu \nu }F^{\mu \nu }+A_\mu ^*\partial
^\mu c\ .
\end{equation}
The reader might wonder if this solution is unique. It will be proven in
the next chapter that the solution in the set of fields $\{A_\mu ,c\}$ (and
antifields) is unique up to canonical transformations. These are
transformations that leave the antibracket invariant, just like in
the Hamiltonian formalism, where canonical transformations leave the
Poisson brackets invariant.

Of course one can introduce extra fields and antifields to the minimal set
of $\{A_\mu ,c\}$. For example, one can introduce a {\it non-minimal
sector} by defining a new field antifield pair $\{b,b^*\}$. Then, we
consider the extended action
\begin{equation}
S=\frac{1}{4}F_{\mu \nu }F^{\mu \nu }+A_\mu ^*\partial
^\mu c+\frac{1}{2}b^{*2}\ .\label{extact}
\end{equation}
One can check that this is still a solution of the classical master
equation. This solution is called {\it non-minimal}.
We will explain now why these non-minimal solutions are useful. Within the
total space of fields and antifields, we can do the following canonical
transformation
\begin{eqnarray}
A'^*_\mu = A^*_\mu+\partial _\mu b\nonumber\\
b'^*=b^*-\partial _\mu A^\mu \ .\label{cantr}
\end{eqnarray}
Substituting into \eqn{extact} gives (after dropping the primes)
\begin{equation}
S=-\frac{1}{2}A_\mu \Box A^\mu+b\Box c +A^*_\mu \partial ^\mu
c+b^*\partial _\mu A^\mu +\frac{1}{2}b^{*2}\ ,
\end{equation}
where $\Box \equiv \partial _\mu \partial ^\mu $. This action
still satisfies $(S,S)=0$. The new action without antifields is
the well known gauge fixed action for electromagnetism. The antifield
dependent part determines the rigid, fermionic gauge symmetry, acting
from the right~:
\begin{equation}
\delta A_\mu =\partial _\mu c \qquad \delta c=0 \qquad \delta b=\partial
_\mu A^\mu\ .\label{transfQED}
\end{equation}
It is the symmetry of the gauge fixed action $S=-\frac{1}{2}A_\mu \Box A^\mu
+b\Box c$. $b$ is called the antighost of $c$. One can check that the gauge
fixed action has no local gauge invariances anymore.

\noindent
To summarise we have that~:
\begin{itemize}
\item Antifields can be used to gauge fix an action with a local symmetry.
This is done by introducing non-minimal sectors and performing canonical
transformations.
\item After the gauge fixing, the local symmetry(ies) is (are) replaced by
a single fermionic rigid symmetry, determined by the antifield dependent
part of the extended action.
\end{itemize}

\section{Equations of motion}
Given a theory (an action), one of course wants to know what its physical
spectrum is. In order to compute this,
we have to give a criteria for what a physical state (or
observable) is. For the classical theory, the field equations determine the
evolution of the system and will therefore play an
important role. Given the classical fields $\phi ^i$ (no ghosts, e.g.
only the $A_\mu $ of the previous section) and the classical action $S^0$
(containing no antifields, e.g. only the term $F_{\mu \nu }F^{\mu \nu }$
of the previous section),
they are defined by
\begin{equation}
y_i\equiv \frac{\dr S^0}{\partial \phi ^i}= 0\ .
\end{equation}
The field configurations which satisfy these field equations form the
stationary surface. A classical observable should then be a gauge invariant
function on
the stationary surface.
Let us for the moment consider a theory without
gauge invariance, so there are no ghosts nor antifields of the ghosts
. We will discuss the case of gauge invariance later on. Whenever
a function is proportional to the
field equations, it is vanishing on the stationary surface, and thus it can
not be an observable. So, in order to find the physical states, one must
find a mechanism to divide out the field equations. This is done by
introducing
antifields and considering the operator, called
{\it Koszul--Tate (KT) differential} \cite{Koszul,Tate} defined as
\begin{eqnarray}
\delta _{KT}\phi ^i=0\nonumber\\
\delta _{KT}\phi ^*_i=y_i\ . \label{KTfister}
\end{eqnarray}
The Koszul-Tate differential is nilpotent and thus it defines a
cohomology problem. It is clear that in this way, the field equations
are cohomologically trivial. For a function $f(\phi )$ to be an
observable, it must be in
the cohomology of this operator $\delta _{KT}$.
So, adding to a classical observable a term proportional to field equations
\begin{equation}
A(\phi )\rightarrow A(\phi )+\lambda ^i(\phi )y_i\ ,
\end{equation}
corresponds to adding a $\delta _{KT}$ exact term, namely $\delta
_{KT}(\lambda ^i\phi ^*_i)$, for arbitrary functions $\lambda ^i(\phi )$.
When one deals with gauge theories, the KT
differential must be extended to the ghost (anti)fields. Indeed, when there
are gauge generators $R^i_a$, we have a $KT$ invariant $\phi ^*_iR^i_a$. In
order to kill this in cohomology, we will introduce ghosts $c^a$ and their
antifields $c^*_a$ such that
\begin{equation}
\delta _{KT}c^a=0 \qquad \delta _{KT}c^*_a=\phi ^*_iR^i_a\ .
\end{equation}
The case of gauge invariance is discussed in more detail in the next
chapter.

One can even generalise this to the quantum theory, where the quantum field
equations are the the Schwinger--Dyson equations, i.e. for any function
$F(\phi )$ we have that
\begin{equation}
<F(\phi )y_i+\frac{\hbar}{i}
\frac{\dr F}{\partial \phi ^i}>=0\ ,
\end{equation}
where the $<>$ symbols mean that it must be evaluated under the path
integral. When studying the quantum
observables, one must of course divide out these Schwinger--Dyson terms.
Again, the antifields will be responsible for this
\cite{Henn,AD,DJSD,DJTH}.
We will see in chapter 6 that one can define a quantum analogue of the
Koszul Tate operator, ${\cal S}_q$, which acts on fields and antifields, is
nilpotent and
has the property that (again, in a theory without gauge invariance)
\begin{equation}
{\cal S}_q (F(\phi )\phi ^*_i)=F(\phi )y_i
+\frac{\hbar}{i}\frac{\dr F}{\partial \phi ^i}\ .
\end{equation}
The first term on the right hand side is the classical part and is
generated
by the (classical) Koszul--Tate operator. The second term is a quantum
correction
proportional to $\hbar$. Again, this equation is understood under the path
integral. Of course, when talking about the path integral, a regularisation
prescription is needed. We will take care of this in chapters 5 and 6. It
does not change the princples explained here.

For a function to be a quantum observable, it must be in the cohomology of
${\cal S}_q $. It is clear now that the antifields and ${\cal S}_q $ are
responsible for dividing out the Schwinger--Dyson equations.

\noindent
So, we remember
\begin{itemize}
\item Antifields and the Koszul--Tate differential help us to remove
equations of motion from the physical spectrum of a theory. This
property holds
in the classical as well as in the quantum theory.
\end{itemize}

\section{The work of Zinn-Justin}
Historically, antifields as sources for BRST transformations were
introduced by
Zinn-Justin in \cite{ZJ}\footnote{At that time, the framework of BRST was
not yet well established. Nevertheless, for non abelian gauge theories, the
BRST rules, then
called "super-symmetry" rules or "supergauge Slavnov transformations"
were already written down.}. He was studying the renormalisation
of non-abelian gauge theories, first in a linear gauge and then
generalised to quadratic gauge fixing functions.
After introducing the usual sources $J$ for each field, he also
introduced sources (antifields) for the BRST transformations rules.
Although he did not introduce an antifield for {\it all} the fields $\Phi
^A$, including the ghosts and auxiliary fields, we will
present his results here in a way that each field has an antifield.
Considering the path integral of the extended action $S(\Phi ,\Phi ^*)$,
depending on fields and antifields, he showed that
the Ward-Takahashi identities could be rewritten as an equation for the
effective action $\Gamma $ in a very compact form,
now-called the Zinn-Justin equation~:
\begin{equation}
\frac{\dr \Gamma }{\partial \Phi ^A}\frac{\dl \Gamma }{\partial \Phi
^*_A}=0\ ,\label{ZJE}
\end{equation}
where a sum over all fields (classical fields, ghosts and antighosts,
multipliers),
labelled by $A$ is understood. As the effective action is in general an
expansion in powers of $\hbar$, $\Gamma =S+\hbar\Gamma ^1+...$, this
equation implies at the classical level
\begin{equation}
\frac{\dr S }{\partial \Phi ^A}\frac{\dl S }{\partial \Phi
^*_A}=0\ .\label{CZJE}
\end{equation}
This is precisely the classical master equation \eqn{SkommaS=0} and it
expresses the gauge invariance of the action $S(\Phi ,\Phi ^*)$.

In his papers, he used the symbol $K_A$ for the source of a BRST
transformation, a notation still used nowadays.

He also studied the effect on the path integral
measure under a change of variables of the type $\Phi ^A \rightarrow \Phi
^A +\delta _{BRST}\Phi ^A$. Because the action is invariant under this
transformation, the path integral will be invariant
provided the measure is invariant. He showed that the invariance of the
measure requires
\begin{equation}
\Delta S\equiv (-)^A\frac{\dl}{\partial \Phi ^A}\frac{\dl}{\partial \Phi
^*_A}S=0 \ ,\label{Delop}
\end{equation}
again summed over $A$, and the statistics of the field is denoted by
$\epsilon (\Phi ^A)=A$. It guarantees the absence of anomalies in a gauge
theory. As we will see, this $\Delta $ operator will play a very important
role in the quantum theory, even when the theory is free from anomalies.

\begin{itemize}
\item {\bf All} the fields $\Phi ^A=\{\phi ^i, c^a,...\}$, including ghosts
and auxiliary fields, have an antifield $\Phi ^*_A=\{\phi
^*_i,c^*_a,...\}$.
\item With the use of these antifields, the theory is subjected to
a set of compact equations, \eqn{ZJE}, \eqn{CZJE} and \eqn{Delop}, that
express the Ward--Takahashi identities
and the invariance of the extended action and path integral measure.
\end{itemize}

\section{Open algebras~: the de Wit - van Holten paper}
As was explained in section (2.2), we showed how antifields can be used to
gauge fix an action with a local gauge symmetry. However, Maxwell's theory
could
also be gauge fixed with the Faddeev--Popov method, so we have not really
gained something new. Even for Yang--Mills or gravity theories one can use
the BRST formalism \cite{BRST}, which is a generalisation of the
Faddeev--Popov method. As the examples become more complicated, like in
supergravity theories or $W$ gravities, the existing
quantisation methods are not applicable anymore, and one has to extend,
or even replace them by another method. In contradistinction with
Yang--Mills or ordinary
gravity, the gauge algebra (commutator of two local gauge transformations)
for these models is not a Lie algebra anymore, but differs in two ways from
it. Firstly,
one has to deal with an algebra with field dependent structure functions
and secondly, the algebra only closes when using the field equations of the
action. Such algebras are called open and are characterised by
the equation
\begin{equation}
\frac{\dr R^i_a}{\partial \phi ^k}R^k_b-(-)^{ab}
\frac{\dr R^i_b}{\partial \phi
^k}R^k_a=R^i_cT^c_{ab}-y_kE^{ki}_{ab}\ ,\label{opalg}
\end{equation}
where $T^c_{ba}$ are the structure functions and $E^{ik}_{ba}$ is a
matrix graded antisymmetric in $(ik)$ and in $(ba)$. For the statistics of
the fields, we use the notation $\epsilon (\phi ^i)=i, \epsilon (c^a)=a+1,
\epsilon (R^i_a)=i+a$.
The way to quantise theories with open algebras was shown in the context
of supergravity \cite{sugraFrPvNFer}, first in the Hamiltonian formalism
\cite{sugraHam},
then using Lagrangian methods \cite{sugraLagr}. It was shown that
there appear quartic ghost terms in the gauge fixed action, determined by
the matrix $E^{ik}_{ba}$. These extra ghost dependent terms play an
essential role in finding the correct Feynman rules for the model.
Supergravity was
the first example of a theory where the usual Faddeev--Popov determinant is
insufficient to construct a consistent quantum theory.

Later on, this method was generalised for an arbitrary gauge theory with an
open algebra \cite{deWitvH}. There, it was proven how to construct a gauge
fixed action that is invariant under an appropriate extension of the BRST
transformations. Both the action and transformation rules are
constructed by expanding it in the ghost fields. All the coefficient
functions in this expansion are determined by requiring invariance of
the action. Then, an extensive calculation is needed to check
that the BRST transformation rules are indeed
nilpotent, upon using the field equations of
ghosts and classical fields.

Although in their paper they did not use the concept of antifields,
their method can be rephrased in a very elegant way using the
antifield formalism. This
was shown in the paper of Batalin and Vilkovisky "{\it Gauge algebra and
quantization}" \cite{BV1}. They explicitly constructed an invariant action,
satisfying \eqn{SkommaS=0},
and BRST transformation based on \eqn{gtr}, namely $\delta \Phi ^A=(\Phi
^A,S)$. This operation is nilpotent because of the Jacobi identity and
the classical master equation, without the use of the field equations. The
action starts as
\begin{equation}
S=S^0+\phi
^*_iR^i{}_ac^a+(-)^b\frac{1}{2}c^*_aT^a_{bc}c^cc^b+(-)^{i+a}\frac{1}{4}\phi
^*_i\phi ^*_jE^{ji}_{ab}c^bc^a+...\ .\label{Sopalg}
\end{equation}
So we see that we need terms quadratic in antifields when the algebra is
open. The dots indicate higher order terms in antifields in the case the
gauge algebra is more complicated. A more systematic treatment of the
construction of this extended action will be given in the next
chapter.

\begin{itemize}
\item The antifield formalism is very useful for quantising theories with
open algebras.
\item Using the antifields, one can construct an extension of the BRST
operator which is nilpotent without using the field equations.
\end{itemize}

\section{Reducible theories}
Apart from the gauge algebra, a gauge theory can also be characterised by
its level of reducibility. By this we mean the following: suppose we have
an action and a set of symmetries $\delta \phi ^i=R^i_a\epsilon ^a$. A
theory is said to be reducible if the gauge generators $R^i_a$ are not
linearly independent. In that case there are relations of the form
\begin{equation}
R^i_aZ^a_{a_1}=0\ ,\label{redrel}
\end{equation}
for some functions $Z^a_{a_1}, a_1=1,...,m$. We are working with the DeWitt
convention where each index also carries a space-time point. When there is
a summation over a certain index, there also is an integration over
space-time. To be precise, \eqn{redrel} must be read as
\begin{equation}
\int dyR^{i(x)}_{a(y)}Z^{a(y)}_{a_1(z)}=0\ .
\end{equation}
Of course, one can always single out a set of linearly
independent generators, but then one will lose either locality or
relativistic covariance.

To illustrate this, we give the example of the antisymmetric tensor field.
The classical action is
\begin{equation}
S^0=B_{\mu \nu \rho }B^{\mu \nu \rho }\ ,
\end{equation}
where
\begin{equation}
B_{\mu \nu \rho }=\partial _\mu B_{\nu \rho }+\partial _\rho B_{\mu \nu }
+\partial _\nu B_{\rho \mu}\ .
\end{equation}
The classical fields are $\phi ^i=B_{\mu \nu }(x)$, antisymmetric in
$(\mu \nu )$, and we have a gauge symmetry that follows from the
transformation rule
\begin{equation}
\delta B_{\mu \nu }=\partial _\mu \epsilon _\nu -\partial _\nu \epsilon
_\mu \ .
\end{equation}
The gauge generators can be read off~:
\begin{equation}
R^i_a=(\partial ^x_\mu \delta _\nu ^\rho -\partial ^x_\nu \delta
_\mu ^\rho )\delta (x-y)\ .\label{nonlocgen}
\end{equation}
The notation is that $i=(\mu \nu ,x)$, antisymmetrised in $\mu $ and $\nu
$, and $a=(\rho ,y)$. The upper index on the derivative indicates in which
point the derivative is taken. It is clear that these generators are
linearly dependent with as "zero mode" function
\begin{equation}
Z^a_{a_1}=\partial ^y_\rho \delta (y-z)\ ,
\end{equation}
where the index $a_1$ runs over only one value. Nevertheless it is written
because it carries the space time index $z$.

Other examples of reducible theories are theories with spin 5/2 gauge
fields \cite{5/2gf,BV2}, or,
more recent, some topological field theories \cite{bvtft,FSbv}.

The quantisation of models with antisymmetric tensor fields was first
discussed in \cite{Siegelgfg}.
There it was shown that one had to introduce extra ghosts, called "ghost
for ghosts", in order to obtain a well defined gauge fixed action, with the
correct number of physical degrees of freedom. Indeed, it is clear that
$\delta \epsilon ^a=Z^a_{a_1}\epsilon ^{a_1}$, or in our case $\delta
\epsilon _\mu =\partial _\mu \epsilon $, leaves $\phi ^i$ invariant.
So, just like we introduced a ghost $c^a$ for the parameter $\epsilon ^a$,
we now introduce a ghost for ghost $c^{a_1}$ for the parameter $\epsilon
^{a_1}$, again with opposite statistics.

All this follows nicely from the antifield formalism. When we
construct the extended action, $S=S^0+\phi
^*_iR^i_ac^a+...$, we see that we have indeed an extra symmetry $\delta
c^a=Z^a_{a_1}\epsilon ^{a_1}$. In the same spirit as for the classical
fields, we introduce an antifield for $c^a$, multiply it with its
transformation rule and add it to the extended action
\begin{equation}
S=S^0+\phi ^*_iR^i_ac^a+c^*_aZ^a_{a_1}c^{a_1}+... \ , \label{extactred}
\end{equation}
where the dots indicate terms  with the structure and non-closure
functions. They are such that \eqn{extactred} satisfies the classical
master equation \eqn{SkommaS=0}.

In their paper, "{\it Quantization of gauge theories with linearly
dependent
generators}" \cite{BV2}, Batalin and Vilkovisky showed how to deal with
this
reducibility using the antifield formalism. The example presented above is
only the first and simplest of a hierarchy of theories. It is called a
first stage theory. Indeed, one can
imagine a theory in which also the functions $Z^a_{a_1}$ are linearly
dependent, so that
there are zero modes $Z^{a_1}_{a_2}$ for which one has to introduce further
ghosts for ghosts. Then the theory is said to be of second stage in
reducibility. The procedure can go on to any order of reducibility. The
gauge theory is then characterised by the level of reducibility, i.e. for
an L-th stage theory one has zero modes $Z^{a_{k-1}}_{a_k}, k=1,...,L$, and
for $k=L$, the generators $Z^{a_{L-1}}_{a_L}$ are linearly independent. So,
in \cite{BV2}, a general prescription was given to quantise L-th
stage theories (with an open algebra). The formalism was even applied for
infinite reducible theories. This typically arises in theories with the
so-called $\kappa $-symmetry \cite{Siegelkappa} like the superparticle or
the superstring.
Recently, we discovered a new example \cite{W52} of an infinite
reducible theory, which we will discuss in chapter 4.

\begin{itemize}
\item The antifield formalism enables us to gauge fix reducible theories,
keeping locality and relativistic covariance manifest.
\end{itemize}

\section{Poisson brackets and antibrackets}
In this section we want to clarify some statements made in the first
section of this chapter. It concerns the analogy between momenta and
antifields. As was explained in section 1, an important difference
between antifields and momenta is
that the latter have the same statistics as their conjugated fields while
the former have opposite statistics. This has
consequences concerning the underlying geometry of the Hamiltonian
and Lagrangian systems, and we will discuss this in chapter 8.
In Hamiltonian theory we can set up Poisson brackets in the phase space~:
\begin{equation}
\{F,G\}=(-)^A\frac{\dr F}{\partial \Phi ^A}\frac{\dl G}{\partial \pi _A}-
\frac{\dr F}{\partial \pi _A}\frac{\dl G}{\partial \Phi ^A}\ ,
\end{equation}
where we have denoted the momenta as $\pi _A$. There are left and right
derivatives, because the fields $\Phi ^A$ can be bosonic or fermionic,
with statistics $\epsilon (\Phi ^A)=A$.
This bracket has the following properties, for any two functions $F$ and
$G$ depending on the fields and their momenta~:
\begin{eqnarray}
&&\epsilon (\{F,G\})= \epsilon (F)+\epsilon (G)\\
&&\{F,G\} = -(-)^{FG}\{G,F\}\nonumber\\
&&\{FG,H\}=F\{G,H\}+(-)^{FG}G\{F,H\}\nonumber\\
&&\{F,\{G,H\}\}+(-)^{F(G+H)}\{G,\{H,F\}\}+(-)^{H(F+G)}\{H,\{F,G\}\}=0
\nonumber \ ,
\end{eqnarray}
where the statistics of a function is denoted by $\epsilon (F)=F$.
The last of these equations is the (graded) Jacobi identity.
In particular, we have, for any bosonic function ($\epsilon (B)=0$), that
$\{B,B\}=0$.

Analogous to Poisson brackets, one can now define {\it antibrackets}, for
any two functions $F$ and $G$ depending on fields and antifields, as
\cite{BV1}
\begin{equation}
(F,G)=\frac{\dr F}{\partial \Phi ^A}\frac{\dl G}{\partial \Phi ^*_A}-
\frac{\dr F}{\partial \Phi ^*_A}\frac{\dl G}{\partial \Phi _A}\
.\label{antibr}
\end{equation}
This antibracket satisfies
\begin{eqnarray}
\label{abprop}
&&\epsilon [(F,G)]= \epsilon (F)+\epsilon (G)+1\\
&&(F,G) = -(-)^{(F+1)(G+1)}(G,F)\nonumber\\
&&(FG,H)=F(G,H)+(-)^{FG}G(F,H)\nonumber\\
&&(F,(G,H))+(-)^{(F+1)(G+H)}(G,(H,F))+(-)^{(H+1)(F+G)}(H,(F,G))=0\nonumber
\ .
\end{eqnarray}
The main difference in these two brackets is in the statistics. For the
antibracket we have, for any fermionic function $(F,F)=0$.
Therefore, Poisson brackets are called even and antibrackets odd.

These properties can be used to check the nilpotency of the BRST
transformations \eqn{gtr}. One finds
\begin{equation}
\delta ^2F=((F,S),S)=\frac{1}{2}(F,(S,S))=0\ ,
\end{equation}
upon using the classical master equation.
Also the
Zinn-Justin equation can be written in terms of the antibracket as
\begin{equation}
(\Gamma ,\Gamma )=0\ .
\end{equation}

As in Hamiltonian mechanics, we can also define canonical
transformations. Here they preserve the antibracket rather than Poisson
brackets. An example was given
in section 2, see \eqn{cantr}. We will discuss this issue in more detail in
the next chapter.

\begin{itemize}
\item Using the antifields, one can define antibrackets, analogously to
momenta and Poisson brackets. The difference lies in the statistics:
antibrackets are odd while Poisson brackets are even.
\item Taking the antibracket of a function $F$ with $S$ generates the gauge
transformation of $F$. This transformation is nilpotent due to the Jacobi
identity of the antibracket and the classical master equation $(S,S)=0$.
\end{itemize}

\section{Geometrical interpretation}
The geometrical meaning of the antifield formalism was first discussed
by Witten \cite{Wittgeom}. We will here briefly explain his ideas.
Let us repeat that we have
introduced a $\Delta $ operator in \eqn{Delop}. This operator is nilpotent
\begin{equation}
\Delta ^2=0\ ,\label{delnil}
\end{equation}
due to the fact that antifields and fields have opposite statistics.
Although being a
second order differential operator, it acts as a linear derivation on the
antibracket \cite{BVcan}~:
\begin{equation}
\Delta (F,G) =(\Delta F,G)+(-)^{F+1}(F,\Delta G)\ .\label{delder}
\end{equation}
On the other hand, we also have that
\begin{equation}
\Delta (FG)=(\Delta F)G+(-)^FF(\Delta G)+(-)^F(F,G)\ .
\end{equation}
Witten took this last equation as a definition for the
antibracket.
It measures the failure of $\Delta $ to be a derivation on the algebra of
functions under pointwise multiplication. However, looking at \eqn{delder},
we see that there is another multiplication, namely the antibracket, for
which it is a derivative. The structure of \eqn{delnil} and \eqn{delder}
is similar to the exterior derivative $d$ on the de Rham complex, as was
shown in \cite{Wittgeom}. Let us sketch the arguments.

Consider a (finite) $n$-dimensional manifold with coordinates $x^i$. Let
$TM$
be the tangent bundle with basis $w^i=\frac{\partial }{\partial x^i}$
and denote by $T^*M$ the cotangent bundle with basis $z^i=dx^i$. There is a
natural bilinear form by pairing tangent and cotangent vectors, namely
$(z^i,w_j)=\delta ^i_j$. Associated with this form one can
construct a Clifford algebra by the relations
\begin{equation}
\{z^i,w_j\}=\delta ^i_j\qquad \{z^i,z^j\}=\{w^i,w^j\}=0\ .\label{CA}
\end{equation}
Now, we can look for representations of this Clifford algebra, by means of
creation and annihilation operators. We to have two obvious choices~:

1) The $R$-picture. We regard the $z^i$ as being creation operators,
and the $w_j$ as annihilation operators. To construct a representation,
we can choose the vacuum to be $|0>_R={\bf 1}$, since a constant
is annihilated by
the $w_j$.
The state space
has as a basis the $2^n$ elements $1,z^i,z^i\wedge
z^j,...$,
working on $|0>_R$ . In this representation $w_j$ acts, using \eqn{CA}, by
differentiation \begin{equation}
w_i=\frac{\partial }{\partial z^i }\ . \label{wdz}
\end{equation}

2) The $R'$ picture. Now, we regard the $w_i$ as creation operators, and
the $z^j$ as annihilation operators. As the vacuum, we can take an $n$-form
$|0>_{R'}=dx^1\wedge...\wedge dx^n$, which is clearly annihilated by the
$z^i$. The state space is build up from the $2^n$ states
$|0>_{R'}, w_i|0>_{R'},w_iw_j|0>_{R'},...$, in which the $z^i$ act as
derivatives
\begin{equation}
z^i=\frac{\partial }{\partial w_i}\ . \label{zdw}
\end{equation}

Both representations are isomorphic. They have the same dimension and one
can write each state in the $R'$ picture as a linear combination of the
states in the $R$ picture. For instance, the vacuum
$|0>_{R'}=z^1\wedge...\wedge z^n|0>_R$, etc. .
Since, the $z^i$ are anticommuting, one should think also about $w_i$ as
being anticommuting. Indeed, because of the structure of the $R'$ vacuum,
one cannot work with more than $n$ $w_i$'s on the vacuum. Also one can
convince one self that the $w_i$'s anticommute on $|0>_{R'}$, and that
their square is zero. Therefore, it is natural to interpret $w_i$ as a
fermionic object, which we call the antifield~:
\begin{equation}
x^*_i=\frac{\partial }{\partial x^i}\ .\label{defaf}
\end{equation}
From \eqn{zdw}, it then follows that, in the $R'$ picture
\begin{equation}
dx^i=\frac{\partial }{\partial x^*_i}\ .         \label{dxx*}
\end{equation}

The $R$ picture is of course the standard de Rham
complex of our manifold. Its exterior derivative
\begin{equation}
d=dx^i\frac{\partial }{\partial x^i}
\end{equation}
is nilpotent and acts as a linear derivative on a product of two functions
in the de Rham complex . So, when $F$ and $G$ are of the form
\begin{equation}
F(x,dx)=f(x)+f_i(x)dx^i+...+f_{i_1...i_n}dx^{i_1}\wedge ...\wedge dx^{i_n}\
,
\end{equation}
we have that
\begin{equation}
d(F\wedge G)=dF\wedge G+(-)^FF\wedge dG\ .
\end{equation}
The operator $(-)^F$ changes the odd order components of $F$.

This can be translated in the $R'$ picture. Using \eqn{dxx*}, we find that
\begin{equation}
d=dx^i\frac{\partial }{\partial x^i}=\frac{\partial }{\partial
x^*_i}\frac{\partial }{\partial x^i}=\Delta \ .
\end{equation}
It does not act as a derivative on a product of two functions
$F(x,x^*)G(x,x^*)$, pointwise multiplicated. However it is a derivative
when replacing the wedge product in $R$ with the antibracket in $R'$, as
was shown in \eqn{delder}.

There are some more interesting consequences that follow from this
interpretation. It concerns the path integral $\int e^{\frac{i}{\hbar}S}$.
For finite dimensional manifolds, one can integrate top-forms over the
manifold. These top-forms are
obviously closed. Applying this for functional integrals \footnote{It is
far from trivial how the above statements can be
generalised to the case of infinite dimensional manifolds.}, the
integrandum
in the Feynman path integral should be closed. Writing this in the $R'$
picture, this becomes
\begin{equation}
\Delta (e^{\frac{i}{\hbar}S})=0 \Leftrightarrow -2i\hbar\Delta S+(S,S)=0\ ,
\end{equation}
which expresses the gauge invariance of the theory at
the quantum level.

\begin{itemize}
\item There is a geometrical interpretation of the antifield formalism in
terms of the de Rham complex.
\item The master equation and the condition for the absence of anomalies
simply follow from the closure of a top-form.
\end{itemize}

\chapter{The heart of classical BV theory}
In this chapter, we will further develop the ideas presented in the
previous chapter, in a more systematic way. By now there are,
besides the original papers of Batalin
and Vilkovisky, many texts and reviews on this subject
\cite{Henn,Fisch,bvsb,Hennbook,bvberk,DJTH,anomw3,JordiRev,bvgursey,bvleuv}
. Therefore, I will sometimes be rather short on some points, and refer to
the literature. On the issues where I obtained new results \cite{anomw3}, I
will be more explicit.

\section{Recap and examples}
We denote by $\{\Phi ^A\}$ the complete set of fields.
It includes the ghosts for all the gauge symmetries, and possibly
auxiliary fields introduced for
gauge fixing.
Then
one doubles the space of field variables by
introducing antifields $\Phi ^*_A$, which play the role of canonical
conjugate variables with respect to the antibracket, whose canonical
structure is
\begin{equation}
(\Phi ^A,\, \Phi ^*_B) =\delta ^A{}_B \ ;\qquad (\Phi ^A,\Phi ^B)=(\Phi
^*_A,\Phi ^*_B)=0\ . \label{canbr} \end{equation}
The antifields $\Phi ^*_A$ and fields $\Phi ^A$ have opposite statistics.
In general the antibracket of $F(\Phi ^A,\Phi ^*_A)$ and
$G(\Phi ^A,\Phi ^*_A)$ was defined in \eqn{antibr}.  We will often use
the shorthand notation
\begin{equation}
\partial _A=\frac{\partial }{\partial \Phi ^A}\ ;\qquad
\partial ^A=\frac{\partial }{\partial \Phi ^*_A},
\end{equation}
to write the antibracket as
\begin{equation}
(F,G) =  \dr_A F\cdot\dl{}^A G - \dr{}^A F\cdot\dl_A G\ .
\label{abracket} \end{equation}
$\dr$ and $\dl$ stand for right and left derivatives
The
separating symbol $\cdot$ is often useful to indicate up to where the
derivatives act, if they are not enclosed in brackets.
Note that this antibracket is a fermionic operation, in the sense that the
statistics of the antibracket $(F,G)$ is opposite to that of $FG$.

We assign {\bf ghost numbers} to fields and antifields. These are
integers such that \begin{equation}
gh(\Phi ^*)+gh(\Phi )=-1\ ,\label{ghfaf}
\end{equation}
and therefore the
antibracket \eqn{abracket} raises the ghost number by 1.

We will often perform canonical transformations in this space of fields and
antifields \cite{BVcan}. These are the transformations such that the
new basis again
satisfies \eqn{canbr}. We will also always respect the ghost numbers. It is
clear that interchanging the name field and antifield of a canonical
conjugate pair ($\phi '=\phi ^*$ and $\phi '^*=-\phi $) is such a
transformation. The new
antifield has the ghost number of the old field. From \eqn{ghfaf} we see
then that there is always a basis in which all fields have positive or zero
ghost numbers, and the antifields have negative ghost numbers. We will
often use that basis. It is the natural one from the point of view of the
classical theory, and therefore we will denote it as the `classical
basis'. We will see below that it is not the most convenient
from the point of view of the path integral.

One defines an `extended action', $S(\Phi ^A,\Phi ^*_A)$, of ghost number
zero, whose
antifield independent part $S(\Phi ^A,0)$ is at this point the classical
action, and which satisfies the {\it master equation}
\begin{equation}
(S,S)=0\ .\label{SS0}
\end{equation}
This equation contains the statements of gauge invariances of the classical
action, their algebra, closure, Jacobi identities, ...~.

A simple example is 2--dimensional chiral gravity. The classical action is
\begin{equation}
S^0=\int d^2x\, \left[ -\half \partial X^\mu \cdot \bar \partial X^\mu
+\half h\,\partial X^\mu \cdot\partial X^\mu \right] \ .\label{S0grav}
\end{equation}
In the extended action appears a ghost $c$ related to the conformal
symmetry based on the transformation rules  $\delta X^\mu =\epsilon
\partial X^\mu , \delta h=
\left(\bar \partial -h\partial +(\partial h)\right)\epsilon $. The fields
are then $\Phi ^A=\{ X^\mu ,h,c\}$ and the extended action is
\begin{eqnarray}
&&S=S^0+ \int d^2x\left[ X ^*_\mu \, c\partial X^\mu   +
h^*\left(\bar \partial -h\partial +(\partial h)\right) c\right.\nonumber\\
&&\hspace{3cm} \left.-c^*\,c\,\partial c\right] \ .\label{SextW2}
\end{eqnarray}
Added to the classical action, one finds here the antifields
multiplied with the transformation rules of the classical fields in their
BRST form. One may then check with the above
definitions
that the vanishing of $\left. (S,S)\right|_{\Phi ^*=0}$ expresses the gauge
invariance. The second line contains in the same way the BRST
transformation of the ghost. It is determined by the
previous line and \eqn{SS0} (BRST invariance). Another example is pure
Yang-Mills theory, for which the extended action is
\begin{equation}
S=Tr\{-\frac{1}{4}F_{\mu \nu }F^{\mu \nu }+A_\mu ^*D^\mu
c+\frac{1}{2}c^*[c,c]\}\ ,\label{YMext}
\end{equation}
with the shorthanded notation $A_\mu =A_\mu ^aT_a, c=c^aT_a, F_{\mu \nu
}=\partial _\mu A_\nu -\partial _\nu A_\mu -i[A_\mu ,A_\nu ]$ and $D_\mu
c=\partial _\mu c+[A_\mu ,c]$. The $T_a$
are the generators of a Lie algebra satisfying $[T_a,T_b]=if_{ab}^cT_c,
Tr(T_aT_b)=\delta _{ab}$.

As mentioned, we use the `classical basis' where antifields have negative
ghost numbers, and fields have zero or positive ghost numbers.
For further use, we now give special names to the fields of each ghost
number. The fields of ghost number zero are denoted by $\phi ^i$. Those of
ghost number 1 are denoted by $c^a$, and those of ghost number $k+1$ are
written as $c^{a_k}$ (In this way the index $i$ can also be denoted as
$a_{-1}$ and for $c^a$ we can also write $c^{a_0}$)~:
\begin{equation}
\{\Phi ^A\}\equiv \{ \phi ^i,\,c^a,\,c^{a_1},\ldots \} \ .
\end{equation}

When one starts from a classical action, one directly obtains the above
structure, where $\phi ^i$ are the classical fields, $c^a$ are the ghosts,
and the others are `ghosts for ghosts'.
The antifields thus have negative ghost number and we define the
{\bf antifield number} ($afn$) as
\begin{equation}
afn(\Phi ^*_A)=-gh(\Phi ^*_A) >0 \ ;\qquad afn(\Phi ^A)=0\ .
\end{equation}
With the above designation of names to the different fields we thus have
\begin{equation}
afn(\phi ^*_i)=1\ ;\qquad afn(c^*_{a_k})=k+2\ .
\end{equation}
Then every expression can be expanded in terms with definite antifield
number. E.g. for the extended action, we can define
\begin{equation}
S=\sum_{k=0} S^k\ ;\qquad S^k=S^k\left(c^*_{a_{k-2}},c^*_{a_{k-3}},\ldots
,\phi ^i,\ldots ,c^{a_{k-1}}\right)\ ,  \label{Skdep}
\end{equation}
where the range of fields and antifields which can occur in each term
follows from the above definitions of antifield and ghost numbers, and the
requirement $gh(S)=0$.

A number of questions naturally arises: we have to find a solution to the
classical master equation \eqn{SS0}. But how do we find it ? What are the
conditions for a solution to exist ? Is the solution unique ? All this will
be answered in the next section.

\section{Strategy for solving the classical master equation}

\subsection{Locality, regularity and evanescent operators}
Before looking for a solution, we must specify some conditions our theory
has to fulfil. The starting point is to give a set of fields $\phi ^i$ and
a classical
action $S^0(\phi )$. We will require that $S^0(\phi )$ belongs to the set
of local functionals. These are integrals over local functions. By the
latter we mean a function of $\phi ^i$ and a finite number of their
derivatives.
In general we need
more restrictions, which depend on the theory. E.g. one should specify
whether a square root of a field is in the set of local functions. This
set
should contain at least the fields themselves, and other functions which
appear in the action and transformation rules. For some applications
one may also consider non--local functions (see e.g. \cite{W2WZ,ARW}).

The mathematical framework to study local functions is called jet bundle
theory \cite{jet}. We define $V^0$ to be the space with coordinates
$\{x,\phi ^i(x)\}$. In general, $V^k$ is the space with coordinates
$\{x,\phi ^i,\partial _\mu \phi ^i,...,\partial _{\mu _1...\mu _k}\phi
^i\}$, and is called the k-th jet bundle. These jet bundles are finite
dimensional spaces. For any smooth function $f$, there exists a $k$ such
that $f\in C^{\infty}(V^k)$. An example is the Lagrangian $\cL (
\phi ^i,\partial _\mu \phi ^i,...,\partial _{\mu _1...\mu _k}\phi
^i)$. $S^0$, the integral over the Lagrangian, is then a local functional.
The equations of motions are then defined as
\begin{equation}
y_i=\frac{\rd S^0}{\delta \phi ^i}=\frac{\partial \cL}{\partial \phi
^i}-\partial _\mu \frac{\partial \cL}{\partial (\partial _\mu \phi
^i)}+...+(-)^s\partial _{\mu _1...\mu _s}\frac{\partial \cL}{\partial
(\partial _{\mu _1...\mu _s}\phi ^i)}=0\ ,
\end{equation}
where all derivatives on the Lagrangian are coming from the right.
Together with their derivatives $\partial _\mu y_i=0,...,\partial _{\mu
_1...\mu _s}y_i=0,...$, these equations determine surfaces $\Sigma ^k$ in
$V^k, \forall k$, called stationary surfaces.

As an example we take 2-dim gravity. The field equations are
\begin{eqnarray}
y_\mu &=&-\bar \partial \partial X^\mu +\partial \left( h(\partial
X^\mu ) \right)\nonumber\\
y_h &=& -\ft12 \partial X^\mu \cdot\partial X^\mu \ .
\end{eqnarray}
Obviously $\Sigma ^0=V^0$, since there are no relations implied
by the field equations in $V^0$. The field equation of $h$, however, is an
equation in $V^1$. For a Euclidean metric in the space of the $X^\mu $'s,
it implies $\partial X^\mu =0$ which determines $\Sigma ^1$. $\Sigma ^2$ is
then determined by the set of equations $y_h=0,\partial y_h=0, {\bar
\partial }y_h=0$ and by $y_{X^\mu }=0$. Acting several times with the
derivatives $\partial $ and ${\bar \partial }$ on the field equations, one
determines the surfaces $\Sigma ^k$.
Suppose the functions $y_i$ are elements of $V^{s_i}$. Then, for a
function $f\in C^{\infty}(V^k)$, we define the symbol (weakly zero) by
\begin{equation}
f\approx 0 \Longleftrightarrow\ f=h^iy_i+h^{i\mu }\partial _\mu
y_i+...+h^{i\mu _1...\mu _n}\partial _{\mu _1...\mu _n}y_i\ , \label{weak}
\end{equation}
where $h^i,h^{i\mu },...$ are functions on
$V^{k-s^i},V^{k-s^i-1},...$.

In jet bundle theory, one is not working with the DeWitt convention. When
an index is written twice, there is no space-time integral but only a sum
over the internal index. It is amusing to translate all expressions to
expressions where the DeWitt convention is used. For instance, the
weakly zero symbol can, in DeWitt language, be defined as
\begin{equation}
f\approx 0 \Longleftrightarrow\ f=H^iy_i\ .
\end{equation}
A space-time integral is now included in the summation over $i$, i.e.
$f(x)=\int dy H^{i(y)}(x)y_{i(y)}$. The
coefficients $H^i$ are in general distributions, like delta functions or
derivatives on delta functions, which are not included in the jet
bundles. Therefore, the DeWitt convention is not appropriate for jet
bundle theory. The relation
between the two conventions in the above example is
\begin{equation}
H^{i(y)}(x)=h^i(x)\delta (x-y)+h^{i\mu }\partial _\mu ^x\delta (x-y)+... \
.
\end{equation}
The functions on the stationary surface are
denoted by $C^{\infty}(\Sigma ^k)$. It can happen that all functions that
vanish on the
stationary surface are weakly zero. Then the theory is said to be regular.
This is however not always the case, like in our example of 2-dimensional
gravity. There we have that (for a Euclidean metric in the space of the
$X^\mu $'s) $\partial X^\mu $ is vanishing on the stationary surface
$\Sigma ^1$, but nevertheless it is not (in a local way) proportional to
field equations in the sense of \eqn{weak}. Operators vanishing on
$\Sigma ^k$ which are not weakly zero will be called
evanescent operators \cite{anomw3}. To construct the extended action, we
will not restrict ourselves to the case of regular
theories. This restriction was imposed in \cite{BVexist,HennKT,Hennloc}.
Instead, we will extend the class of theories to those that also
contain evanescent operators, which will be added to the functions on the
stationary surface, see \cite{anomw3}.

\subsection{Completeness and properness}

There is another requirement which has to be satisfied, called
completeness. It is needed in order to guarantee a solution to the master
equation that can be used to construct a gauge fixed action. In words, it
means that we have to take into account {\bf all} gauge symmetries of the
classical action $S^0$. Indeed, $S^0$ itself obviously satisfies the master
equation $(S^0,S^0)=0$. But when $S^0$ has gauge symmetries, it can
not be used for the path integral. Therefore, we require completeness.

In jet bundle space, the symmetry transformations of the fields can be
written as
\begin{equation}
\delta \phi ^i=r^i_a\epsilon ^a+r^{i\mu }_a\partial _\mu \epsilon
^a+...+r^{i\mu _1...\mu _t}\partial _{\mu _1...\mu _t}\epsilon ^a\ ,
\end{equation}
for some fixed value of $t$. The functions $r^i_a,r^{i\mu }_a,...$ depend
on the fields and their derivatives up to some finite order.
Using the DeWitt notation, this takes the more compact form
$\delta \phi ^i=R^i_a\epsilon ^a$. Gauge invariance is expressed by the
fact that the field
equations $y_i$ are not all independent. In jet bundle space, this means
\begin{equation}
y_ir^i_a-\partial _\mu (y_ir^{i\mu _a})+...+(-)^t\partial _{\mu _1...\mu
_t}(y_ir^{i\mu _1...\mu _t})=0\ ,
\end{equation}
or, written in DeWitt notation
\begin{equation}
y_iR^i_a=0\ .
\end{equation}
The condition of completeness can now be formulated as~: if for any
set of local functions $T^i(x)$
\begin{equation}
y_i T^i(x)=0\ \Longrightarrow \ T^i(x) = R^i{}_a \mu ^a(x)
+y_j v^{ji}(x) \label{allsym}\ ,
\end{equation}
where $\mu ^a(x)$ and $v^{ij}(x)$ are local functions, the latter being
graded antisymmetric
\begin{equation}
v^{ij}(x)=(-)^{ij+1}v^{ji}(x) \ . \label{vantis}
\end{equation}
To avoid misunderstanding in the notation, I will once more write explicit,
e.g. $y_iT^i(x)=\int dz y_{i(z)}T^{i(z)}(x)$, etc. .

Another requirement on the extended action $S(\Phi ,\Phi ^*)$ is the
{\bf properness condition}, which we will now explain.
The master equation $(S,S)=0$ implies relations typical for a general gauge
theory. In the collective notation of fields and antifields, $z^\alpha
=\{\Phi ^A,\Phi ^*_A)\}$, the master
equation takes the form
\begin{equation}
 \dr_\alpha S\cdot\omega ^{\alpha \beta }\cdot\dl_\beta S = 0 \ \mbox{
with }\ \omega ^{\alpha \beta }=(z^\alpha ,z^\beta )\ .\label{masteral}
\end{equation}
We also introduce the Hessian
\begin{equation}
Z_{\alpha \beta }\equiv P\left( \dl_\alpha  \dr_\beta S\right)\ ,
\label{defZal}
\end{equation}
where $P$ projects onto the surface  $\{\Phi ^*_A=c^{a_k}=0\}$.
Because $S$ has zero ghost number, $Z_{\alpha
\beta }$ is only non--zero if $gh(z^\alpha
)+gh(z^\beta )=0$. This implies that its non--zero elements are $
Z_{ij}= S^0_{ij}=\dl_i \dr_jS^0$, $Z^i{}_{a}$, $Z^a{}_{a_1}$, ...,
$Z^{a_k}{}_{a_{k+1}}$.
Note that upper indices of $Z$ appear here because derivatives are taken
w.r.t. antifields $\Phi ^*_A$. Of course also elements as
$Z_{a_{k+1}}{}^{a_k}$ are non--zero, being the supertransposed of the above
expressions.

{}From the ghost number requirements we can determine that $S$
is of the form
\begin{equation}
S=S^0+\phi ^*_i Z^i{}_{a}(\phi ) c^a+\sum
_{k=0}c^*_{a_k}Z^{a_k}{}_{a_{k+1}}c^{a_{k+1}}+\ldots \ ,\label{SinZ}
\end{equation}
where $\ldots$ stands for terms cubic or higher order in fields of
non--zero ghost number.
Considering the master equation at antifield number zero, we obtain
\begin{equation}
0=\ft12\left.(S,S)\right|_{\Phi ^*=0}=\dr_i S^0\cdot
\dl{}^i S^1=y_iZ^i{}_{a}c^a\ ,
\end{equation}
from which it follows that $Z^i_a=R^i_a$.

Taking two derivatives of \eqn{masteral}, and applying the projection
$P$, we get
\begin{equation}
Z_{\alpha \gamma }\omega ^{\gamma \delta } Z_{\delta \beta }=(-)^{\alpha
+1}y_i \ P\left(\dl{}^i\dl_\alpha  \dr_\delta S\right) \ .  \end{equation}
This says that the Hessian $Z_{\alpha \beta }$ is weakly nilpotent.
Explicitly, we obtain (apart from the derivative of $y_iR^i_a=0$)
\begin{eqnarray}
&&R^{i}{}_{a}Z^a{}_{a_1}=2 y_j f^{ji}{}_{a_1}
\approx 0 \label{RZisf}\\
&&Z^{a_k}{}_{a_{k+1}}Z^{a_{k+1}}{}_{a_{k+2}}\approx 0\ ,\label{ZZa0}
\end{eqnarray}
where
\begin{equation}
f^{ji}{}_{a_1}=\ft12(-)^i \ P\left( \dl{}^j \dl{}^i\dr_{a_1}S\right)\ .
\label{deffa1}\end{equation}
In the first relation the r.h.s. is written explicitly because it exhibits
a graded antisymmetry in $[ij]$.

As the latter is weakly nilpotent its maximal (weak) rank is
half its dimension. The properness condition is now the requirement that
this matrix is
of maximal rank, which means that for any (local) function $v^\alpha (z)$
\begin{equation}
Z_{\alpha \beta }v^\beta \approx 0 \Longrightarrow v^\beta \approx \omega
^{\beta \gamma }Z_{\gamma \delta }w^\delta\ ,\label{proper}  \end{equation}
for a local function $w^\delta $.

The properness conditions \eqn{proper} can now be written explicitly as
\begin{eqnarray}
S^0_{ij}v^j\approx 0 &\Longrightarrow& v^j\approx
R^j{}_{a}w^a\nonumber\\
R^i{}_{a}v^a\approx 0 & \Longrightarrow& v^a\approx
Z^a{}_{a_1}w^{a_1}\nonumber\\
Z^{a_k}{}_{a_{k+1}}v^{a_{k+1}}\approx 0 &\Longrightarrow &
v_{a_{k+1}}\approx Z^{a_{k+1}}{}_{a_{k+2}}w^{a_{k+2}} \nonumber\\
&\mbox{and}&\nonumber\\
v_i R^i{}_{a}\approx 0 &\Longrightarrow& v_i \approx w^jS^0_{ji}\nonumber\\
v_a Z^a{}_{a_1}\approx 0 &\Longrightarrow& v_a\approx w_i
R^i{}_{a}\nonumber\\
v_{a_k} Z^{a_k}{}_{a_{k+1}}\approx 0 &\Longrightarrow& v_{a_k}\approx
w_{a_{k-1}} Z^{a_{k-1}}{}_{a_k} \ .
\label{properZ}\end{eqnarray}
The second group of equations follows from the first group, using that the
right and left ranks of matrices are equal. The first one implies
\eqn{allsym} if there are no non--trivial symmetries
which vanish at stationary surface. By the latter we mean that there would
be relations
\begin{equation}
y_i y_j T^{ji}=0 \ \mbox{ where }y_jT^{ji}\neq R^i{}_{a}\epsilon ^a
\end{equation}
and $T^{ij}$ is graded symmetric. If such non--trivial symmetries would
exist, then \eqn{allsym} is an extra requirement with $T^i$ replaced by
$y_j T^{ji}$.
This subtlety has been pointed out in \cite{anomw3}. We will come back
to this in the following section and in chapter 4.

\subsection{The acyclicity and nilpotency of the Koszul-Tate operator}

As mentioned in the previous chapter, the Koszul-Tate (KT) operator was
introduced to kill the field equations in cohomology. As a consequence, the
KT operator provides a resolution of the functions on the stationary
surfaces $C^{\infty}(V^k), \forall k$. In \eqn{KTfister}, the KT operator
was only defined on the antifields $\phi ^*_i$ (similar to the construction
of Koszul \cite{Koszul}), but not yet on the antifields of the ghosts
$c^*_{a_k}$ (in the same spirit of Tate \cite{Tate}). The purpose of
this section is to define the KT operator in the full space of antifields
such that it is nilpotent
\begin{equation}
\delta _{KT}^2=0\ ,
\end{equation}
and acyclic on the space of local functions. Acyclicity here is defined as
\begin{equation}
\delta _{KT}F(\phi ,\Phi ^*)=0 \Longrightarrow F(\phi ,\Phi ^*)=\delta
_{KT}H(\phi ,\Phi ^*) +f(\phi )\ ,\label{acyclKT}
\end{equation}
for some $H$ and where $f$ is a function on the stationary surface or an
evanescent operator. From now on, when we talk about functions on the
stationary surface, evanescent operators will always be included.
These two properties (nilpotency and acyclicity) will guarantee the
existence and uniqueness (modulo
canonical transformations) of the solution of the master equation. The
proof was given in different steps. First, in \cite{BVexist}, existence and
uniqueness theorems were proven for irreducible gauge algebras,
without using the KT differential.
The shortcoming of these proofs is that
they did not prove the locality and Lorentz covariance of the solution.
Second, in \cite{HennKT},
the KT operator was first introduced in the antifield formalism. They
proved the existence and uniqueness of the extended action for reducible
theories. Also here, the question of locality
and Lorentz covariance of the solution was not addressed.
The problem of locality was solved in \cite{Hennloc}, using the KT operator
and jet bundle theory. A proof where locality and Lorentz covariance is
manifest was given in \cite{anomw3,bvleuv}.

The proof of the nilpotency and acyclicity goes
perturbatively in the level of antifields. We will not give all the proofs
here since this is quite tedious. We refer therefore to the above mentioned
papers. Instead, to gain some insight in the construction, we will give
some examples and clarify some points, not well discussed in the
literature so far.

Let us start with functions at antifieldnumber zero, i.e.
without antifields. The KT operator is clearly nilpotent since it does not
work on the fields, $\delta _{KT}\phi ^i=0$. It is not acyclic however,
because the field equations are not $\delta _{KT}$-exact. For acyclicity,
see \eqn{acyclKT}, they should
be. Therefore, we introduce the first level of antifields, $\phi ^*_i$, and
kill the field equations in cohomology by defining
\begin{equation}
\delta _{KT}\phi ^*_i=y_i\ .
\end{equation}
Since we have now introduced these antifields, we can study the nilpotency
and acyclicity for functions at antifieldnumber 1, i.e. linear in
antifields $\phi ^*_i$. In general, such a function takes the form $F=\phi
^*_iK^i$.
$\delta _{KT}$ is nilpotent on these functions, so only acyclicity has to
be checked. In order that $\delta _{KT}F=0$, one must have that $y_iK^i=0$.
There are two types of solutions to this equation. The first is that $K^i$
is proportional to the gauge generators, i.e. $K^i=R^i_a\epsilon ^a$. The
second is that $K^i=
y_jv^{ji}$, with $v^{ji}$ graded antisymmetric. When
$K^i=y_jv^{ji}$, then $F$ is KT exact, namely $\phi
^*_iy_jv^{ji}=\frac{1}{2}\delta
_{KT}(\phi ^*_i\phi ^*_jv^{ji})$, so this solution does not spoil
acyclicity.
The term $\phi ^*_iR^i_a$ can however not be written as the KT of
something. Therefore, to restore acyclicity, we have to introduce a new
field-antifield pair $\{c^a,c^*_a\}$ and we define
\begin{equation}
\delta _{KT}c^*_a=\phi ^*_iR^i_a\ .
\end{equation}
The $c^a$ are called ghosts, and there are as many ghosts as there
are gauge generators $R^i_a$. The above analysis also shows that
symmetries, graded antisymmetric in the field equations, do not need a
ghost. However, what happens with symmetries graded symmetric in the field
equations ? According to our KT analysis, one should introduce a ghost for
it, since these symmetries should be included in the set of $R^i_a$. This
result also follows from requiring completeness, i.e. \eqn{allsym}. Notice
that it is NOT implied by requiring only properness. This we mentioned
already at the end of the previous section. Unfortunately, we have not yet
found a good example in which there are symmetries, graded symmetric in the
field equation. We will come back to this point in chapter 4.

Now, we go to functions of antifieldnumber 2. Candidate invariants are
of the form
\begin{equation}
F=c^*_aZ^a+\phi ^*_i\phi ^*_jf^{ji}\ ,
\end{equation}
where $f^{ij}$ is graded
antisymmetric in $i$ and $j$. For simplicity, I will assume all the
classical fields
are bosonic and all the ghosts are fermionic. First, the nilpotency of
$\delta _{KT}$ on $F$ follows from the construction at
antifieldnumber 1. Second, for $F$ to be KT-invariant, we should have that
\begin{equation}
R^i_aZ^a=2y_jf^{ji}\approx 0 \ .\label{closeq}
\end{equation}
We will label the solutions to this equation with the index $A_1$.
The space of all pairs $(Z^a_{A_1},f^{ij}_{A_1})$ satisfying \eqn{closeq}
determines the space of all KT invariants at antifieldnumber two. The
question is now~: which of these invariants is KT exact by using only
functions of $\phi ^i,\phi ^*_i,c^a,c^*_a$ ? To answer this, we must
solve
\begin{equation}
F(Z^a_{A_1},f^{ij}_{A_1})=
\delta _{KT}[c^*_a\phi ^*_iK^{ia}+\frac{1}{3}\phi ^*_i\phi ^*_j\phi
^*_kX^{ijk}]\ ,
\end{equation}
for some functions $K^{ia}, X^{ijk}$.
The functions $X^{ijk}$ are of course
antisymmetric in the three indices. We will split our KT invariant solution
space into two spaces, namely the exact ones and the non exact ones. Then
our index $A_1$ runs over the KT exact invariants, which we label by
$\alpha _1$, and the non-exact ones, which we label by $a_1$.
The
exact pairs
are of the form\footnote{Antisymmetrisation is done with weight 1/2, i.e.
$[ij]=\frac{1}{2}(ij-ji)$.}
\begin{eqnarray}
Z^a_{\alpha _1}&=&y_iK^{ia}_{\alpha _1}\nonumber\\
f^{ij}_{\alpha _1}&=&R^{[j}_aK^{i]a}_{\alpha _1}+y_kX^{kij}_{a_1}\
.\label{expair} \end{eqnarray}
For all the invariant pairs that are not in this subspace, we introduce
ghosts for ghosts $c^{a_1}$ and antifields $c^*_{a_1}$ with
\begin{equation}
\delta _{KT}c^*_{a_1}=c^*_aZ^a_{a_1}+\phi ^*_i\phi ^*_jf^{ji}_{a_1}\ .
\end{equation}
Does all this implies that on shell vanishing zero modes do not need ghosts
for
ghosts ? Not in general~: \eqn{expair} shows that if $Z^a_{a_1}$ does not
vanish on shell, the zero mode is necessarily not exact, so it needs a
ghost for ghosts. However, it can happen that we find a KT invariant with
a weakly vanishing zero mode matrix $Z^a_{a_1}$, but the $f^{ij}_{a_1}$ are
not of the form \eqn{expair}. Again, this property does not follow
from the properness condition. For a weakly vanishing null vector $v^a$ of
$R^i_a$, the properness condition is always satisfied. So, $v^a$ does
not need to be proportional to $Z^a_{a_1}$, see \eqn{properZ}.
Examples of this will be given in the next chapter.

Finally, we will discuss functions at antifieldnumber three. They are of
the form
\begin{equation}
F=c^*_{a_1}Z^{a_1}+c^*_a\phi ^*_if^{ia}+\frac{1}{3}\phi ^*_i\phi ^*_j\phi
^*_kX^{ijk}\ ,
\end{equation}
where $X^{ijk}$ is antisymmetric in its three indices. Requiring that
$\delta _{KT}F=0$ leads to two equations~:
\begin{eqnarray}
Z^a_{a_1}Z^{a_1}&=&-y_if^{ia}\approx 0 \label{Z1Z2weakzero}\nonumber\\
f^{ji}_{a_1}Z^{a_1}+R^{[i}_af^{j]a}&=&-y_kX^{ijk}\approx 0\
.\label{eqclos3} \end{eqnarray}
The first one is that we indeed have a zero mode. The second condition is a
new
one. It is a condition on the functions $f^{ia}$ and can be compared
with the condition at antifieldnumber two, namely that
$f^{ij}_{a_1}$ is antisymmetric in $i$ and $j$. If the second condition is
not satisfied, we have no KT invariant and so, we certainly need not to
introduce ghosts for ghosts for ghosts\footnote{It is not clear yet if one
can find a basis in which this equation is automatically satisfied.
Also at higher antifieldnumbers, one would find more conditions to
make
KT invariants. It is not clear if these are really extra conditions, or if
they can automatically be satisfied. This problem is overlooked in the
literature and is currently
under study, in collaboration with K. Thielemans.}. Again, we can label the
space of solutions with the index $A_2=\{a_2,\alpha _2\}$. Here $a_2$ runs
over the non-KT-exact invariants, and $\alpha _2$ runs over the KT exact
ones.
As an exercise one can find out which of these invariants are KT exact.
This will give further conditions on the $Z^{a_1}$ and
$f^{ia}$.

It is clear how this construction can be continued at higher
antifieldnumbers. Each time one encounters a KT
invariant which is not exact, one must kill it by
introducing a new field antifield pair.
The KT operator then takes the form
\begin{equation}
\delta _{KT}c^*_{a_{k+1}}=c^*_{a_k}Z^{a_k}_{a_{k+1}}+M_{a_{k+1}}(\phi ,\phi
^*,c^*_a,...,c^*_{a_{k-1}})\ ,
\end{equation}
where the function $M_{a_{k+1}}$ must be such that $\delta _{KT}$ is
nilpotent.
Again, it must be emphasised that the general proofs follow a different
strategy than the above examples. There, the acyclicity is proved for
functions of fields and antifields up to a certain $c^*_{a_k}$, and they
can have arbitrary antifieldnumber.

The acyclicity applies for local ($x$--dependent) functions. It
does not apply in general for integrals. Indeed, consider the following
example \begin{eqnarray}
S&=&\int d^dx\, \ft12\partial _\alpha  X^\mu \cdot \partial ^\alpha  X^\mu
\ ,\ \mbox{ with }\mu =1,2\  \  \alpha =1,...,d\nonumber\\
F&=&\int d^dx\, \left( X^*_1 X^2 -X^*_2 X^1\right)\ \rightarrow \ \dkt F=0\
. \end{eqnarray}
Nevertheless, $F$ can not be written as $\dkt G$. The violation of
acyclicity for integrals is due to
rigid symmetries.
However, for $F$ a local integral (integral of a local function) of
antifield number 0 (and thus obviously $\dkt F=0$),
which vanishes on the stationary surface, we have by definition $F=\int y_i
F^i$ and $F=\dkt\int (\phi ^*_i F^i)$.

The acyclicity was proven here for functions independent of ghosts.
If one considers local functions $F(\Phi ^*,\phi ,c)$ depending
on ghosts, then the acyclicity can be used when we first expand in
$c$. Therefore the modified acyclicity statement is then
\begin{eqnarray}
\dkt F(\Phi ^*,\phi ,c) =0 &\Rightarrow & F=\dkt H + f(\phi,c
)\nonumber\\ \mbox{and if }f(\phi ,c)\approx 0&\Rightarrow & f=\dkt
G(\Phi ^*,\phi ,c)\ , \label{acyclc} \end{eqnarray}
where, as mentioned before, $\approx $ stands for using the field
equations of $S^0(\phi
)$ for $\phi $, while the ghosts $c$ remain unchanged.
If $F$ is an integral, where the
integrand contains ghosts, then we apply the acyclicity to the
coefficient functions of the ghosts which are local functions. If
$gh(F)>0$ (or even if just $puregh(F)>0$), then each term can be treated in
this way and the statement
\eqn{acyclc} holds even when $F$ is a local functional.
Also if $gh(F)=0$ then any term has either a ghost, or it just depends on
$\phi ^i$ in which case the acyclicity statement also applies for integrals.
So, to conclude, the KT operator is only not acyclic
on local functionals of negative ghostnumbers.

\subsection{Construction of the extended action~: $W_3$ gravity as an
example}
In this section we will summarise the proof of the
solution of the master
equation $(S,S)=0$, as given in the second paper of \cite{HennKT} and also
in \cite{Fisch,bvleuv} . The general technique will be illustrated
in an example, namely that of chiral $W_3$ gravity \cite{Hull}. Its
classical action is \footnote{We
will use the notations $\partial =\partial _+$ and $\bar \partial
=\partial _-$, where $x^\pm=\rho (x^1\pm x^0)$, and we leave the factor
$\rho $ undetermined.\label{fn:convrho}}
\begin{equation}
S^0=-\ft{1}{2}\pl \bpl +\ft{1}{2}h\pl
\pl +\ft{1}{3}\ds B\pl\plb \plg \,\label{S0W3} \end{equation}
where $\ds$ is a symmetric tensor satisfying the nonlinear identity
\begin{equation}
d_{\mu (\nu \rho }d_{\sigma )\tau  \mu }=\kappa \delta _{(\nu
\sigma }\delta _{\rho )\tau  }\ .\label{dsym}
\end{equation}
The $()$ indicate symmetrisation, and $\kappa $ is some arbitrary,
but fixed parameter. The general solution
of this equation was found in \cite{dsymb}. We will discuss
more about this in section 6.3.3 .
The model contains $n$ scalar fields $X ^\mu
, \mu =1,...,n$ and two gauge fields $h$ and $B$, which imply the
existence of two gauge symmetries
\begin{eqnarray}
\delta X^\mu&=&\pl \epsilon +\ds \plb \plg \lambda \nonumber\\
\delta h&=&(\na ^{-1}\epsilon )+\ft{\kappa }{2}\pl \pl
(D^{-2}\lambda )\nonumber\\
\delta B=&=&(D^{-1}\epsilon )+(\na ^{-2}\lambda )\ ,
\end{eqnarray}
for local parameters $\epsilon ,\lambda $. We have made the shorthand
notation
\begin{eqnarray}
\na ^j=\bar{\partial }-h\partial -j(\partial h)\nonumber\\
D^j=-2B\partial -j(\partial B)\,.
\end{eqnarray}

The general idea is now to expand the master equation according to its
antifield number~:
\begin{equation}
B^n\equiv (S,S)^n=\delta _{KT}S^{n+1}+D^n(S^0,...,S^n)\ , \label{split}
\end{equation}
i.e. one can split it into two pieces~: one that contains $S^{n+1}$ and the
other that contains terms only depending on $S^0,...,S^n$. One can show
that the first term in this split indeed involves the KT operator, as
introduced in the previous subsection. One can
also show that the terms $S^k;k\geq n+2$ do not contribute to the master
equation at antifieldnumber $n$. E.g.
\begin{equation}
(S,S)^1=2(S^0,S^2)+2(S^1,S^2)+(S^1,S^1)\ .
\end{equation}
There is e.g. no term $(S^0,S^3)$, since one has to take the derivative in
$S^0$ w.r.t. a field $\phi ^i$, so then one must take the derivative in
$S^3$ w.r.t $\phi ^*_i$. The result is of
antifieldnumber two, which does not contribute to $B^1$.

At the lowest level, $B^0=0$, we find $D^0=0$ and $y_i\dl ^iS^1=0$ which
gives as a solution
\begin{equation}
S^1=\phi ^*_iR^i_ac^a\ .
\end{equation}
For $W_3$ this gives
\begin{eqnarray}
S^1\ =\ \phi ^*_iR^i{}_ac^a&=&
X _\mu ^*\left[\pl c+\ds \plb \plg u\right]\nonumber\\
& &+h^*\left[(\na ^{-1}c)+\ft{\kappa }{2}\pl \pl
(D^{-2}u)\right]\nonumber\\ &&+B^*\left[(D^{-1}c)+(\na ^{-2}u)\right]\ ,
\label{S1W3}\end{eqnarray}
where $c$ and $u$ are the ghosts for the $\epsilon $ and $\lambda $
symmetries, and are both fermionic.

The rest of the proof goes by induction. We assume that the master equation
is solved up to antifieldnumber $n$. This has determined the extended
action up to $S^{n+1}$. To solve $B^{n+1}=0$ for $S^{n+2}$, one needs to
know that $\delta _{KT}D^{n+1}=0$. This can be proven from the Jacobi
identity $(a,(a,a))=0$, with $a=\sum S^k$, where $k$ runs from $0$ to
$n+1$. Then one can use the Koszul-Tate acyclicity for local functionals
containing at least one ghost to write
\begin{equation}
D^{n+1}=\delta _{KT}U^{n+2}\ .
\end{equation}
There can not be a function on the stationary surface, since $D$ has
antifieldnumber greater than zero. Doing so, the master equation reduces to
\begin{equation}
\delta _{KT}(S^{n+2}+U^{n+2})=0\ .
\end{equation}
The solution is
\begin{equation}
S^{n+2}=\delta _{KT}(c^*_{a_{n+1}}c^{a_{n+1}})-U^{n+2}+\delta
_{KT}V^{n+3}\ .\label{Sn} \end{equation}
Adding different trivial terms $\delta _{KT}V^{n+3}$ gives
different solutions for the extended action, which are
related by canonical transformations. The first term on the
r.h.s. in \eqn{Sn} must also be add if further zero modes
exist.

In our example, the next step is to solve $B^1=0$. This corresponds to
the equation $\delta _{KT}S^2+D^1=0$. Now, we can compute
\begin{eqnarray}
D^1&=&\phi ^*_i\dr _jR^i_a R^j_bc^bc^a(-)^{a+1}\nonumber\\
&=&\frac{1}{2}\phi
^*_i[R^i_cT^c_{ab}-y_kE^{ki}_{ab}]c^bc^a(-)^{a+1}\nonumber\\
&=&\delta
_{KT}[\frac{1}{2}c^*_aT^a_{bc}c^cc^b(-)^{b+1}]-(-)^{k+a}\frac{1}{4}\delta
_{KT}[\phi ^*_k\phi ^*_iE^{ik}_{ab}c^bc^a]\ ,
\end{eqnarray}
which gives back the form \eqn{Sopalg} without the dots.
For $W_3$ one can compute the gauge algebra and finds
\begin{eqnarray}
S^2 &=&c^*\left[(\partial c)c+\kappa \pl \pl (\partial
u)u\right]+u^*\left[2(\partial c)u-c(\partial u)\right]\nonumber\\
& &-2\kappa X_\mu ^*h^*(\partial u)u\pl \ .
\label{S2W3}
\end{eqnarray}
This form of the extended action was already found in
\cite{Hullmatter,SSvNMiami}.
One can check that $D^2=0$, and so $S^3=0$. The total extended action is
then
\begin{equation}
S=S^0+S^1+S^2\ .
\end{equation}

\section{Antibracket cohomology}
Since we now have that $(S,S)=0$, we can define a nilpotent operator on any
function $F(\Phi ,\Phi ^*)$~:
\begin{equation}
{\cal S} F=(F,S)\ ,
\end{equation}
which raises the ghost number with one.
This defines a cohomology problem, called antibracket cohomology.
To compute it, we must find
the functions $F(\Phi ,\Phi ^*)$ that satisfy $(F,S)=0$,
modulo a part $F=(G,S)$.

First we have to consider which functions are invariant under the
${\cal S}$ operation. Again, we can make a split analogous to \eqn{split}~:
\begin{eqnarray}
\left( {\cal S}F\right) ^n&=&(-)^F\dkt  F^{n+1}
+D^nF(S^1,\ldots S^{\tilde n},F^0, \ldots, F^n)  \label{cSFexp}\\
\dkt F &=& \sum_{k=0}(S^k,F)_{k+1}=
\sum_{k=-1}S^{k+1}\dr_{a_k}\cdot\dl{}^{a_k}F \ , \label{defdkt0}\\
D^nF&\equiv&
\sum_{k=0}^{n} \sum_{m=1}^{\tilde k}(F^k,S^{n-k+m})_{m}\
,\label{defDn} \end{eqnarray}
where $f$ is the ghost number of $F$ and $\tilde k=k$ if $f< 0$ and $\tilde
k =k+f+1$ for $f\geq
0$. One must distinguish between negative and non-negative ghost numbers.
For negative ghost numbers, it can be proven that there is no
cohomology for local functions, see the references
given in previous sections . For  non--negative ghost numbers the situation
is more complicated.
The equation ${\cal S}F=0$ at zero antifield number is by \eqn{cSFexp}
\begin{equation}
-(-)^F\dkt F^1 =D^0F^0=\sum_{m=1}^{f+1} (F^0,S^m)_m \ .
\label{cSF00} \end{equation}
$D^0 $ is a fermionic right derivative operator,
which acts on fields only, and is given by
\begin{equation}
D^0  F^0  =
\left. (F^0,S)\right|_{\Phi ^*=0}
\ .\end{equation}
For antifield number 0, the KT differential is
acyclic on functions which vanish on the stationary surface.
Therefore $F^1$ exists if $D^0 F^0\approx 0$, and one can prove that then
also the full $F$ can be constructed perturbatively
in antifield number such that ${\cal S}F$=0. Again, for the proofs, we
refer to the literature.

The operator $D^0 $ raises the pureghost number \\ ($pgh(\phi ^i)=0,
pgh(c^{a_k})=gh(c^{a_k})=k+1, pgh(\Phi ^*_A)=0$) by 1. It is nilpotent
on the classical stationary surface: $D^0D^0 F^0\approx 0$. One can define
a (weak) cohomology of this operator on functions
of fields only, and this is graded by the pureghost
number $p$. The main result is that this weak cohomology is equivalent to the
(strong)
cohomology of ${\cal S}$ for functions of ghost number $p$~\footnote{Even
for
closed gauge algebras one has to use the field equations in order to have
equivalence between the two cohomologies.}~:\\
{\bf Theorem} : Any local function $F$ of negative ghost number which
satisfies ${\cal S}F=0$ can be written as $F={\cal S}G$. \\
For functions of non--negative ghost number, the following statement holds.
For a local function or a local integral $F^0(\Phi )$ (not containing
antifields)
\begin{equation}
D^0 F^0\approx 0\ \Leftrightarrow  \ \exists F(\Phi,\Phi ^*)\ : {\cal
S}F=0 \label{mainth1}\end{equation}
where $F(\Phi ,0)=F^0$. Further
\begin{equation}
F^0\approx D^0G^0\  \Leftrightarrow \ \exists G(\Phi ,\Phi ^*)\ : F={\cal
S}G\ ,
\end{equation}
where again $G^0=G(\Phi ,0)$, and $F$ is a function determined by
\eqn{mainth1}. The ghost numbers of $F$ and $G$ are equal to the
pureghost numbers of $F^0$ and $G^0$.

The inclusion of local integrals for the second part of the theorem follows
from the fact that for non--negative ghost numbers we could at the end of
subsection 3.2.3 include these in the acyclicity statement,
and this was the only ingredient of the proof. There are no
general statements for functionals at negative ghost number. However,
for ghost number minus one, it can be proven that
the cohomology is isomorphic to the space of constants of motion
\cite{BBHcom}.

At ghost number zero the antibracket cohomology gives the
functions on the stationary surface, where two such functions
which differ by gauge transformations are
identified. These are physically meaningful quantities. Indeed, the KT
cohomology reduced the functions to those on the stationary surface.
$D^0$ acts within the stationary surface, and its cohomology reduces
these functions to the gauge invariant ones, as
for (pure) ghost number 0,
\begin{equation}
D^0  F^0=\dr_i F^0\cdot R^i{}_{a} c^a\ ,
\end{equation}
gives the gauge transformation of $F^0$. Here we clearly see how
the antibracket and BRST formalisms are connected.

At ghost number 1, we can apply the theorem for the analysis of anomalies.
We will see in chapter 6
that anomalies ${\cal A}$ are local integrals of ghost number 1. They
satisfy the Wess--Zumino consistency relations \cite{WZcc} in the
form\footnote{We consider here only 1--loop effects.} ${\cal
S}{\cal A}=0$, while anomalies can be absorbed in local counterterms if
${\cal A}={\cal S}M$. The anomalies are thus in fact elements of
the cohomology of ${\cal S}$ at ghost number 1 in the set of local
integrals. We have found here that in the classical basis,
these anomalies are completely determined by their part ${\cal
A}^0$ which is independent of antifields and just contains 1 ghost of ghost
number 1. On this part there is the consistency condition $D^0 {\cal
A}^0\approx 0$.  If this equality is strong, then ${\cal A}$ does not need
antifield--dependent terms for its consistency. If it is weak, then
these
are necessary, but we know that they exist. If for an anomaly ${\cal
A}^0\approx D^0 M^0$, then we know that it can be cancelled by a local
counterterm. Consequently the
anomalies (as elements of the cohomology) are determined by their
part without antifields, and with ghost number one, and can thus be
written as
\begin{equation}
{\cal A}={\cal A}_a (\phi )c^a+\ldots\ ,
\end{equation}
where the written part determines the $\ldots$.
Therefore we can thus
split the anomalies in parts corresponding to the different symmetries
represented by the index $a$.
Indeed, people usually talk about anomalies in a certain symmetry (although
this can still have different forms according to the particular
representant of the cohomological element which one considers), and we show
here that this terminology can always be maintained for the general gauge
theories which the BV formalism can describe.

For recent examples of computing antibracket cohomology, at different
ghost numbers, we refer the reader to \cite{Brandt}.

\section{Gauge fixing and canonical transformations}

In previous sections, we always have worked in the so called classical
basis, where the fields have non-negative ghost numbers and the antifields
have negative ghost numbers. In this basis we have the
classical limit $S(\Phi ,\Phi ^*=0)=S^0(\phi )$. This basis is however not
useful when going to the gauge fixed action. In the previous chapter,
section 2.2, we have already seen the example of Maxwell theory. The
classical basis of nonnegative ghost-numbers are the fields $\Phi
^A=\{A_\mu ,c,b^*=\lambda \}$, where we have introduced the field $\lambda
$, because in the classical basis, fields do not carry a star-index. On the
other hand, the basis $\Phi ^A=\{A_\mu ,c,b\}$ is the one to use for doing
the path integral. Therefore we will call this the gauge fixed basis, since
in this basis, the action without antifields contains no more gauge
invariances. This is the general strategy \cite{Siegelgf}~: starting from
the non-minimal solution of
the classical master equation, we perform a canonical transformation that
mixes fields and antifields, such that in the new basis, called the
gauge-fixed basis, the extended action takes the form
\begin{equation}
S(\Phi ,\Phi ^*)=S_{gauge-fixed}(\Phi )+\mbox{   antifield--dependent
terms}\ .
\end{equation}
In this new basis some antifields will have
positive or zero ghost numbers, so it is not any more of the type mentioned
above.
`Gauge fixed' means that in the new definitions of fields the matrix of
second derivatives w.r.t.  fields, $S_{AB}$, is non--singular when setting
the field equations equal to zero. We have seen in the example of Maxwell
theory that one has to introduce extra fields, before doing the canonical
transformation. Indeed, it is not possible, starting from the "minimal
solution" $S=\frac{1}{4}F_{\mu \nu }F^{\mu \nu }+A_\mu ^*\partial _\mu c$,
to perform a canonical transformation with only these fields such that one
obtains a gauge fixed action. So, in general one needs to add "non-minimal"
sectors (like the $\frac{1}{2}b^{*2}$ term), that are cohomologically
trivial in
the antibracket sense (indeed, ${\cal S}b=b^*, {\cal S}b^*=0$), to be able
to do the canonical transformation.
For more details on this procedure we refer to
\cite{BV1,BV2,disp,bvsb,bvgursey,bvleuv}.

In the example of 2d chiral gravity, gauge fixing is obtained by the
canonical transformation where $h$ and $h^*$ are replaced by $b$ and $b^*$:
\begin{equation}
b=h^*\ ;\qquad b^*=-h\ .\label{bhcantr}
\end{equation}
One checks then that the part of $S$ depending only on the new `fields',
i.e. $X^\mu $, $b$ and $c$, has no gauge invariances. The gauge fixed
action is
\begin{equation}
S_{gf}=- \half \partial X^\mu \cdot \bar \partial X^\mu  + b\bar
\partial c\ ,
\end{equation}
and $b$ is called the antighost.

Let us take a closer look at canonical transformations. Just like for
Poisson brackets, they can be obtained from a generating function (in BV
theory, this is a fermion) $f(\Phi ,\Phi '^*)$, for which we have that
\cite{BVcan,anombv,bvleuv}
\begin{equation}
\Phi '^A=\Phi ^A +\frac{\partial }{\partial \Phi '^*_A}f(\Phi ,\Phi '^*)\
;\qquad
\Phi ^*_A=\Phi '^*_A+\frac{\partial }{\partial \Phi ^A}f(\Phi ,\Phi '^*)\ .
\label{cantrf}
\end{equation}
These type of transformations are called "infinitesimal transformation".
For generating "finite canonical transformations", of which $\phi =
\phi '^*, \phi ^*=-\phi $ is the simplest example, see the appendix of
\cite{anombv}.
For Maxwell theory,
the corresponding generating fermion for \eqn{cantr} is
\begin{equation}
f=b\partial _\mu A^\mu \ ,
\end{equation}
where an intergal is understood. This corresponds to the gauge choice
$\partial _\mu A^\mu =0$.
Choosing different generating functions in \eqn{cantrf}
corresponds to
different gauge choices. Let us illustrate this in the case of chiral 2-d
gravity. Starting from the minimal solution \eqn{SextW2}, we add a
non-minimal sector $S_{nm}=-b^*\pi _h$. Now, we can perform a canonical
transformation generated by
\begin{equation}
f_1=bh\ .
\end{equation}
In this case, we are in the "background gauge". Indeed, after the canonical
transformation, we can intergate out the field $\pi _h$ which gives a delta
function that fixes $h$ to a general background field, played by the
antifield $b^*$, i.e. $h=-b^*$. This was the gauge presented above. On the
other hand, instead of doing $f_1$, we could also have done the canonical
transformation generated by
\begin{equation}
f_2={\bar \partial }b h\ ,
\end{equation}
which gives, after dropping the primes
\begin{eqnarray}
S&=&-\frac{1}{2}\partial X^\mu {\bar \partial }X^\mu +\frac{1}{2}h\partial
X^\mu \partial
X^\mu \nonumber\\ &&+X^*_\mu \partial X^\mu c+h^*\nabla c+c^*\partial
cc\nonumber\\
&&-b^*\pi _h+{\bar \partial }h\pi _h+{\bar \partial }b\nabla c\ .
\end{eqnarray}
This brings us to the temporal gauge ${\bar \partial h}=b^*$.
In order to go from this second order action to a first order
action, we introduce another trivial system
$S_{enm}=-\pi _b\pi _c$.
Both fields are fermionic, $\pi _b$ has ghost number one, $\pi _c$ ghost
number minus one. We then do a canonical transformation generated by
\begin{equation}
f_3={\pi '}^*_c({h '}^*+{\bar \partial} b)-{\pi '}^*_b\nabla c\ ,
\end{equation}
Then the extended action takes the form
\begin{eqnarray}
S&=&-\frac{1}{2}\partial X^\mu {\bar \partial }X^\mu +\pi _h{\bar \partial
}h+\pi _b{\bar \partial }b+\pi _c{\bar \partial }c-\pi _b\pi _c
+h(T_{mat}+T_{gh})\nonumber\\
&&+X^*\partial Xc-h^*\pi _b-\pi ^*_c[T_{mat}+T_{gh}]\nonumber\\
&&+c^*\partial cc-b^*\pi _h\ ,
\end{eqnarray}
with
\begin{equation}
T_{mat}=\frac{1}{2}\partial X\partial X \qquad T_{gh}=-2\pi _c\partial
c-\partial \pi _cc \ .
\end{equation}
The quantisation of chiral $W_2$ and $W_3$ gravity in these gauges
was done in \cite{tempgauge}.

Another application of canonical transformations is rewriting of the gauge
algebra and the extended action before
gauge fixing. Consider the example of chiral $W_3$ gravity, of which we
have given $S^0$ and $S^1$ in \eqn{S0W3} and \eqn{S1W3}. $S^2$ could then
be determined by computing the gauge algebra, i.e. the structure functions
$T^a_{bc}$ and field equation coefficients $E^{ij}_{ab}$. One first
needs to compute the left hand side of \eqn{opalg}. This leads to an
expression that has to be split in a part with the gauge generators $R^i_a$
and a part with the field equation. This split is not unique when some of
the gauge generators are proportional to field equations. In our example,
the transformation of $h$ involves a term proportional to its own field
equation.
The arbitrariness of this split reflects itself into different
possibilities for $S^2$, and one finds solutions
\begin{eqnarray}
S^2&=&c^*\left[(\partial c)c+\kappa (1-\alpha ) \pl \pl (\partial
u)u\right]+u^*\left[2(\partial c)u-c(\partial u)\right]\nonumber\\
&&-2\kappa\alpha  h^*(D^3B^*+\na^2h^*)(\partial u)u -
2\kappa(\alpha+1)  X_\mu ^*h^*(\partial u)u\pl \ ,
\end{eqnarray}
for arbitrary $\alpha $.
For $\alpha =0$, the algebra closes off-shell on $B$,
while for $\alpha =-1$ the algebra closes off-shell on $X^\mu $.
Remark also the choice
$\alpha =1$, which gives the simplest structure functions. The relation
between these
actions is given by the canonical transformation: starting from the action
with $\alpha =0$ the transformation with generating function
\begin{equation} f=2\kappa\alpha  h'^*(\partial u)uc'^*\,.
\label{cantrlambda}\end{equation}
gives in the primed coordinates the action with this arbitrary parameter
$\alpha $ \cite{anomw3}.

The non-uniqueness of the extended action has its origin in adding the
exact term
$\delta _{KT}V^{n+3}$ in \eqn{Sn}. It can be shown that this
corresponds to the canonical transformation generated by
\begin{equation}
f=-V^{n+3}(\Phi ,\Phi ^*)\ .
\end{equation}

Let us finally end with an important remark. In the previous section, we
have introduced a BRST operator $D^0$, whose weak cohomology was equivalent
to the strong cohomology of the antibracket. The BRST operator was defined
on functions depending
on fields only, and one had to take the bracket with $S$ and then put all
antifields equal to zero. When gauge fixing, however, the role of fields
and antifields can be interchanged, so that, in this basis, one has to put
other fields equal to zero after taking the bracket with $S$. This leads to
a different operator, which is still called the BRST operator.
One can check that this operator is weakly nilpotent, but now using
the field equations of the fields in the gauge fixed basis. The
weak cohomology of this operator is again isomorphic to the antibracket
cohomology, but only for local functions. For local functionals one can not
make any statements anymore, since the proof of the equivalence heavily
relies on the acyclicity of the Koszul-Tate operator. In the gauge fixed
basis, fields can have negative ghost numbers, and so, acyclicity does not
hold on local functionals.

\chapter{New examples of infinitly reducible theories}
\section{Motivation and introduction}
In chapter 2, the concept of reducible gauge theories was explained. We
gave the example of the antisymmetric tensor field, which was an example of
a first order reducible theory. We also mentioned theories with the so
called $\kappa $-symmetry as examples of infinitly reducible theories.

Here, we will give 2 new examples of infinitly reducible theories and show
how the BV formalism can be applied to such models. These examples can be
seen in the context of first order actions with relations between the
generators. They have actions of the form
\begin{equation}
S=K_i(\phi ){\bar \partial }\phi ^i+\psi ^aT_a(\phi )\ ,\label{FOS}
\end{equation}
where we call $\phi ^i$ the matter fields, $\psi ^a$ are gauge
fields and $T_a$ are first class constraints in the Hamiltonian language,
generating $a=1,...,n$ gauge symmetries. The time derivative ${\bar
\partial }$ is always explicitly written and is not understood in the
DeWitt notation. These theories can be quantised in
the Hamiltonian approach using the Batalin-Fradkin-Vilkovisky (BFV) method
(see
\cite{Hennrep,Govboek} for reviews), or in the Lagrangian approach using
the BV method, see section 3 of \cite{disp}, or \cite{bvleuv}.
Examples of this are a large class of conformal field
theories \cite{BPZ}, like ordinary chiral gravity and $W_3$ gravity,
and the bosonic relativistic particle.
In the Hamiltonian language one can define Dirac brackets under which
the constraints $T_a$ form the (current) algebra
\begin{equation}
[T_a,T_b]=f_{ab}^cT_c\ .
\end{equation}

Now, the generators $T_a(\phi )$ can be linearly dependent. In that case
one has relations between the generators of the form
\begin{equation}
Z^a_{a_1}T_a=0\ ,\label{redcon}
\end{equation}
and one calls the constraints (in Hamiltonian language) reducible.  These
systems can also be quantised using the BFV or BV approach, and we will use
the latter. It is clear that, in this case, we have extra symmetries of the
form
\begin{equation}
\delta \psi ^a={\bar{\epsilon}}^{a_1}Z^a_{a_1}\ .
\end{equation}
Therefore, one should also introduce, besides the ghosts $c^a$ coming from
the gauge symmetries generated by $T_a$, extra ghosts ${\bar {c}}^{a_1}$
\cite{ToineMoskou}. On the other hand, \eqn{redcon} means that the gauge
generators are not linearly independent, and so, we need ghosts for
ghosts $c^{a_1}$. The
quantisation of such systems, under suitable assumptions of the gauge
algebra, is discussed
in \cite{disp} (section 3) and in \cite{bvleuv}, in the context of the
superparticle and the Green-Schwarz superstring. In these theories,
there are even further zero modes
\begin{equation}
Z^a_{a_1}Z^{a_1}_{a_2}=0\ ,
\end{equation}
such that one has to introduce further ghosts for ghosts.
In fact, these models are infinitly
reducible, such that one has to work with an infinite tower of
ghosts for ghosts.

In this chapter, we will give new examples of these systems, in the
context of conformal field theory. They do not
completely follow from the description in \cite{disp} because the
assumptions there are not satisfied anymore. As a warm--up, we discuss a
toy model in the next section to show the general idea and the basic
principles. After that, we give the example of the $W_{5/2}$ algebra, which
is a new type of gauge algebra, not discussed in the literature
so far. New in the sense that there are symmetries which are proportional
to symmetric combinations of the field equations.
As explained at the end of section 3.2.2, this
makes that the properness condition is not equivalent to completeness.

The following sections are based
on work in collaboration with K. Thielemans \cite{W52}.

\section{Toy model~: a
fermion in a gravitational field}
\subsection{Some generalities}
We have already seen the example of 2d chiral gravity, which is a conformal
field theory  with energy momentum tensor $T=\frac{1}{2}\partial X^\mu
\partial X^\mu $. The classical action is of the form
\begin{equation}
S^0=S_0+hT\ ,
\end{equation}
where $S_0$ depends only on the matter fields, in this case the $X^\mu $'s.
$S_0$ and $T$ transform under the conformal symmetry as
\begin{equation}
\delta S_0=-{\bar \partial }\epsilon T \qquad \delta T=\epsilon \partial
T+2\partial \epsilon T\ . \label{confsys}
\end{equation}
The derivatives only work on the first object behind the $\partial $ or
${\bar \partial }$. The model can be considered as being a scalar matter
field
coupled to a chiral gravitational background field $h$ in 2 dimensions.
The
construction of the extended action was straightforward and given in
\eqn{SextW2}. However, \eqn{confsys} suggests that one can build the
extended action in terms of the current $T$ and the gauge field $h$,
without specifying the realisation. In our example, the extended
action can be rewritten as
\begin{equation}
S=S^0+T^*[c\partial T+2\partial cT]+h^*[\nabla^{(-1)}c]+c^*\partial cc \ .
\end{equation}
This way of writing the extended action is only to show that our results
are realisation independent. One can not simply treat $T$ and $T^*$ as
elementery fields, since we can not express e.g. the $X^\mu $ in terms of
$T$. However, as a working definition, one could use the rule
$\frac{\partial S^0}{\partial X^\mu }\frac{\partial S^1}{\partial X^*_\mu }
=\frac{\partial S^0}{\partial T}\frac{\partial S^1}{\partial T^*}$.
The advantage of this approach is
that this extended action can be used for any realisation satisfying
\eqn{confsys}.

However, one can also find realisations with fermions as matter
fields.
For instance, a two fermion model with anticommuting fields $\psi ,{\bar
\psi }$, a classical action $S_0=\psi {\bar \partial }{\bar \psi }$ and
$T=\frac{1}{2}\partial \psi {\bar \psi }-\frac{1}{2}\psi \partial {\bar
\psi
}$. The conformal symmetry is the invariance under the transformation
$\delta \psi
=\epsilon \partial \psi +\frac{1}{2}\partial \epsilon \psi $, and analogous
for ${\bar \psi}$.
Even more simple, one can take a model with a single fermion
$\psi $, a classical action $S_0=\frac{1}{2}\psi {\bar \partial }\psi $ and
$T=\frac{1}{2}\partial \psi \psi $, with the same transformation rule as
above.

It can happen that there appear
extra constraints between the generators. E.g. in our one
fermion model, the extra constraint is $T^2=0$. This is
due to
the fact that fermions anticommute and square to zero. As mentioned in the
previous section, these constraints generate extra
gauge symmetries. In the theory where $T$ is used to work in a realisation
independent way, which we will call the macroscopic theory from now on,
this extra
symmetry is $\delta h=T\epsilon $. In the theory where the realisation is
explicitly specified, which we will call the microscopic theory, we have
two extra gauge symmetries, namely $\delta h=\psi \epsilon _1;
\delta h=\partial
\psi \epsilon _2$. So, in the quantisation, we must distinguish between
these two cases.  We will first discuss the macroscopic theory and comment
on the microscopic theory later.

\subsection{The macroscopic theory}

In this theory, we will assume there is some realisation that gives a
constraint
\begin{equation}
T^2=0\ .\label{T2=0}
\end{equation}
To quantise this theory, we start with an action $S_0$ and an energy
momentum tensor that satisfy
\eqn{confsys}. The gauge symmetries can be written in terms
of the current $T$, coming from some transformation rules on the
(unspecified) matter fields, i.e. coming from the microscopic theory.
Because of the completeness, we also have to
include the extra gauge symmetry on the gauge field $h$, due to the
constraint \eqn{T2=0}. The complete set of symmetries is
\begin{eqnarray}
\delta _{\epsilon _1}T=\epsilon _1\partial T+2\partial \epsilon _1T
&\qquad& \delta _{\epsilon _2} T=0\nonumber\\
\delta _{\epsilon _1}h=\nabla^{(-1)}\epsilon _1 &\qquad& \delta _{\epsilon
_2}h=T\epsilon _2\ .
\end{eqnarray}
The $\epsilon _1$ symmetry is the Virasoro symmetry, for which we introduce
a ghost $c^1$. Remark that the $\epsilon _2$ symmetry on the fields is a
symmetric combination of the field equations, i.e. it is of the form
$\delta \phi ^i=y_jS^{ji}$, where the $S$ matrix is symmetric in $i$ and
$j$, which label the fields $T$ and $h$. This implies that properness
does
not imply completeness\footnote{However, vice versa, completeness always
implies properness.}. As mentioned in the previous chapter, the question
arises wether we should introduce a ghost for this symmetry or not.
Equivalently, should we require properness or
completeness ? The answer is given by the requirement of the acyclicity
of the Koszul Tate operator. We have in our case a KT invariant~: $\delta
_{KT}[\phi ^*_iy_jS^{ji}]=0$. This invariant is however not exact since it
cannot be written as $\delta _{KT}[\phi ^*_i\phi ^*_jE^{ji}]$. Here, the
matrix $E$ is always (graded) antisymmetric in $i$ and $j$. So, in order to
guarantee the acyclicity of the KT operator, we will introduce a ghost
$c^2$ with $\delta _{KT}c^*_2=h^*T$. This means that, if there is a
difference between properness and completeness, one should require
completeness.

The extended action at antifieldnumber one is then
\begin{eqnarray}
S^1&=&T^*[c^1\partial T+2\partial c^1T]\nonumber\\
&&+h^*[\nabla^{(-1)}c^1+Tc^2]\ .
\end{eqnarray}
The transformation matrix $R^i_a$ is
\begin{equation}
R^i_a=\pmatrix{\partial T+2T\partial & 0 \cr \nabla^{(-1)} & T \cr}
\ , \label{R}
\end{equation}
where this matrix is working on an additional delta function $\delta
(x-y)$, and the derivatives in this matrix are w.r.t. $x$.
Remark that the $a$ index is different from the one in the previous
section. Here, $a=1,2$. It includes the $\epsilon ^a$ and the ${\bar
\epsilon }^{a_1}$ symmetries. As can be easily seen,
the gauge generators are not all independent, because of the relation
$T^2=0$. For instance, one could take as a zero mode the column
$v^a=\pmatrix{0\cr T\cr}$ which clearly satisfies $R^i_av^a=0$. Whether
this is a "good" zero mode or not will be discussed in the next subsection.

\subsection{Zero modes dictated by Koszul-Tate}
For each zero mode that one finds,
one expects to introduce a ghost for ghosts. This follows in
fact from the properness condition \eqn{properZ}.
It says that, whenever there is a zero mode $v^a$, it should be weakly
proportional to the reducibility matrix $Z^a_{a_1}$, for which one has to
introduce a ghost for ghosts.
However, there are some subtleties when the zero modes are
vanishing
on shell, because the properness condition is satisfied for any $v^a\approx
0$. We have discussed this subtlety in section 3.2.3 for arbitrary gauge
theories. Let us illustrate this in our example.
We have the following
equations~:
\begin{eqnarray}
&&\delta _{KT}h^*=T\nonumber\\
&&\delta _{KT}c^*_1=-T^*\partial T-2\partial T^*T-\nabla h^*\nonumber\\
&&\delta _{KT}c^*_2=h^*T\ .
\end{eqnarray}
First of all, let us look at functions at antifieldnumber 1. For instance
we have that $\delta _{KT}(h^*T)=0$, for which we have introduced
$\{c^2,c^*_2\}$ that kills this cycle. Then observe that
$\delta _{KT}(\partial h^*T)=0$ and also $\delta _{KT}(h^*\partial T)=0$.
But these can be written as $\delta _{KT}[1/2(\partial c_2^*\pm h^*\partial
h^*)]$, so that we do not have to include the symmetry $\delta h=\partial
T\epsilon $. In fact, this symmetry can be written as a combination of our
true gauge symmetry and an antisymmetric combination of the field equation
of $h$. At antifieldnumber two we have further $KT$ invariants, e.g. $\delta
_{KT}(c^*_2T)=0$. This invariant is vanishing on shell, so it corresponds
to a zero mode on shell, namely $v^a=\pmatrix{0\cr T\cr}$.
However, this cycle is $KT$ exact and we do not need
further ghosts. Indeed, $(c^*_2T)=\delta _{KT}(c^*_2h^*)$. Looking at the
general conditions for KT exactness \eqn{expair}, one can check that these
equations are satisfied.

On the
other hand we also have that $\delta _{KT}(c^*_2\partial T)=0$, which is
also vanishing on shell, but it is not NOT $KT$ exact. For this cycle,
although also vanishing on
shell, we define a new field-antifield pair $c^{2_1}$
(bosonic) and $c^*_{2_1}$ (fermionic) such that \begin{equation}
\delta _{KT}c^*_{2_1}=c^*_2\partial T\ .
\end{equation}
Another non $KT$ exact invariant is
$-[\nabla ^{4}c^*_2+2c_1^*T+h^*\nabla ^{2}
h^*]$, for which we introduce the pair $c^{1_1}$ (bosonic)
and $c^*_{1_1}$ (fermionic), with \begin{equation}
\delta _{KT}c^*_{1_1}=-[\nabla^{4}c^*_2+2c_1^*T+h^*\nabla ^{2}h^*]\ .
\end{equation}
One can check that these are all the non-trivial zero modes. They form the
reducibility matrix
\begin{equation}
Z^a_{a_1}=\pmatrix{-2T & 0 \cr \nabla^{-3} & \partial T \cr}
\ . \label{Z1}
\end{equation}
This determines the extended action up to antifieldnumber 2~:
\begin{eqnarray}
S^2&=&-2c^*_1Tc^{1_1}-h^*\nabla h^*c^{1_1}\nonumber\\
&&+c^*_2[\nabla ^{(-3)}c^{1_1}+\partial Tc^{2_1}]\nonumber\\
&&+c^*_1\partial c^1c^1+c^*_2[\partial c^2c^1-3c^2\partial c^1]\ .
\end{eqnarray}

\subsection{The infinite tower of zero modes}
Having found \eqn{Z1}, one can now look for further zero modes, starting
at antifieldnumber 3. After
some analysis, one finds again $KT$ non-exact invariants. The first one
contains the $\nabla$ operator (or time derivative)~:
\begin{eqnarray}
\delta _{KT}c^*_{1_2}&&=-\nabla ^{7}c^*_{2_1}+c^*_{1_1}\partial T+T\partial
c^*_{1_1}-c^*_2\nabla \partial h^*\nonumber\\
&&+2\nabla c^*_2\partial h^*+\nabla (\partial
c^*_2h^*)+2c^*_2\partial \nabla h^*+\partial \nabla c^*_2h^*\ ,
\end{eqnarray}
where $\{c^{1_2},c^*_{1_2}\}$ is the new ghost for ghosts pair to kill this
cycle.
We also find two zero modes, corresponding to the constraint $T^2=0$. They
are
\begin{eqnarray}
\delta _{KT}c^*_{2_2}&=&c^*_{2_1}T\nonumber\\
\delta _{KT}c^*_{3_2}&=&\partial ^2Tc^*_{2_1}+\frac{1}{3}Tc^*_1\partial
^3h^*\ .
\end{eqnarray}

This leads to the second level reducibility matrix
\begin{equation}
Z^{a_1}_{a_2}=\pmatrix{-T\partial  & 0 & 0 \cr \nabla^{-6} & T &
\partial ^2 T \cr} \ . \label{Z2}
\end{equation}
It determines terms in the extended action proportional to antifieldnumber
3~:
\begin{equation}
S^3=-c^*_{1_1}T\partial c^{1_2}+c^*_{2_1}[\nabla
^{(-6)}c^{1_2}+Tc^{2_2}]+...\ ,
\end{equation}
where the dots now indicate terms quadratic or more in antifields.

Intuitively, it is now clear that this procedure repeats itself ad
infinitum.
We will always have two kinds of zero modes, which we call dynamical and
algebraic zero modes. The dynamical zero modes contain a time
derivative and ensure that the ghosts for ghosts are propagating after gauge
fixing. Indeed, after
this procedure is completed, the gauge fixing is analogous to ordinary 2d
gravity, namely $h^*=b_1, c^*_2=b_{1_1}, c^*_{2_1}=b_{1_2}, ...$ etc. .
The gauge fixed action then takes the form
\begin{equation}
S_{gf}=\frac{1}{2}\psi {\bar \partial }\psi +b_1{\bar \partial }c^1+b_{1_1}
{\bar \partial }c^{1_1}+b_{1_2}{\bar \partial }c^{1_2}+... \
.\label{Sgfmac} \end{equation}
On the other hand there are the algebraic zero modes. These
follow directly from the constraint $T^2=0$, and from the derivatives on
it. One can convince oneself
that the terms in the extended action coming from the algebraic zero modes
disappear after gauge fixing.
The general structure is that the algebraic zero modes at
antifield number $n$ determine the dynamical zero modes at
antifield
number $n+1$. The antifields of the ghosts for ghosts for the algebraic
zero
modes become, after gauge fixing, the antighosts of the ghosts for ghosts,
coming from the dynamical zero modes.
This was, without using
the terminology
of algebraic and dynamical zero modes, already pointed out in \cite{disp}.

Let us finally comment on the extra condition for finding KT invariants,
namely eqn. \eqn{eqclos3}. We will show here that in the macroscopic
theory, this equation is
not automatically satisfied. Consider the zero mode $Z^{a_1}=\pmatrix{0 \cr
A\cr}$, for any function $A$. This zero mode clearly satisfies
\eqn{Z1Z2weakzero}, with \begin{equation}
f^{ia}=\pmatrix{0 & 0 \cr 0 & A\partial \cr}\ .
\end{equation}
However, it does not automatically satisfies \eqn{eqclos3} for arbitrary
$A$. One can check that $A=T$ and $A=\partial ^2T$ satisfy the extra
condition. They indeed correspond to the zero modes in \eqn{Z2}.

\subsection{The microscopic theory}
In this section, we will briefly comment on the difference between the
micro- and macroscopic theories.
Let us work with the simple one fermion model for which
the action is
\begin{equation}
S^0=\frac{1}{2}\psi {\bar \partial }\psi -\frac{1}{2}h\psi \partial \psi \
. \end{equation}
We now have as a complete set of symmetries
\begin{eqnarray}
\delta _{\epsilon _1}\psi =\epsilon _1\partial \psi +\frac{1}{2}\partial
\epsilon _1\psi \qquad &\delta _{\epsilon _2}\psi =0& \qquad \delta
_{\epsilon _3}\psi =0\nonumber\\
\delta _{\epsilon _1}h=\nabla^{(-1)}\epsilon _1 \qquad &\delta _{\epsilon
_2}h=\psi \epsilon _2& \qquad \delta _{\epsilon _3}h=\partial \psi \epsilon
_3\ .
\end{eqnarray}
We have one more gauge symmetry than in the case of previous sections. This
is of course due to the fact that we know the (microscopic) details of the
energy momentum tensor.
The transformation matrix is now
\begin{equation}
R^i_a=\pmatrix{\partial \psi +1/2\psi \partial   & 0 & 0 \cr \nabla^{-1} &
\psi & \partial \psi  \cr} \ . \label{Rmic}
\end{equation}
In the microscopic theory, we got rid of the symmetries, graded symmetric
in the field equations. Indeed, the symmetry $\delta h=T\epsilon $ is in
the microscopic theory a combination of the $\epsilon _2$ and $\epsilon _3$
symmetries. The latter are however not vanishing on the stationary surface.

We have to introduce three
ghosts~:
$c^1$, fermionic, $c^2$ and $c^3$, both bosonic. Using these fields, we
can construct the extended action up to antifield number one~:
$S^1=\phi ^*_iR^i_ac^a$. As for the macroscopic theory, the generators
$R^i_a$
are not linearly independent. It turns out one can find five zero modes,
corresponding to the following invariants~: $\nabla c^*_2+c^*_1\psi
-h^*\psi ^*; \nabla c^*_3+c^*_1\partial \psi -\frac{1}{2}c^*_2\partial
^2h-h^*\partial \psi ^*; c^*_1\psi ; c^*_2\partial \psi $ and
$c^*_1\partial \psi +c^*_2\psi $. They form the first level reducibility
matrix
\begin{equation}
Z^a_{a_1}=\pmatrix{\psi  & \partial \psi & 0 & 0 & 0 \cr \nabla^{-3/2} &
-1/2\partial ^2h & \psi  & 0 & \partial \psi \cr 0 & \nabla^{(-5/2)} & 0 &
\partial \psi & \psi \cr} \ . \label{Z1mic}
\end{equation}
There are two dynamical and three algebraic zero modes.
One can again imagine that this leads to an infinitly reducible theory.
Nevertheless, one can construct the
extended action and gauge fix it. It will again lead to an
action analogous to \eqn{Sgfmac}. But now, there are more ghosts for
ghosts per level of reducibility, and they have different statistics. Also
the conformal spins of the ghosts are different. It is not clear to what
extend the  micro- and macroscopic theories are equivalent\footnote{It is
surprising that, if one forgets about the extra symmetries (both in the
micro- and macroscopic theory), one can still
obtain a gauge fixed action $S=\frac{1}{2}\psi {\bar \partial }\psi +b{\bar
\partial }c$ with the standard BRST rules $\delta \psi =\partial \psi c
+\frac{1}{2}\psi \partial c; \delta c=\partial cc; \delta b=-T+2b\partial
c+\partial bc$. Therefore, it is not excluded that all the
ghosts for ghosts cancel out each other.}. We leave this as an open
problem.

\section{A new type of gauge theory~: the $W_{2,\frac{5}{2}}$
algebra}

\subsection{The current algebra}
The \W52--algebra was one of the first
\WA s constructed, see \cite{zamoW3} where it is presented in the
quantum case with Operator Product Expansions \cite{BPZ}. We need it here
as a classical \WA, \ie using Dirac brackets. The algebra consists of two
currents~: $T$, the Virasoro generator, and a primary dimension $\ft52$
(fermionic) current $G$. They satisfy
\bea
\left[T(z),T(w)\right]&=& -2T(w) \dz\delta(z-w)+\del T(w) \delta(z-w)\nonu
\left[T(z),G(w)\right]&=& -\ft 5 2 G(w) \dz\delta(z-w)+\del G(w)
\delta(z-w)\nonu
\left[G(z),G(w)\right]&=& T^2(w) \delta(z-w)\ .\label{PBalgebra}
\eea
Here, these brackets are only defined formally, and one should look for
systems with fields and their momenta that realise this algebra. Taking
\eqn{PBalgebra} as a definition, the Jacobi identities are only satisfied
modulo a ``null field''
\be
N_1\equiv 4T\, \del G -5 \del T\, G \ . \label{nullfield}
\ee
In this context, we call ``null fields'' all the combinations of $T$ and
$G$ which should be put to zero such that the Jacobi identities are
satisfied. We can check by repeatedly computing Dirac brackets with
$N_1$ that the null fields are generated by $N_1$ and
\be
N_2 \equiv 2 T^3 - 15 \dz G\, G\ . \label{nullfield2}
\ee
More precisely, all other null fields are of the form~:
\be
f_1(T,G) \del^n N_1 + f_2(T,G) \del^m N_2
\ee
with $f_i$ differential polynomials in $T$ and $G$.

A two fermion realisation, like in section 4.2.1,  for this algebra was
found in \cite{TAMhs}~: \bea
T&=& -\ft12 \psi\, \del\bar\psi +\ft12 \del\psi\, \bar\psi \ ,\nonumber\\
G&=& \ft12\left(\psi+ \bar\psi\right)\, T \ ,
\eea
where $\psi$ is a complex fermion satisfying the Dirac bracket
$[\psi(z),\bar\psi(w)]=\delta(z-w)$.  One can easily verify for this
realisation that the null fields $N_i$ vanish.

In fact, for {\sl any} realisation in terms of fields with associated
Dirac brackets (\eg free fields),
the null fields will
vanish identically. Indeed, they appear in the \rhs\ of the Jacobi
identities, which are automatically satisfied when Dirac brackets are used.
This means that in any realisation, the generators $T,G$ are not independent.
They satisfy (at least) the relations $N_i=0$. In the following section, we
will see that these relations have important consequences for the gauge
algebra.

\subsection{The gauge algebra}
In order to construct a gauge theory based on this algebra, one must
work in a certain realisation, \ie one must specify
matter fields $\phi^i$ and an action $S^0$, as in \eqn{FOS}
. The generators $T(\phi )$, $G(\phi )$
satisfy the algebra \eqn{PBalgebra} with $N_1=N_2=0$, because we are
now using Dirac brackets. They generate, being first class constraints,
the gauge transformations on the fields via
\begin{equation}
\delta_{\epsilon^a} \phi = \int \epsilon^a [ T_a, \phi ]
\label{phitrans}
\end{equation}
where the index $a$ runs over the number of generators, and there is no
summation in the \rhs .
In the same spirit of the previous sections, we
will not make a choice for the realisation and use
only the information contained in the algebra of the generators to construct
a gauge theory. Instead, we will treat $T$ and $G$ as elementary fields. In
that
case the transformation rules can be written on the currents, and one finds
\begin{eqnarray}
\delta _\epsilon T=\epsilon \partial T+2\partial \epsilon T&\qquad&
\delta _\alpha T=\frac{3}{2}\alpha \partial G+\frac{5}{2}\partial \alpha
G \nonumber\\
\delta _\epsilon G=\epsilon \partial G+\frac{5}{2}\partial \epsilon
G&\qquad& \delta _\alpha G=\alpha T^2
\end{eqnarray}

Hence, we will assume that there is some action $S_0$ that transforms
under conformal resp. supersymmetry with parameters $\epsilon $ resp.
$\alpha $ as
\begin{equation}
\delta _\epsilon S_0=
  -\bd\epsilon T \qquad  \delta_\alpha S_0=-\bd\alpha G\ .
\label{Noether}
\end{equation}
The commutators between two symmetries can be computed using the Jacobi
identities~:
\begin{eqnarray*}
[\delta_{\epsilon^a}, \delta_{\epsilon^b}] \phi &=&
  \int \epsilon^a \int \epsilon^b(-)^{ab}
  \left(\{ T_a, \{ T_b, \phi \}\} - \{ T_b, \{ T_a, \phi \}\}\right)\\
&=&
 -\int \epsilon^a \int \epsilon^b
  \{ \{ T_a,  T_b\}, \phi \} \ .
\end{eqnarray*}
We find~:
\begin{eqnarray}
{[}\delta_{\epsilon _1},\delta _{\epsilon _2}] &=&
   \delta _{\tilde\epsilon=\epsilon _2\partial \epsilon _1-
        \epsilon _1\partial \epsilon _2}\nonumber\\
{[} \delta _\epsilon,\delta _\alpha ] &=&
   \delta _{\tilde \alpha=
   -\epsilon \partial \alpha +3/2\alpha \partial \epsilon}
    \nonumber\\
{[} \delta _{\alpha _1},\delta _{\alpha _2}] &=&
   \delta _{\tilde\epsilon = 2\alpha _2\alpha _1T}\,.
\label{chiralcommutators}
\end{eqnarray}

Now, we can gauge these symmetries by introducing gauge fields
$h$ (bosonic) and $f$ (fermionic) for the
conformal and susy symmetries. The action is then
\begin{equation}
S^0=S_0 +h T+f G\ .\label{fullS}
\end{equation}
The transformation rules that make the action invariant are
\begin{eqnarray}
\delta _\epsilon h =\nabla ^{-1}\epsilon&\qquad &
\delta _\alpha h =\alpha f T\nonumber\\
\delta _\epsilon f =\epsilon \partial f -\frac{3}{2}f \partial
\epsilon &\qquad&
\delta _\alpha f =\nabla ^{-(\frac{3}{2})}\alpha \ ,\label{gtr2}
\end{eqnarray}
These
rules enable us to study the gauge algebra. Computing the commutators of
the gauge symmetries on the gauge fields, we see that they close only
after using equations of motion. In the usual case for open algebras
\cite {deWitvH} one has the usual structure functions $T^a_{bc}$ and a
graded antisymmetric field equation matrix $E^{ij}_{ab}$.
In the case of \W52 however, the commutator of two supersymmetries gives
us something unexpected. Computing the commutator
\eqn{chiralcommutators} on the gauge fields, we find~:
\begin{eqnarray}
\Bigl(\left[ \delta _{\alpha _1},\delta _{\alpha _2}\right] -
\delta _{\tilde\epsilon = 2\alpha _2\alpha _1T}\Bigr) h
&=& -[\nabla^{-3}(\alpha _2\alpha _1)T+2\alpha _2\alpha
_1\nabla^{2}T]\nonumber\\
&&-\frac{5}{2}\partial (\alpha _2\alpha _1)f G-3\alpha _2\alpha _1f
\partial G\nonumber\\
\Bigl( \left[ \delta _{\alpha _1},\delta _{\alpha _2}\right] -
      \delta _{\tilde\epsilon = 2\alpha _2\alpha _1T} \Bigr) f
&=&-3
\alpha _2\alpha _1\partial (Tf )-\frac{1}{2}\partial (\alpha _2\alpha
_1)f T\nonumber\\
&&+9\alpha _2\alpha _1f \partial T+\partial (\alpha _2\alpha _1f )T\ ,
\end{eqnarray}
The terms between square
brackets on the first line is an antisymmetric combination of the field
equation of $h ,\ y_h =T$. The last two terms of the
\rhs\ for $h$ together with the first and second term of the \rhs\ for
$f$ again form trivial field equation symmetries. However, the two
terms on the last line of the \rhs\ for $f$ remain. As they arise from
the commutator of two symmetries, they must leave the action invariant too.
So, they generate
a new fermionic symmetry, given by ~:
\begin{eqnarray}
\delta_{n} \phi^i&=&0\nonumber\\
\delta_nh&=&0\qquad
\delta_{n} f\ =\ 9n\partial T+4\partial n T\ .\label{eps1gtr}
\end{eqnarray}
Note that it acts only on the gauge fields, and hence leaves $S_0$
invariant. Remark that this symmetry is proportional to field equations.
However it is not a graded antisymmetric combination, in which case it
could
be neglected in the quantisation.  Here, this extra symmetry is essential
for constructing the extended and gauge fixed action.

Of course, in hindsight it is obvious there is a
corresponding symmetry associated with a null field. However, if one
tries to quantise the action without knowing the algebra of the previous
section, one is surprised that the gauge transformations \eqn{gtr2} do
not form a closed algebra, even after using trivial (i.e. graded
antisymmetric combinations of the field equations) equation of motion
symmetries.

Completely analogous, one can find a second new symmetry. This symmetry
will appear in the commutator $[\delta _\alpha ,\delta _n]$, on the
gauge fields again. It does not close using trivial field equation
symmetries, but gives rise to terms proportional to the field equations
$y_h=T$ and $y_f=-G$. The second
symmetry, with bosonic parameter $m$ can be written as
\begin{eqnarray}
\delta _m\phi^i&=&0\nonumber\\
\delta _mh &=&2mT^2\qquad\delta _mf
=-15m\partial G\ , \label{eps2gtr}
\end{eqnarray}
which indeed leaves the action \eqn{fullS} invariant when using $N_2=0$,
see \eqn{nullfield2}.

The two new symmetries are proportional to field equations itself. The way
they are written down is not unique. For instance, one could change
\eqn{eps1gtr} to $\delta _n h=-4n \partial G; \delta
_n f =5n \partial T$. This choice is however
equivalent, since it corresponds to \eqn{eps1gtr} by adding graded
antisymmetric combinations of the field equations. This does not change
the theory and
its quantisation. In any case, we have here an example of weakly
vanishing symmetries  which are not antisymmetric combinations of the field
equations, but contain a
symmetric part in the field equations of the gauge fields. For these models
the conditions of properness and completeness are not equivalent, and now
the statement of completeness requires to add these field equation
symmetries to the other two (the conformal and supersymmetry). In total we
thus have four symmetries.
One can then check that the gauge algebra of these four symmetries are of
the form \eqn{opalg}, with $E^{ab}$ graded-antisymmetric.

\subsection{Reducibility}
Having these null fields, we are in the situation of reducible constraints,
and so, we need to know all the relevant zero modes satisfying
\begin{equation}
R^i_aZ^a_{a_1}=2y_jf^{ji}_{a_1}\ ,     \label{zeromodes}
\end{equation}
where now the index $a=1,...,4$ and $i$ runs over the matter (or
realisation independent, $T$ and $G$) plus gauge fields. By relevant
we mean that they correspond to nontrivial cycles under the Koszul-Tate
differential. This point was explained in section 3.2.3 and illustrated in
our toy-model of the previous section.
The matrix of gauge generators is given by
\begin{equation}
R^i_a=\pmatrix{\partial T+2T\partial &-3/2\partial G-5/2G\partial & 0 & 0
\cr \partial G+5/2G\partial  & T^2 & 0 & 0 \cr \nabla^{(-1)} & -fT & 0 &
2T^2 \cr \partial f-3/2f\partial & \nabla^{(-3/2)} & 9\partial T+4T\partial
& -15\partial G}\ ,
\end{equation}
so we have four ghosts $c^a;a=1,...,4$.
We expect that the zero modes
are related to the relations $N_i$. Let us first look to the
transformations of the matterfields $\phi^i$. Consider the
transformations generated by taking Poisson brackets with the $N_i$.
Because the relations contain only the generators $T,G$, we can use
Leibniz rule and some partial integrations to rewrite the transformation
as a combination of those generated by $T,G$. For example~:
\bea
\int \zeta^1 [N_1, \phi^i] &=&
\left(\delta_{\epsilon=9 \zeta^1 \del G  + 5 \del \zeta^1 G}-
       \delta_{\alpha=9 \del T \zeta^1 + T \del \zeta^1}
         \right) \phi^i \ .\label{symrelation1}
\eea
However, because $N_1=0$, the previous equation gives us a relation
between the transformations of the matter fields (valid for every
realisation). Similarly, via $N_2$ we can find another relation between
the gauge transformations acting on the matterfields.

These two relations satisfy eqn.~\eqn{zeromodes} off shell, i.e.
with vanishing $f$ coefficients, for the $i$--index running
over the matter fields, . However, a zero mode must have four entries. The
above relations only determine the first two, because the $R^i_a$ matrix
has zeroes in the right upper corner. The two other entries are determined
by requiring that the relation \eqn{zeromodes} also holds when the index
$i$ runs over the gaugefields. Then we find
\be
\begin{array}{cc}
-{9\over 4} \del G - {5\over 4} G \del  & 3 T^2\\
-{9\over 4} \del T - T \del  & -15 \del G - {15\over 2} G \del\\
\nabla^{-{9\over 2}} & -{1\over 8} f T\\
{3\over 2} \del f -{1\over 2}f \del & \nabla^{-5}
\end{array} \label{Z1a}
\ee
together with some nonvanishing $f^{ij}_{a_1}$'s. These zero
modes
are relevant because, as one can check, they correspond with non exact $KT$
invariants for which we now introduce ghost for ghosts~:
\begin{eqnarray}
\delta _{KT}c^*_{1_1}&=&-\nabla c^*_3-5c^*_2\partial T+4\partial
c^*_2-4\nabla h^* \partial f^*+5\partial \nabla h^*f^*\nonumber\\
&&+\frac{1}{2}\partial c^*_4f+2c^*_4\partial f-4c^*_1\partial G+5\partial
c^*_1G-20f^*\partial f^*\partial f\nonumber\\
&&-\frac{25}{4}f^*f^*\partial ^2f-\frac{15}{2}f^*\partial
^2f^*f-\frac{3}{4}\partial f^*\partial f^*f+3h^*\partial h^*fT\nonumber\\
\delta _{KT}c^*_{2_1}&=&-\nabla c^*_4-6T^2c^*_1+15c^*_2\partial
G-15\partial c^*_2G-\frac{1}{2}c^*_3fT\nonumber\\
&&-4h^*\nabla h^*T+7h^*\partial f^*fT-\frac{25}{2}h^*f^*f\partial
T\nonumber\\
&&+15h^*f^*\partial fT+15\partial h^*f^*fT+\frac{15}{2}f^*\nabla\partial
f^*+\frac{15}{2}\partial \nabla f^*f^*\ .
\end{eqnarray}
Using the terminology of previous section, these zero modes are called
dynamical. One sees that computations become rather involved, especially
when going to higher antifieldnumber. The calculations can be done using
Mathematica
\cite{KrisproBV}. Apart from these dynamical zero modes, we also have
algebraic zero modes, corresponding to the non-exact $KT$ invariants
\begin{eqnarray}
\delta _{KT}c^*_{3_1}&=&c^*_4\partial T+2T^2h^*\partial h^*-15f^*\partial
G\partial h^*\nonumber\\
\delta _{KT}c^*_{4_1}&=&c^*_3T^2-15\partial f^*\partial
f^*G+\frac{25}{2}f^*\partial ^2f^*G\ ,
\end{eqnarray}
for which we need two further ghost for ghosts $c^{3_1}$ and $c^{4_1}$.

With this information we can continue the
computation of the extended action one step further. At the next level, we
have to look for zero modes of the $Z$ matrix~:
\bea
Z^{a}_{a_1} Z^{a_1}_{a_2} = y_i f^{ia}_{a_2}\ .
\label{zzeromodes}
\eea
Again, this is only a necessary condition, and we have to compute the
cohomology of $\delta_{KT}$ (now at antifield level 3). The results is
\be
Z^{a_1}_{a_2}=\pmatrix{
0 & 0  & 0 & 0 & 0 & 0 \cr
0 & 0  & 0 & 0 & 0 & 0 \cr
\nabla^{-8} & 0 & 0 & T^2 & 0 & 0 \cr
4\partial  f -\frac{3}{4}f \del & \nabla^{(-\frac{17}{2})}& \del T & 0 &
T^2 & TG \cr}  \label{ZZ2}
\ee
They again consist of dynamical and algebraic zero modes.
In the same way, we will find zero modes for
$Z^{a_1}_{a_2}$ and so on. This means
a gauged \W52 system is reducible with an infinite number of stages.

\subsection{Gauge fixing and BRST operator}
In this section, we show briefly how the gauge fixing can be performed and
how the resulting BRST charge will look like. We assume that we have
succeeded in finding all zero modes at all stages. So, we have
introduced ghosts $c^a$ for every symmetry $R^i_a$, ghosts-for-ghosts
$c^{a_1}$ for every zero mode $Z^a_{a_1}$ and so on. The extended action
takes the form
\bea
S &=& S^0 + \phi^*_i R^i_a c^a +
    c_{a_i}^* Z^{a_i}_{a_{i+1}} c^{a_{i+1}} + \ldots
\eea
where $\phi ^*_i=\{T^*,G^*,h^*,f^*\}$ and the ellipsis denotes terms at
least quadratic in antifields or ghosts. They
are determined by the master equation (e.g. the field equation matrices
$E^{ij}_{ab}$ and $f^{ij}_{a_1}$ appear in the part quadratic in
antifields, and the structure functions $T^a_{bc}$ appear in the part
quadratic in the ghosts). It turns out we can always choose a basis such
that there are no terms with $T^*$ nor $G^*$, except in $S^1$. Therefore,
there are no difficulties in taking derivatives w.r.t. $T$ and $G$ and
their antifields.

We will denote $c^{a_i}$ corresponding to a dynamical zero mode in
$Z^{a_{i-1}}_{a_i}$  collectively as $c^{\{a_i\}}$, and
those corresponding to algebraic zero modes as $\tilde{c}^{\{a_i\}}$.
The gauge fixing is can easily be done analogous to the toy-model.
Again, the algebraic zero modes at antifield number $n$ determine the
dynamical zero modes at antifield number $n+1$. The antifields of the ghosts
for ghosts coming from the algebraic zero modes become, after gauge
fixing, the
antighosts of the ghosts for ghosts for the dynamical zero modes
\cite{disp}.
So, we will chose the gauge corresponding to the
canonical transformation
\begin{eqnarray}
h^*=b_{\{1\}} &\qquad& h=-b_{\{1\}}^*\nonumber\\
f^*=b_{\{2\}} &\qquad& f=-b_{\{2\}}^*\nonumber\\
\tilde{c}^*_{\{a_i\}} =b_{\{a_{i+1}\}} &\qquad&
\tilde{c}^{\{a_i\}}=-b^{*\{a_{i+1}\}}\ .
\end{eqnarray}
It consists of putting the
gauge fields
$h,f$ and all the ghosts $\tilde{c}^{\{a_i\}}$ to zero.
The gauge fixed action can then be brought to the from of a free field
theory
\be
S_{gf}= S_0 + \sum b_{\{a_i\}} \db c^{\{a_i\}}\ ,
\ee
where sum is over all "dynamical" ghost at each
stage and then summed over all (infinitly many) stages.
Moreover, the extended action is linear in the new antifields. This can be
proven using dimensional arguments. We leave this as an exercise for the
reader. This means that the
BRST transformations $\delta \Phi ^A=(\Phi ^A,S)$ in this gauge choice are
nilpotent off shell. The BRST
operator that generates these transformations can be derived from
\begin{equation}
\delta \Phi ^A=[\Phi ^A,Q]\ .
\end{equation}
Applied to our case this gives a BRST current that starts
like~:
\begin{eqnarray}
Q&=&c^{\{1\}} T + c^{\{2\}} G +
(5 b_{\{1\}}\del G - 4 \del b_{\{1\}} G) c^{\{1_1\}}\nonumber\\
&&+ (15 b_{\{1\}} T^2 - 2 b_{\{2\}}\del G) c^{\{2_1\}} + \ldots
\end{eqnarray}

\chapter{Anomalies and Pauli-Villars regularisation}
\section{The basic philosophy}
In this section we review the basic philosophy behind the Pauli-Villars
(PV) regularisation procedure \cite{PV} that we will use and we point out
what the main
steps are in calculating anomalies with this scheme. It was first shown in
\cite{Diaz} how this method can be used to obtain consistent Fujikawa
regulators for anomaly calculations. The advantage of Pauli-Villars
regularisation is that it gives a regularised expression for the
complete one-loop path integral and not only for specific diagrams.
The one-loop regularised path integral provides a regularised expression
for the Jacobian of the measure under transformations of the fields.
For a more detailed description of this set-up, we
refer to \cite{Diaz,measure} and to
\cite{DJTH,RuudTH,bvleuv} for more pedagogical expositions and
applications. For a review on other regularisation schemes, see
\cite{Bonneau}.

Suppose that we start from an action $S[\Phi^A]$ that has some rigid
symmetries given by $\delta_\epsilon \Phi^A = \delta _\alpha
\Phi^A \epsilon^\alpha $. This can be the gauge fixed action of a certain
theory, or simply an action without local symmetries.
One introduces for every field $\Phi^A$ a
Pauli-Villars partner $\Phi^A_{PV}$ of the same Grassmann parity. These PV
fields have the action
\begin{eqnarray}
  S_{PV} & = & \frac{1}{2} \Phi^A_{PV} (\frac{\dl}{\partial \Phi^A}
           \frac{\dr}{\partial  \Phi^B} S) \Phi^B_{PV} \nonumber \\
         &\equiv& \frac{1}{2} \Phi^A_{PV}
         S_{AB}\Phi^B_{PV}\ .   \label{SPV}
\end{eqnarray}
From this, it follows that $S+S_{PV}$ is invariant under the following
transformation rules~:
\begin{eqnarray}
\delta _\epsilon \Phi ^A&=&[\delta _\alpha \Phi^A
+\frac{1}{2}\Phi ^B_{PV}(\frac{\dl}{\partial \Phi^B}
           \frac{\dr}{\partial  \Phi^C}
\delta _\alpha \Phi^A )\Phi ^C_{PV}]\epsilon ^\alpha \nonumber\\
\delta _\epsilon \Phi ^A_{PV}&=&\frac{\dr \delta_\alpha
\Phi^A}{\delta \Phi^B} \Phi^B_{PV}\epsilon^\alpha ,
   \label{linPV}
\end{eqnarray}
up to fourth order in PV fields. These terms can be neglected since we only
work at one loop.

We also have to choose a mass-term $S_M$ for the PV fields,
which generically takes the form
\begin{equation}
  S_M =-\frac{1}{2} M^2 \Phi^A_{PV} T_{AB} [\Phi]  \Phi^B_{PV} .
  \label{defT}
\end{equation}
The matrix $T_{AB}$ has to be invertible, such that all PV fields have well
defined propagators. The mass term is in general not invariant under
\eqn{linPV}. This depends on the choice of the $T$ matrix.

The one-loop regularised path integral is then given by
\begin{equation}
  \cZ_R  = \int [d\Phi][d\Phi_{PV}]  e^{\ihbar(S+S_{PV}+S_M)}.
\end{equation}
The Gaussian integral over the PV fields is defined in such a way that the
PV fields actually regularise. The PV fields itself generate extra
diagrams which have to be added to the original ones. The sum of the two
diagrams can be made finite if the divergences of the PV diagram cancel the
divergences of the original diagram. This cancellation of divergences can
be achieved when the PV loops produce an
extra minus sign w.r.t. the corresponding loops of the original
particles.

To do this properly, one should introduce several copies
$\Phi^A_i;i=1,...,N$ of PV fields, each copy carrying a (different)
mass $M_i$ and a number $c_i$. The total PV action is then (dropping the PV
index)
\begin{equation}
S_{PV}+S_M=\frac{1}{2} \sum _{i=1}^{N}\Phi^A_i
         [S_{AB} -M^2_iT_{AB}(\Phi)]\Phi^B_i\ .
\end{equation}
The integration over the PV fields is defined by
\begin{equation}
\int [d\Phi
_i]e^{\frac{i}{\hbar}(S_{PV}+S_M)}=[det(S_{AB}-M^2_iT_{AB})]^{-\frac{1}{2}c
_i}\ .
\end{equation}
Diagrams can now be regulated if we impose (at least) the conditions
\begin{equation}
\sum_{i=0}c_i=0 \qquad \sum_{i=0} c_iM^2_i=0\ ,\label{PVcond}
\end{equation}
where $c_0\equiv1$. As an example, one could take three PV fields with
$c_1=c_2=-1, c_3=1$ and masses $M_3^2=2M_2^2=2M_1^2$. When the PV fields
are bosonic, it is hard to imagine that we can take its $c$ number equal to
minus one. To make it more realistic, one can replace each such boson by
two ordinary fermions and an extra boson, together with the usual
integration rules.

The fundamental principle of PV regularisation is that the total measure is
invariant under BRST transformations or other rigid symmetries, i.e. under
\eqn{linPV}. From the definition of PV integration it follows
that the PV measure transforms with Jacobian
\begin{equation}
J_{PV}=[\det (\frac{\partial}{\partial \Phi ^B}\delta _\alpha \Phi
^A)\epsilon ^\alpha ]^{\sum _{i=1}c_i}\ .
\end{equation}
The Jacobian of the PV fields then cancels the Jacobian
of the ordinary fields when imposing the condition $\sum_{i=0}c_i=0$.
So at one loop, the total measure is invariant. The only
possible non-invariance in the regularised path integral is the mass
term $S_M$.

The non-invariance of the mass term
could cause anomalies. In this scheme we then have that possible
anomalies\footnote{A small
remark about our terminology is needed here. We often speak of `anomaly'
when we really mean the regularised Jacobian of the measure under a
specific symmetry transformation. Sometimes we use the term `genuine
anomaly' when an anomaly -a regularised Jacobian- can not be cancelled by
the addition of a local counterterm.}
are given by
\begin{equation}
   \cA_{\epsilon}[\Phi ]= \int [d\Phi_{PV}] \ihbar \delta_\epsilon
   S_M  e^{\ihbar(S_{PV}+S_M)},
\end{equation}
in the limit $M\rightarrow \infty$.
If we define a matrix $Z_{AB}$ by
\begin{equation}
    \delta_\epsilon S_M = -M^2 \Phi^A_{PV} Z_{AB}
[\Phi,\epsilon ] \Phi^B_{PV} \  ,   \label{defZ}
\end{equation}
we obtain the expression in matrix notation
\begin{equation}
  \cA_{\epsilon} = -\str \left[ M^2 Z \frac{1}{S - M^2 T}
  \right]
\end{equation}
for the anomaly in the $\epsilon$-symmetries in this regularisation
scheme\footnote{The supertrace here is that $str K=(-)^{A(K+1)}K^A{}_A$}.
Using the cyclicity of the trace, we have
\begin{equation}
   \cA_{\epsilon} =  \str \left[ J \frac{1}{
   1-\frac{\cR}{M^2}}\right] \ ,  \label{SSMPV}
\end{equation}
with the definition of the jacobian $J = T^{-1} Z$ and with $\cR =
T^{-1} S$.

It is now easy to make the connection with the approach of Fujikawa
\cite{FujiAll} for calculating anomalies. Using
\beq
   \int_0^\infty d\lambda \,\, e^{-\lambda} e^{\lambda X} =
   \frac{1}{1-X} \, ,
\eeq
the expression for the anomaly may be written as
\beq
  \cA_{\epsilon} = \lim_{M \rightarrow \infty} \int_0^\infty
   d\lambda \,\, e^{-\lambda}  \str \left[ J e^{\frac{\lambda}{M^2}
   \cR} \right] \, .
   \label{deze}
\eeq
Notice that we have carelessly interchanged the integral over $\lambda$
and the $\str$. When computing only the finite part of the anomaly, this
causes no problem. This is because $\lambda $ is always appearing in
combination with $M^2$. So, the $M^2$ independent part of the supertrace
will also be $\lambda $ independent, as we will see below. Once we know
this, we can perform the
$\lambda $ integral, which gives one. For the infinite part of the anomaly,
one cannot simply commute the trace and the $\lambda $ integral.  This is
clearly explained in \cite{bvleuv}. A careful
analysis leads to extra logarithmically diverging terms, like $M^2\log
M^2$, which have to be absorbed by the renormalisation before the limit
$M^2\rightarrow \infty $ is taken.

To evaluate the traces, one can put the operator between a basis of
plane waves. So,
\begin{eqnarray}
\str\left[J\exp{t{\cal R}}\right]&=& \int d^dx\,\int d^dy \delta
(x-y)J(x) e^{t{\cal R}_x} \delta (x-y)\nonumber\\
&=&
\int d^dx\int \frac{d^dk}{(2\pi )^d}e^ {-ikx}J e^{t{\cal R}_x}
e^{ikx}\ ,
\end{eqnarray}
where $t=\lambda /M^2$.
Then one pulls the $e^{ikx}$ to the left, replacing derivatives by
$\partial +ikx$, and one takes the trace. However, this whole procedure is
included in the results of the heat kernel method \cite{heatk}. The
heat kernel is the expression
\begin{equation}
e^{t{\cal R}_x}\delta (x-y)= G(x,y;t;\Phi ) \ .\label{defGdel}
\end{equation}
This has been considered for
second order differential operators ${\cal R}$ of the form
\begin{equation}
{\cal R}_x[\Phi ]= {1\over {\sqrt g}}\left( \partial_\alpha \unity +{\cal
Y}_\alpha  \right)
{\sqrt g}\, g^\alpha {}^\beta  \left( \partial_\beta \unity +{\cal Y}_\beta
\right) + E  \ , \label{formcR}
\end{equation}
where $\Phi =\{g^{\alpha \beta  },{\cal Y}_\alpha  ,E\}$. The latter two
can be matrices in an internal space. The restriction here is that
the part which contains second derivatives
is proportional to the unit matrix in internal space. Further, in principle
$g$ should be a positive
definite matrix. For Minkowski space, we have to perform first a Wick
rotation, which introduces a factor $-i$ in the final expression, and thus
we have to use
\begin{equation}
\str\left[J\exp{\frac{{\lambda \cal R}}{M^2}}\right]=-i \int d^dx\,\int
d^dy \delta (x-y)J(x) G\left(x,y;\frac{\lambda }{M^2}\right) \ .
\end{equation}

The `early time' expansion of this heat kernel is as follows
\begin{equation}
G(x,y;t;\Phi )= \frac{\sqrt{g(y)}\,\Delta ^{1/2}(x,y)}{ (4\pi t )^{d/2}}
e^{-\sigma (x,y) /2t } \sum_{n=0}a_n(x,y,\Phi ) t^n\ ,\label{Gexpbn}
\end{equation}
where $g=|det\ g_{\alpha \beta }|$, and
 $\sigma (x,y)$ is the `world function', which is discussed at length
in \cite{Synge}. It is half the square of the geodesic distance between $x$
and $y$
\begin{equation}
\sigma (x,y)=\frac{1}{2}g_{\alpha \beta }(y-x)^\alpha
(y-x)^\beta  +{\cal O}(x-y)^3\ .
\end{equation}
Further, $\Delta (x,y)$ is defined from the `Van Vleck--Morette
determinant' (for more information on this determinant, see
\cite{VisservV})
\begin{equation}
{\cal D}(x,y) =\left|det\left( -\frac{\partial ^2\sigma }{\partial x^\alpha
\partial y^\beta  }\right) \right| \end{equation}
as
\begin{equation}
{\cal D}(x,y) =\sqrt{g(x)}\,\sqrt{g(y)}\,\Delta (x,y)\ .
\end{equation}
At coincident points it is 1, and its first derivative is zero.

The `Seeley--DeWitt' coefficients $a_n(x,y,\Phi )$ have been obtained using
various methods for the most important cases. In two dimensions the
relevant coefficients are $a_0$ for the infinite part and
$a_1$ for the finite part, while in four dimensions these are $a_1$ and
$a_2$. For most applications we only
need their value and first and second derivatives at coincident points.
Remark that, when interchanging the trace and the $\lambda $
integral, one would never obtain logarithmic terms in $M^2$. Instead, the
diverging terms one now obtains are quadratic, which vanish upon using
the conditions $\sum c_iM^2_i=0$. This leads to wrong results.

It is clear that when regularising the theory in this way, one has the
freedom to choose different mass terms, i.e. different choices of the
matrix $T_{AB}$. When we have a set of rigid symmetries, it is not always
possible to choose a $T$ matrix that preserves all of them. Instead,
choosing different $T$ matrices can correspond to keeping different
symmetries manifestly invariant. The question then arises whether changing
$T$ will change the quantum theory. As was conjectured in \cite{anombv},
and proven in \cite{hideanom}
the two theories, determined by taking two different $T$ matrices, are
related by a local counterterm $M_1$.
The argument goes as follows. We want to find the counterterm such that
\begin{equation}
  \cZ_R  = \int [d\Phi][d\Phi_{PV}]  e^{\ihbar(S+S_{PV}-\frac{1}{2}
  M^2 \Phi^A_{PV} T_{AB}(\alpha ) \Phi^B_{PV}+\hbar M_1(\alpha ) )},
\end{equation}
is independent of $\alpha $, i.e. $\frac{d\cZ_R}{d\alpha }=0$. This
leads to a differential equation for $M_1(\alpha )$, whose solution
is \begin{equation}
M_1(\alpha )=\int ^\alpha dt\,
\str[T^{-1}\frac{dT}{dt}e^{\frac{\cR(t)}{M^2}}]\ ,\label{interp}
\end{equation}
again, in the limit $M^2\rightarrow \infty$.

In the next two sections we will illustrate these ideas in two examples. In
the first example, we will compute BRST anomalies and show how to work with
a field dependent $T$ matrix. As a second example, we will compute
anomalies
in different rigid symmetries and comment on the interpolation formula
\eqn{interp}.

\section{Strings in curved backgrounds}
This section is based on \cite{strbackg}.
We will investigate the anomaly structure coming from matter loops
of the bosonic string in a nontrivial background.
Consider the classical action
\begin{equation}
S^0=-\frac{1}{2}\int d^2x\sqrt hh^{\alpha \beta}\partial_\alpha X^\mu
\partial_\beta X^\nu G_{\mu \nu}(X) \ ,
\end{equation}
where $\mu =1,...,D$ and $G_{\mu \nu}(X)$ is a general metric on space
time. This action has both local reparametrisation and
Weyl invariance. These symmetries can be recognised in the extended
action at
antifieldnumber one~:
\begin{equation}
S^1=X^*_\mu c^\alpha \partial_\alpha X^\mu+h^{*\alpha \beta}
[c^\gamma\partial_
\gamma h_{\alpha \beta}+(\partial_\alpha
c^\gamma)h_{\gamma\beta}+(\partial_\beta c^\gamma)h_{\alpha \gamma}+c
h_{\alpha \beta}]\ ,
\end{equation}
where $c^\alpha $ and $c$ are the ghost for reparametrisation resp. Weyl
invariance. Using the techniques of chapter 3, one can find the extended
action at antifield number two~:
\begin{equation}
S^2=-c^*_\beta c^\alpha \partial_\alpha c^\beta -c^*c^\alpha \partial_\alpha
c.
\end{equation}
One can check that there are no further terms at higher antifieldnumbers,
i.e. $S=S^0+S^1+S^2$ satisfies the classical master equation $(S,S)=0$.
A gauge fixed
action is then obtained by doing the canonical transformation
\begin{equation}
h_{\alpha \beta}=-b^*_{\alpha \beta}\qquad h^{*\alpha \beta}=
b^{\alpha \beta} \ ,
\end{equation}
where $b^*_{\alpha \beta}$ is now treated as an external source and
$b^{\alpha \beta}$ as the antighost. The BRST rules can be computed as
$\delta \Phi ^A=(\Phi ^A,S)|_{\Phi ^*=0}$ in the gauge fixed basis.

Now we regulate the theory using Pauli-Villars regularisation. Since
we will
only deal with the matter loops, we only introduce PV partners for them,
say $X^\mu _{PV}$. To construct an action for the PV fields, we have to
choose a mass term. Here we will choose a field dependent one
\begin{equation}
S_M=-\frac{1}{2}M^2h^aX^\mu _{PV}G_{\mu \nu}(X)X^\nu _{PV}\ .
\end{equation}
So the mass matrix we have chosen is
\begin{equation}
T_{\mu \nu}=h^aG_{\mu \nu}(X)\ .
\end{equation}
The regulator then is
\begin{equation}
{\cal R}^\mu{} _\nu=(T^{-1})^{\mu \rho}S_{\rho\nu}\ ,
\end{equation}
where $S_{\rho\nu}$ is the matrix of second derivatives of the action
w.r.t. the fields.
Remark that the regulator is reparametrisation invariant if $a=1/2$,
resp. Weyl invariant if $a=0$. So, we will have a Weyl anomaly for $a=1/2$,
and an Einstein anomaly for $a=0$. To compute the anomaly we have to
write the regulator as a second order differential operator~:
\begin{equation}
{\cal R}^\mu{}_\nu=\frac{1}{\sqrt g}(\partial_\alpha \unity +\cY_\alpha )
\sqrt gg^{\alpha \beta}(\partial_\beta \unity +\cY_\beta)+E\ ,
\end{equation}
where the derivatives keep on working to the right.
The objects $\cY_\alpha $ and $E$ are matrices in the internal space with
indices $\mu ,\nu $. Now, we can read off these objects from our
computation~: \begin{eqnarray}
g_{\alpha \beta}&=&h_{\alpha \beta}h^{(a-\frac{1}{2})}\nonumber\\
(\cY_\alpha )^\mu {}_\nu&=&\partial_\alpha X^\sigma\Gamma^\mu
{}_{\sigma\nu}\nonumber\\
E^\mu {}_\nu&=&g^{\alpha \beta}G^{\mu
\lambda}\partial_\beta X^\sigma\partial_\alpha X^\omega G_{\sigma\rho}
R^\rho{}_{\lambda \omega \nu}\nonumber\\&&+G^{\mu \lambda}\Box _g
X^\sigma \Gamma _{\sigma \lambda \nu}\nonumber\\&&+g^{\alpha \beta}
G^{\mu \lambda}\partial_\beta X^\sigma \partial_\alpha
X^\omega \Gamma _{\rho\sigma\omega}\Gamma^\rho{}_{\lambda \nu}\ ,
\end{eqnarray}
where $\Box _g=g^{-1/2}\partial _\alpha g^{1/2}g^{\alpha \beta }\partial
_\beta $. The expression for $E$ can also be written as
\begin{equation}
E^\mu {}_\nu =g^{\alpha \beta}
\partial_\beta X^\sigma\partial_\alpha X^\omega
R^\mu {}_{\sigma \nu \omega }+g^{-1/2}G^{\mu \lambda}y_\rho \Gamma^\rho{}
_{\lambda\nu}\ ,
\end{equation}
where $y_\rho$ is the field equation for $X^\rho$. Since terms
proportional to field equations in anomalies can always be cancelled (see
at the end of section 3.3) , we will drop this term.

The jacobian is
\begin{eqnarray}
J^\mu {}_\nu&=&\frac{1}{2}a\delta^\mu _\nu[h^{\alpha
\beta}c^\gamma\partial_\gamma h_{\alpha \beta}+2\partial_\alpha c^\alpha
+2c]\nonumber\\&&+\Gamma^\mu {}_{\nu\sigma}c^\alpha
\partial_\alpha X^\sigma+\delta^\mu _\nu c^\alpha \partial_\alpha
\nonumber\\&=&\frac{1}{2}a\delta^\mu _\nu[h^{\alpha
\beta}c^\gamma\partial_\gamma h_{\alpha \beta}+2\partial_\alpha c^\alpha
+2c]\nonumber\\&&+c^\alpha (\partial_ \alpha \unity +\cY_\alpha)\ .
\end{eqnarray}
To compute the anomalies, we need the following Seeley-De Witt coefficients
\begin{eqnarray}
a_0|=1 &\qquad& \nabla _\alpha a_0(x,y)|=0\\
a_1|=E-\frac{1}{6}R(g) &\qquad & \nabla _\alpha a_1(x,y)|=\frac{1}{2}
\nabla _\alpha (E-\frac{1}{6}R(g))+\frac{1}{6}\nabla ^\beta W_{\alpha \beta
}\nonumber\\ \ , \end{eqnarray}
where $|$ stands for the value at coincident points $x=y$ (after taking the
derivatives), the covariant derivative is with connection $\cY _\alpha $
and $W_{\alpha \beta }=\partial _\alpha \cY_\beta -\partial _\beta \cY
_\alpha +[\cY _\alpha ,\cY _\beta ]$.

For the divergent part of the anomaly, taking into account the
logarithmically diverging terms we mentioned before, we find
\begin{equation}
\cA(a)=M^2\log M^2\frac{D}{4\pi}\int h^a[(a-\frac{1}{2})\partial_\alpha
c^\alpha +ac] \ .
\end{equation}
Miraculously, the terms with the connection coefficients have
cancelled each other.
This means the renormalisation procedure is the same as in a flat
bacground.
Moreover, one sees that for $a=0$, that is a Weyl invariant regulator,
there is
no divergent part of $\Delta S$ (dropping boundary terms), while for
$a=1/2$, the $c^\alpha $ terms
drop out (because the regulator is now reparametrisation invariant) and one
gets
\begin{equation}
\cA_c(a=\frac{1}{2})=M^2\log M^2\frac{D}{8\pi}\int \sqrt h c\ .
\end{equation}
This term however can be absorbed by adding a counterterm proportional to
the cosmological constant
\begin{equation}
\cA_c(a=\frac{1}{2}) =(S,M_1) \qquad M_1=\frac{1}{8\pi}M^2\log M^2\int
\sqrt h\ .
\end{equation}
Remark that for $a=0$ the cosmological
constant is not renormalised, so we did not have to introduce it.
However, for $a=1/2$ it will be renormalised. This is a clear difference
between the two regularisation schemes.
Now we will study the finite part of the anomaly.
We find
\begin{eqnarray}
\cA(a)&=&-\frac{D}{24\pi}\int h^a[(a-\frac{1}{2})\partial_\alpha
c^\alpha +ac]R(g)\nonumber\\&&+\frac{1}
{4\pi}\int \sqrt h
h^{\beta \gamma}\partial_\gamma X^\mu \partial_\beta X^\nu
R_{\mu \nu}(X)[(a-\frac{1}{2})\partial_\alpha c^\alpha +ac]\ ,
\end{eqnarray}
where we made use of the identity $\Gamma^\mu {}_{\mu \sigma,\rho}
-\Gamma^\mu {}_{\mu \rho,\sigma}=0$, which can be proven by writing the
Christoffel symbols in terms of the metric.
Let us consider the two cases $a=0$ and $a=1/2$. In the first case the
anomaly must be proportional to $c^\alpha $, while in the second case it
must be proportional to $c$. First we have to write the Ricci scalar,
defined by the "regulator metric" $g_{\alpha \beta }$
in terms of the world sheet metric $h_{\alpha \beta }$.
The relation for $a=0$ is
\begin{equation}
R(g)=\sqrt hR(h)-\frac{1}{2}\sqrt h \Box _h \log h\ .
\end{equation}
For $a=1/2$, we of course have that $R(h)=R(g)$.
The results are~:
\begin{eqnarray}
\cA _c(a=1/2)&=&-\frac{D}{48\pi}\int c\sqrt hR(h)\nonumber\\
&&+\frac{1}{8\pi}\int c\sqrt hh^{\alpha \beta}\partial_\alpha X^\mu
\partial_\beta X^\nu R_{\mu \nu}(X)\ ,
\end{eqnarray}
and
\begin{eqnarray}
\cA _{c^\alpha }(a=0)&=&\frac{D}{48\pi}\int \partial_\alpha c^\alpha
[\sqrt hR(h)-\frac{1}{2}\sqrt h \Box _h ln h]\nonumber\\
&&-\frac{1}{8\pi}\int \partial_\alpha c^\alpha
\sqrt hh^{\alpha \beta}\partial_\alpha X^\mu
\partial_\beta X^\nu R_{\mu \nu}(X) \ .
\end{eqnarray}
One can compute the counterterm $M_1$ that is needed to shift the
anomaly from the Weyl sector to the Einstein sector, using \eqn{interp}.
For a flat background, this was done in \cite{anombv}. In the case of a
curved background, we have an extra term such that
\begin{eqnarray}
M_1&=&\frac{1}{8\pi }[M^2\log M^2-\frac{1}{12}\log h\sqrt hR(h)
+\frac{1}{48}\log h \Box _h\log h\nonumber\\
&&+\frac{1}{2}\log h \sqrt hh^{\alpha \beta}\partial_\alpha X^\mu
\partial_\beta X^\nu R_{\mu \nu}(X)]\ .
\end{eqnarray}

What are the conditions for the absence of anomalies ~?
The first part, independent of the chosen background metric,  of both
($a=0$ and $a=1/2$) anomalies can be cancelled
by the ghost loops (which we have not treated here), provided one takes the
dimension to be $D=26$.
The part coming from the background vanishes if the Ricci tensor vanishes~:
\begin{equation}
R_{\mu \nu}(X)=0\ ,\label{Ricciflat}
\end{equation}
which are precisely Einstein's field equations in empty space.
One can easily generalise this technique to more complicated
backgrounds,  e.g. when including an antisymmetric tensor, dilatons,
etc. .
The above results, and the extensions to more general backgrounds, were
already found in \cite{anombgr}, using dimensional regularisation.
There \eqn{Ricciflat} arises as a condition for the vanishing of the
$\beta $ function.

\section{Fermions and $b-c$ systems}
In this section we will investigate anomalies in rigid symmetries.
As an example we take a very simple two fermion model
\begin{equation}
S=2i{\bar \psi}\partial _-\psi +2i{\bar \xi} \partial _+\xi \ ,\label{tfact}
\end{equation}
where the ${\bar \psi},{\bar \xi}$ is the complex conjugate of $\psi ,\xi $. The
factor $i$ makes the action real. The (Lorentz) spins $s$ of the
fields
are $\lambda ,\gamma $ for $\psi ,\xi $ and  $-1-\lambda,1-\gamma $ for
${\bar \psi},{\bar \xi}$. So, we have a rigid symmetry
\begin{equation}
\delta _L\phi ^i=s\phi ^i\ ,
\end{equation}
where $\phi ^i$ denotes the collective set of fields.
We are especially interested in the following three cases:
\begin{eqnarray}
\mbox{ Fermion phase : } && \lambda =-1/2 \qquad \gamma =1/2\nonumber\\
\mbox{ A phase : } && \lambda =0 \qquad \gamma =1\nonumber\\
\mbox{ B phase : } && \lambda =-1 \qquad \gamma =1
\end{eqnarray}
The interest in these special cases comes from topological field theory
\cite{TFT,TFTrep}. It is known that one can twist \cite{TFT,EY} an
$N=2$ supersymmetric
field theory, with the above spin 1/2 fermions, into a so called A or B
topological field theory \cite{WittAB,BF}. Here, we consider only the
fermionic subsectors of these models. After twisting, the fermions are
declared to be ghosts with ghostnumbers :
\begin{eqnarray}
\mbox{ Fermion phase : } && gh(\psi )=0 \qquad gh(\xi ) =0\nonumber\\
\mbox{ A phase : } && gh(\psi ) =1 \qquad gh(\xi ) =-1\nonumber\\
\mbox{ B phase : } && gh(\psi ) =-1 \qquad  gh(\xi )=-1
\end{eqnarray}
So, in the A and B phase, the fermions can be interpreted as $b-c$ systems.
These models can have, as we will see, anomalies in ghosts number or
Lorentz symmetry. In fact, the topological twist procedure must be done
in a regularised way. Choosing different mass terms will respect either
ghost number or Lorentz symmetry. The results presented here are based on
\cite{RegTwist}.

To regularise the theory, we have to choose a mass term. There are three
obvious choices
\begin{eqnarray}
S_M^0&=&2M[\psi {\bar \psi}+\xi {\bar \xi}]\nonumber\\
S_M^1&=&2M[\psi {\bar \xi}+\xi {\bar \psi}]\nonumber\\
S_M^2&=&2M[\psi \xi +{\bar \xi}{\bar \psi}]\ .
\end{eqnarray}
The first one we will never use, since it never preserves Lorentz
invariance explicitly. So we concentrate on $S_M^1$ and $S_M^2$.
Let us first consider the fermion phase. One sees that both mass terms
respect Lorentz as well as ghost number symmetry. So, there are no
anomalies in the fermion phase. Consider now the A phase. The first mass
term is Lorentz invariant, so there is no Lorentz anomaly. However, the
first term in $S_M^1$
has ghost number 2 and the second term has ghost number -2. We thus expect
a possible anomaly in the ghost number current. For the second mass term,
it is vice verca: it has ghost number zero but we expect a Lorentz anomaly.
An analogous reasoning can be made for the B phase.
When prefering Lorentz invariant results, $S_M^1$ is useful for the A phase
and $S_M^2$ can be used for the B phase.

However, in the context of twisting $N=2$ theories, only $S_M^2$ will
preserve supersymmetry explicit, both for the A and B
phases\footnote{After the twist, a certain combination of the
supersymmetry charges defines the BRST charge.}. Of course, to understand
this, one must add the bosonic sector to the theory. It is beyond
the scope of this chapter to explain this. For more details, see
\cite{RegTwist}. So, we imagine this fermion model to be embedded in a
larger theory where we want to regularise in a manifesly
supersymmetric
(for the fermion phase) or BRST (in the A and B phases) invariant way.
Therefore, we can only use the mass term $S_M^2$.

So, if we want to regulate the A phase, we will start from the mass term
$S_M^2$, and produce the Lorentz anomaly. To obtain Lorentz invariant
results, we have to compute the counterterm that moves the
anomaly from the Lorentz to the ghost number current. To do this, we
need the interpolation formula \eqn{interp}.
The mass term that interpolates in a continuous way between Lorentz
$\alpha =1$ and ghost number invariant $\alpha =0$ mass terms is
\begin{equation}
S_M(\alpha )=\alpha S_M^1+(1-\alpha )S_M^2\ .\label{intpolmass}
\end{equation}

Before starting the computation, one must change a little bit the formulas
for computing anomalies, when dealing with fermions. Indeed, one sees that
the regulator one finds is linear in derivatives, for which the heat kernel
method is not applicable. To solve this problem, we do the following trick.
First we notice that the mass term is linear in $M$, which follows from
dimensional arguments. Then we have that
\begin{eqnarray}
   \cA_{\epsilon} &=&  \str \left[ J \frac{1}{
   1-\frac{\cR}{M}}\right]=
    \str \left[ J(1+\frac{\cR}{M})\frac{1}{(1+\frac{\cR}{M})} \frac{1}{
   1-\frac{\cR}{M}}\right]\nonumber\\
   &=&\str \left[ J(1+\frac{\cR}{M}) \frac{1}{
   1-\frac{\cR^2}{M^2}}\right]=
   \str \left[ J(1+\frac{\cR}{M})\exp^{\frac{\cR ^2}{M^2}}\right]\ .
\end{eqnarray}
Now, we have obtained a regulator quadratic in derivatives and we can use
the heat kernel method. However, there is an extra term in the jacobian
proportional to $1/M$. As we expand the regulator, using the Seeley-DeWitt
coefficients, in inverse powers of $M^2$, we will have no contribution
of this extra $1/M$ term to the finite part of the anomaly, in the limit
$M^2\rightarrow \infty$.

It is now clear that, by taking simply \eqn{tfact} as an action, one will
produce no anomalies. Indeed, the matrices $T$ and $J$ are field
independent (they are just numbers), and the regulator consists of only
derivatives, no fields. Applying the heat kernel method, there will be no
$E$ matrix in the $a_1$ coefficient and the scalar curvature $R$ is also
vanishing. This is not inconsistent, since our path integral is just a
number. We have no external fields (e.g. background fields, antifields, ...)
which we can vary.  Therefore, we will
change the action \eqn{tfact}, by introducing background gauge fields~:
\begin{equation}
S=2i{\bar \psi}\nabla _-^{0,1}\psi +2i{\bar \xi} \nabla _+^{1,-1}\xi \
,\label{tfactgf} \end{equation}
where the covariant derivatives are defined by
\begin{equation}
\nabla^{s,gh}_{\pm}=\partial _{\pm}+s\omega _{\pm}+gh A_{\pm}\ .
\end{equation}
$\omega _{\pm}$ and $A_{\pm}$ can be seen as gauge fields
for which the symmetries can be made local. Then, one must transform
the gauge fields in the appropriate way, i.e. $\delta _{\epsilon
^{s}}\omega
_{\pm}=\partial _{\pm}\epsilon ^{s}$ for local Lorentz transformations
\footnote{For local Lorentz transformations, one must of course work in
the vielbein formalism. This is implicit in our notation in the sense
that $\partial _{\pm}=e_{\pm}^\mu \partial _\mu $ and $\omega _{\pm}=
e_{\pm}^\mu \omega  _\mu $.}, and
$\delta _{\epsilon ^{gh} }A_{\pm}=\partial _{\pm}\epsilon ^{gh} $, for
local ghost number
symmetry. However, we keep them external in the sense that there is no
path integral over the gauge fields and so, they do not appear in loops.
Doing so, we do not need the ghosts and the gauge fixings for these
symmetries.

We will now compute the 2 different anomalies in the A phase, with
\eqn{intpolmass} as a mass term. For the inverse mass matrix we find
\begin{equation}
T^{-1}=\frac{1}{2(2\alpha -1)}\pmatrix{0 & 0 & \alpha -1
& \alpha \cr 0 & 0 & -\alpha & 1-\alpha
\cr 1-\alpha  & \alpha  & 0 & 0 \cr -\alpha  & \alpha -1 & 0 & 0} \ .
\end{equation}
Remark that we go through a singularity when passing $\alpha =1/2$. At this
point the $T$ matrix is not invertible and there are no propagators for the
PV fields anymore. This point is of no danger for
our computations, because we can solve this problem by slightly deforming
the path we take. For instance, we could go around the singularity by
taking $\alpha $ complex.

The jacobian of the Lorentz symmetry is (we dropped the $\epsilon ^s$
parameter for simplicity)
\begin{equation}
J_s(\alpha )=\frac{\alpha -1}{2(2\alpha -1)}\pmatrix{1-\alpha & -\alpha & 0
& 0 \cr \alpha  & \alpha -1 & 0 & 0
\cr 0 & 0 & 1-\alpha  & \alpha  \cr 0 & 0 & -\alpha  & \alpha -1} \ .
\end{equation}
This matrix indeed vanishes for $\alpha =1$, since our mass term
then is Lorentz invariant.
The jacobian for the ghost number symmetry is
\begin{equation}
J_{gh}(\alpha )=\frac{\alpha }{(2\alpha -1)}\pmatrix{-\alpha & 1-\alpha & 0
& 0 \cr \alpha -1 & \alpha  & 0 & 0
\cr 0 & 0 & \alpha  & 1-\alpha  \cr 0 & 0 & \alpha -1 & -\alpha } \ .
\end{equation}
In this case the jacobian vanishes for $\alpha =0$, in agreement with the
ghost number invariance of the mass term at this value of $\alpha $.

The regulator for arbitrary $\alpha $ takes a complicated form.
One can
check that the piece quadratic in derivatives is only diagonal for $\alpha
=0$ or $\alpha =1$. To treat the general case one has to use an extra trick
\footnote{One uses the identity $\str[J\exp({\cR ^2}/M^2)]=
\str[AJA^{-1}\exp(A{\cR ^2}A^{-1}/M^2)]$ for arbitrary matrices $A$. Then
one looks for a matrix $A$ such that the new regulator $\cR '=A{\cR
^2}A^{-1}$ is diagonal in second derivatives.}.
To keep things technically simple, we only give the results for
$\alpha =0$ and $\alpha =1$. The regulators are
\begin{equation}
\cR ^2(\alpha =0)=-\pmatrix{\nabla ^{0,1}_+\nabla ^{-1,-1}_- & 0 & 0 & 0
\cr 0 & \nabla ^{1,-1}_+\nabla ^{0,1}_- & 0 & 0 \cr
0 & 0 & \nabla ^{0,1}_-\nabla ^{1,-1}_+ & 0 \cr
0 & 0 & 0 & \nabla ^{-1,-1}_-\nabla ^{0,1}_+} \ ,
\end{equation}
and
\begin{equation}
\cR ^2(\alpha =1)=-\pmatrix{\nabla ^{1,-1}_+\nabla ^{-1,-1}_- & 0 & 0 & 0
\cr 0 & \nabla ^{0,1}_+\nabla ^{0,1}_- & 0 & 0 \cr
0 & 0 & \nabla ^{-1,-1}_-\nabla ^{1,-1}_+ & 0 \cr
0 & 0 & 0 & \nabla ^{0,1}_-\nabla ^{0,1}_+} \ .
\end{equation}
So, for $\alpha =0$, the Lorentz anomaly is given by
\begin{equation}
\cA _{\epsilon ^{s}}(0)=\str [{\epsilon ^{s}}
J_s(0)\exp ^{\frac{\cR^2(0)}{M^2}}]\ , \end{equation}
where ${\epsilon ^{s}} $ is the parameter of the Lorentz transformation.
One can now compute the anomaly using the heat kernel method.
Using the vielbein formalism, one finds the following result~:
\begin{equation}
\cA _{\epsilon ^{s}}(0)=\int \epsilon ^{s}\partial _\mu (g^{\mu \nu
}\omega _\nu +\epsilon ^{\mu \nu }A_\nu )\ ,
\end{equation}
where $g^{\mu \nu }=\frac{1}{2}(e_\mu ^+e_\nu ^-+e_\mu ^-e_\nu ^+)$ is the
metric on the Riemann surface where the action
is integrated over and $\epsilon ^{\mu \nu }$ is the permutation symbol.
Because we have regulated our theory in a manifest ghost number invariant
way, this expression for the Lorentz anomaly is ghost number invariant.

It is a subtle point what to do with this expression if $\epsilon $ is
taken to be constant. In the computation, we did not need to specify
whether $\epsilon $ is taken to be local or constant. This is because the
fermions transform without derivatives and the gauge fields are taken to be
external. When $\epsilon $ is a constant, the anomaly is the integration of
a total derivative. This
gives a boundary term, which one usually drops. However, as we will see for
the ghost number anomaly, boundary terms can not be neglected. The ghost
number anomaly  will be proportional to
the Euler characteristic, which is also the integral of a total derivative.
But we can not simply drop it since it is a
topological invariant.

So for constant $\epsilon $, the anomaly is still present.
It is then very surprising that for local $\epsilon $ the anomaly can be
removed by adding a counterterm of the form
\begin{equation}
M_1=\frac{1}{2}\omega _\mu \omega _\nu g^{\mu \nu }+\epsilon ^{\mu \nu
}\omega _\mu A_\nu +F(A)\ .\label{Mcount}
\end{equation}
This can only be done for a local parameter since for a constant we have
that $\delta _{\epsilon ^{s}}\omega _{\pm}=\partial _{\pm}\epsilon
^{s}=0$. The function $F(A)$ can be chosen arbitrary since $A_\mu $ does
not transform under local Lorentz transformations. We will fix it later on.

One can now compute the ghost number anomaly by following completely the
same strategy as for the Lorentz anomaly. The expression at $\alpha =1$ is
given by \begin{equation}
\cA _{\epsilon ^{gh}}(1)=\str [{\epsilon ^{gh}}J_{gh}(1)\exp ^{\frac{\cR ^2
}{M^2}}]\ .
\end{equation}
As a result we find~:
\begin{equation}
\cA _{\epsilon ^{gh}}(1)=\int \epsilon ^{gh}\partial _\mu (\epsilon ^{\mu
\nu }\omega _\nu +g^{\mu \nu }A_\nu )\ .
\end{equation}
The first term of this expression can be rewritten as $\sqrt g R(g)$. This
is because we can express the spin connection in terms of the metric and
the vielbeins. For
a constant parameter this gives the Euler characteristic. This part of the
anomaly was already discovered in \cite{ghanom}. Also here the same remarks
about constant or local parameters can be made. For a local parameter
the anomaly can be absorbed by our counterterm \eqn{Mcount}, for a certain
choice of the function $F(A)$ . It is easy to see that
\begin{equation}
M_1=\frac{1}{2}\omega _\mu \omega _\nu g^{\mu \nu }+\epsilon ^{\mu \nu
}\omega _\mu A_\nu +\frac{1}{2}A_\mu A_\nu ,\label{M1count}
\end{equation}
gives the correct answer. We now see that this counterterm moves the
anomaly from the Lorentz sector to the ghost number sector, for a local
$\epsilon $ parameter. For constant $\epsilon $, one can not interpolate,
since the gauge fields do not transform. We can also
obtain this result directly from \eqn{interp}. One then has to deal with
the extra technical complication as described in the footnote.

To conclude this section and chapter, let us repeat that we can move
anomalies
into different sectors. One must however be very careful with rigid
symmetries, where one has to take care of total derivatives and global
aspects of the theory.

\chapter{Quantum BV theory}
\section{The quantum master equation}
In this section we will discuss the BV formalism at the quantum level and
show how the quantum master equation can be derived. From the classical
theory we have learned that there is an extended action $S(\Phi ,\Phi ^*)$
satisfying the classical master equation $(S,S)=0$. We will now see that
this is only the first of a tower of equations determined by the quantum
theory. From the previous chapter, we have learned that sometimes one
needs counterterms that absorb the anomaly. In the antifield formalism,
such counterterms can be antifield dependent. In fact,
the full quantum (extended) action $W(\Phi ,\Phi ^*)$ can be
expanded in powers of $\hbar$:
\begin{equation}
W=S+\hbar M_1 +\hbar ^2M_2 +...\ .  \label{QA}
\end{equation}
For a local field theory we require the $M_i$ to be local functionals. The
antifield dependence of these counterterms generate
quantum corrections to the
transformation laws. An example of this will be given in section 3.
The expansion \eqn{QA} is the
usual one, but we will see later on that terms of order
$\sqrt{\hbar }$ also can appear.

This quantum action $W$ appears in the path integral
\begin{equation}
 Z(J, \Phi ^*)=
\int {\cal D} \Phi \exp \left( \frac{i}{\hbar }W(\Phi,\Phi^*)
+J(\Phi)\right) \ ,\label{pathint}
\end{equation}
where we have introduced sources $J(\Phi )$, and in this subsection we work
in the gauge--fixed basis. The gauge fixing which we discussed, can be seen
as the procedure to select out of the $2N$ variables, $\Phi ^A$ and $\Phi
^*_A$, $N$ variables over which one integrates (i.e. one has to choose a
`Lagrangian submanifold'). To define the path integral
properly, one has to discuss regularisation, which can be seen as a way to
define the measure. In gauge theories, one can
not always find a regularisation that respects all the gauge symmetries.
This means that symmetries of the classical theory are not
preserved in the quantum theory. Anomalies are the expression of this
non--invariance. If there are no anomalies, then the quantum theory does
not depend on the chosen gauge. This property does not hold when there are
anomalies.
In that case the quantum theory will have a different content than the
classical theory. One can obtain in this way induced theories,
which from our point of view are theories where antifields become
propagating fields (this point is nicely explained in \cite{bvberk}).

We have seen that gauge fixing can be done by a canonical
transformation. Choosing another gauge amounts to performing another
canonical
transformation. The consistency on the path integral is its
independence of the chosen gauge. Stated otherwise,
it must be invariant under canonical transformations. More geometrically ,
the path integral must be invariant under a continuous deformation of the
chosen Lagrangian submanifold. This leads to a condition on the quantum
extended action $W(\Phi ,\Phi ^*)$, well discussed in the literature,
e.g. \cite{BV1,BV2,bvleuv,DJTH,Schwarz1}~:
\begin{equation}
{\cal A}\equiv \Delta \exp^{\frac{i}{\hbar}W}=0 \Leftrightarrow
(W,W)=2i\hbar \Delta W\ , \label{ME} \end{equation}
where
\begin{equation}
\Delta =(-)^A\dl _A\dl ^A\,. \label{BOX}
\end{equation}
In powers of $\hbar$ the first two equations (zero--loop and one--loop)
are \begin{eqnarray}
&&(S,S) = 0\nonumber\\
&&i{\cal A}_1\equiv i\Delta S-(M_1,S)=0\ . \label{MEC}
\end{eqnarray}
The first one is the classical master equation discussed before. The
second one is an equation for $M_1$. In a local field theory we will
moreover demand that $M_1$ is a local functional.
If there does not exist such an $M_1$, then
${\cal A}_1$ is called the anomaly. It is clearly not uniquely defined,
as $M_1$ is arbitrary. The anomaly satisfies
\begin{equation}
({\cal A}_1,S)=0\ ,\label{WZccS}
\end{equation}
which is a reformulation of the Wess--Zumino
consistency conditions \cite{WZcc}. This can be proven in a formal way by
acting with $\Delta $ on the classical master equation.
But as mentioned, we need a regularisation procedure.
In the expressions
above, divergences arise when acting with the $\Delta $ operator
on a local functional. In general this leads to terms
proportional to $\delta (0)$.

In section 4, we will
prove the consistency condition in a regularised way. Motivated by the
previous chapter, we will use Paul-Villars regularisation. Let us however
mention that also dimensional regularisation can be used in the context of
BV quantum theory \cite{Tonin}. The way to
implement the Pauli-Villars regularisation scheme into the BV formalism
was shown in \cite{anombv,bvsb}. The first step is to construct the
generalised PV action, depending on the fields and antifields. Denoting
$z^\alpha =\{\Phi ^A,\Phi ^*_A\}$ and $w^\alpha =\{\Phi ^A_{PV},\Phi
^*_{A,PV}\}$, the regularised PV action is
\begin{equation}
S^{reg}=S(z)+\frac{1}{2}w^\alpha S_{\alpha \beta }w^\beta
-\frac{1}{2}M^2\Phi ^A_{PV}T^{AB}\Phi ^B_{PV}\ ,
\end{equation}
where $S(z)$ satisfies the classical master equation. The first term of the
PV action is a generalisation of \eqn{SPV}. Together with $S(z)$ it
satisfies, up to fourth order in PV fields, the master equation where the
antibracket
now is in the space of fields and PV fields, and their antifields.

The Delta operator is modified such that
\begin{equation}
\Delta =(-)^A[\dl _A\dl ^A-\frac{\dl}{\partial \Phi ^A_{PV}}
\frac{\dl}{\partial \Phi ^*_{A,PV}}]\ .
\end{equation}
We then satisfy
\begin{equation}
\Delta S^{reg}=0\ ,
\end{equation}
due to a cancellation between fields and PV fields, based on \eqn{PVcond}.
Of course, one should
first sum over all fields and antifields before one integrates over
the internal momenta.

Anomalies then correspond to a violation of the master equation
$(S^{reg},S^{reg})\neq 0$, which we identify with
$\Delta S$. It can be shown that, after
integrating out the PV fields, analogous to the derivation of section 5.1,
one has
\begin{equation}
\Delta S=\lim_{M^2\rightarrow \infty}Tr\left[J\,\exp({\cal R}/M^2)\right]\
, \label{BOXS} \end{equation}
where the jacobian $J$ is given in terms of the transformation matrix $K$,
the derivative of the extended action w.r.t. an antifield and a field, as
\begin{equation}
J^A{}_B=K^A{}_B+\ft12 (T^{-1})^{AC}\left(T_{CB},S\right)(-)^B\ ;\qquad
K^A{}_B = S^A{}_B\ .\label{defK}
\end{equation}
Note that in general $S_{AB}$ contains antifields.
Also the other
matrices may be antifield dependent, and thus $\Delta S$
will in general contain antifields. For open gauge algebras we will give
an explicit example of this
antifield dependence. But also for closed algebras this might happen, as
was argued in \cite{antdep}.
It can again be evaluated using the heat kernel
of the previous chapter. In section 4,  we explicitly check that this
expression for $\Delta S$ satisfies the consistency condition
\begin{equation}
(S,\Delta S)=0\ .          \label{consdelS}
\end{equation}
Alternative formulations of the expression for the anomaly are given in
\cite{jordi}. They show that there
is always an $M_1$ such that \eqn{MEC} is satisfied. However,
this expression is in general non--local. Anomalies appear if
no local expression $M_1$ can be found satisfying that
equation. This is also reflected in the Zinn-Justin
equation for the effective action, see e.g. \cite{bvleuv}
\begin{equation}
(\Gamma ,\Gamma )=\frac{2\hbar}{i}<\cA>\ ,\label{anZJ}
\end{equation}
which are nothing but the anomalous Ward identities for gauge theories.

If the calculations are done in a specific gauge, one can not see at the
end whether
one has obtained gauge--independent results. Therefore one often includes
parameters in the choice of gauge, or background fields. The question
then remains whether this choice was `general enough' to be able to trigger
all possible anomalies. E.g. for the example of the bosonic string
theory (for simplicity in a flat space time background) the gauge $h=0$
would
not show the anomaly, which is of the form $\int d^2x\, c\sqrt h R(h)$.
Instead, a
gauge choice $h=H$, where the latter is a background field, is sufficient.
More general, one cannot gauge fix a classical symmetry, if
this symmetry does not survive the quantum theory. In the BV formalism, no
background fields are necessary, but one keeps
the antifield--dependent terms through all calculations. The anomalies are
reflected at
the end by dependence on these antifields. In our example, we would obtain
an anomaly $\int d^2x\, c\sqrt b^* R(b^*)$. In this formalism it
is then also clear how anomalies change by going to other gauges. The
relation is given by canonical transformations. It was shown in the
appendix of \cite{anombv} that under canonical transformations,  for any
object $X(\Phi ,\Phi ^*)$, one has that
\begin{equation}
\Delta X-\Delta 'X=\frac{1}{2}(X,\log J)\ ,\label{delcan}
\end{equation}
where $J$ is the jacobian of the canonical transformation.

Let us finally mention that the statement on choosing different mass
terms $T_{AB}$ still holds in the BV formalism. The counterterm that
connects two regularisation schemes is still given by \eqn{interp}.
The anomalies which arise by choosing different mass terms are related
by the formula
\begin{equation}
i\Delta S(1)=i\Delta S(0)+(S,M_1)\ ,\label{interanom}
\end{equation}
Because the regulator can
be antifield dependent, so can $M_1$ be antifield dependent.
Comparing
with \eqn{delcan}, \eqn{interanom} expresses that choosing a different mass
term is equivalent with doing a canonical transformation.

\section{The regularised jacobian for pure YM}
We will now illustrate the techniques presented above in the example of
Yang-Mills theory. This computation was done in \cite{regYM}, using
the background field method, and in \cite{anomYM} using the BV formalism.
We will here
closely follow the latter. The extended action was already given
in \eqn{YMext}.
Before going to the anomaly computation, we first have to gauge fix the
theory by adding the non-minimal sector $S_{nm}=b^{*2}/2$, and
performing a canonical transformation with generating fermion
$f=A^\mu \partial _\mu b$. We will work in the matrix notation as
explained in section 3.1, so a trace is understood. This way of
gauge fixing is completely
analogous to the one we did in the example of Maxwell's theory of section
2.2. After the canonical transformation, the extended action becomes
\begin{eqnarray}
S=&&-\frac{1}{4}F_{\mu \nu }F^{\mu \nu }-\frac{1}{2}(\partial _\mu A^\mu
)^2 +b\partial _\mu D^\mu c\nonumber\\
&&+A^*_\mu D^\mu c+\frac{1}{2}c^*[c,c]-b^*\partial _\mu A^\mu
-\frac{1}{2}b^{*2}\ .
\end{eqnarray}
One can check that, when putting the antifields equal to zero, this indeed
leads to a good gauge fixed action, i.e. the Hessian (the matrix of second
derivatives of the action w.r.t. the fields) is non-singular.

The next step is to choose a mass matrix for the PV action $T_{AB}$. In the
basis of increasing ghost number $\{b,A^\mu ,c\}$ we choose its inverse to
be
\begin{equation}
(T^{-1})^{AB}=\pmatrix{0 & 0 & -1 \cr 0 & g^{\mu \nu } & 0 \cr 1 & 0 & 0}\ .
\end{equation}
Remember that there is still the index of the Lie algebra. So the 1 is in
fact the unit matrix in the internal space of Lie algebra valued objects.
The jacobian
is easily seen to be (since the mass matrix is field independent, $J=K$)
\begin{equation}
J^A_B=\pmatrix{0 & -\partial _\nu  & 0 \cr 0 & -c\delta ^\mu _\nu  & D^\mu
\cr 0 & 0 & c}\ .
\end{equation}
Notice that this matrix contains derivatives, which makes things more
complicated. Indeed, this leads to computing the Gilkey
(Seeley-DeWitt) coefficients
with a derivative on it. This is possible, but quite tedious in four
dimensions. There is however a way out by using a trick. It is based on the
footnote of section 5.3. It is the identity
\begin{equation}
\str [Je^{(T^{-1}S/M^2)}]=\str [J_se^{(T^{-1}S/M^2)}]\ ,
\end{equation}
where the symmetrised jacobian is defined as
\begin{equation}
J_s=\frac{1}{2}(J+T^{-1}J^t T) \ .
\end{equation}
We have
used here conventions for transposing
supermatrices and used the cyclicity of the supertrace. These conventions
can be found in section 4 of this chapter, because there they will be used
extensively. The symmetrised jacobian turns out to be free of derivatives~:
\begin{equation}
(J_s)^A_B=\frac{1}{2}\pmatrix{ -c & A_\nu &  0 \cr 0 & 0 & A^\mu
\cr 0 & 0 & c} \ .
\end{equation}
The regulator is found to be
\begin{equation}
\cR=\pmatrix{ D_\rho \partial ^\rho & -(\partial _\nu b)+A^*_\nu & c^* \cr
c\partial ^\mu & R^\mu _\nu &(\partial ^\mu b)-A^{*\mu } \cr
0 & -\partial _\nu c & \partial _\rho D^\rho }\ ,
\end{equation}
where a derivative keeps on working to the right, unless it is between
brackets. Remember that there is still a delta function on which this
matrix works. We also have abbreviated the matrix $R^\mu _\nu =D_\rho
D^\rho
\delta ^\mu _\nu -D^\mu D_\nu +\partial ^\mu \partial _\nu +2F_{\mu \nu }$.
Now we have to find, according to \eqn{formcR}, the connection $\cY
_\alpha$ and the matrix $E$ in $d$ dimensions, which can be read off from
the regulator~:
\begin{equation}
\cY _\alpha =\frac{1}{2}\pmatrix{A_\alpha & 0 & 0 \cr
c\delta ^\mu _\alpha & 2A_\alpha -A^\mu g_{\nu \alpha }-A_\nu \delta ^\mu
_\alpha & 0 \cr
0 & -cg_{\nu \alpha } & A_\alpha }\ ,
\end{equation}
and
\begin{equation}
E=\pmatrix{\frac{1}{4}A^2-\frac{1}{2}\partial ^\rho A_\rho  & -(\partial
_\nu b)+A^*_\nu & -c^* \cr
V^\mu  & E^\mu _\nu  & (\partial ^\mu b)-A^{*\mu } \cr
\frac{d}{4}c^2 & (V^t)^\mu _\nu  & \frac{1}{4}A^2+\frac{1}{2}\partial
^\rho A_\rho }\ ,
\end{equation}
with
\begin{eqnarray}
E^\mu _\nu &=&-\frac{1}{4}A^2\delta ^\mu _\nu -3\partial ^{[\mu }A_{\nu
]}+(\frac{3}{2}-\frac{d}{4})A^\mu A_\nu  -A_\nu A^\mu \nonumber\\
V^\mu &=& -\frac{1}{2}\partial ^\mu c+\frac{1}{4}[(d-1)A^\mu c-cA^\mu
]\nonumber\\
(V^t)^\mu _\nu &=&-\frac{1}{2}\partial _\mu c+\frac{1}{4}[A_\mu
c-(d-1)cA_\mu ]\ .
\end{eqnarray}
The antisymmetrisation is $[\mu \nu ]=(\frac{1}{2}\mu \nu -\nu \mu )$.
We
will now compute $\Delta S$ in two and four dimensions. Let us start in 2
dimensions. Here we need the Seeley-DeWitt coefficient $a_1=E$ (the
curvature term is absent here, since we work in a flat background).
We find
\begin{equation}
\Delta S(d=2)=-\frac{1}{4\pi }\tr [A^\mu D_\mu c]\ ,
\end{equation}
which can be absorbed in a variation of a local counterterm
\begin{equation}
M_1(d=2)=-\frac{1}{8\pi }\tr [A_\mu A^\mu]\ .
\end{equation}
In fact, one must add here a coupling constant $g^2$, since the
engineering dimensions are not correct. In two dimensions, the coupling
constant has dimension 1, because we have chosen that $A_\mu $ has
dimension zero. To derive this
properly from the start, one has to insert the coupling constant into the
covariant derivative. Notice that in four dimensions, the coupling constant
is dimensionless and can be put equal to 1.

Now we repeat the calculation in four dimensions. For this, we need the
Gilkey coefficient
\begin{equation}
a_2= \left(
 \frac{1}{12}  W_{\alpha \beta} W^{\alpha \beta} +
\frac{1}{2}  E^{2} +
 \frac{1}{6}\Box E \right) \ ,
\end{equation}
where
\begin{eqnarray}
W_{\alpha \beta}&=&\partial _\alpha {\cal Y}
_\beta -\partial _\beta {\cal Y}_\alpha
+[{\cal Y}_\alpha,{\cal Y}_\beta]\nonumber\\
\Box E&=&\nabla _\alpha \nabla ^\alpha E \nonumber\\
\nabla_\alpha X&=&\partial _\alpha X + [{\cal Y}_\alpha ,X]\ .
\end{eqnarray}
The computation is quite tedious, and we will only give the results. As a
first step one can check that the antifield dependent part is zero. The
part without antifields is given by
\begin{eqnarray}
\Delta S(d=4)&=&\frac{1}{(4\pi )^2}\frac{1}{12}\tr \{ (\partial ^\nu
c)\left[ 4A_\mu A_\nu A^\mu -8A^\mu \partial _{[\mu }A_{\nu
]}\right.\nonumber\\
&&\left.-4A_\nu \partial _\mu A^\mu +\partial _\mu \partial ^\mu
A_\nu -3\partial _\nu \partial _\mu A^\mu \right] \} \ .
\end{eqnarray}
Again, there exists a counterterm that absorbs the anomaly. In four
dimensions it is
\begin{eqnarray}
M_1(d=4)&=&\frac{1}{(4\pi )^2}\frac{1}{12}\tr [\frac{3}{2}(\partial _\mu
A^\mu
)^2-\frac{1}{2}(\partial _\mu A_\nu )(\partial ^\nu A^\mu )\nonumber\\
&&-2A^\mu (\partial _\mu A_\nu )A^\nu+\frac{3}{2}A_\mu A_\nu A^\mu A^\nu
-\frac{1}{2}A^2A^2]\ .
\end{eqnarray}
This shows, as was of course already known long ago \cite{thooft} , that
Yang-Mills theory is free from anomalies.

\section{Anomalies in chiral $W_3$ gravity}
As we have seen in the classical analysis in chapter 3, section 3.2.4,
chiral $W_3$ gravity is an example of an open algebra. In the previous
section, we have seen an example of an anomaly computation for closed
algebras. We will now compute in detail the anomalies for an open gauge
algebra of symmetries. Partial results were already obtained in
\cite{partanomw3}. A
complete (one loop) treatment, which we will present below, in the context
of the BV formalism was first
given in \cite{anomw3}, using PV regularisation. Later, the calculation was
extended to higher loops \cite{jordiw3,jorwal}, by using nonlocal
regularisation in the BV formalism.
Very recent, the anomaly structure was analysed
using BPHZ regularisation \cite{bphzw3}. We will comment on this
later on.

Before going to the quantum theory, let us first give the extended action
in the gauge fixed basis. The gauge fixing is done by performing the
canonical transformation from $\{h,h^*,B,B^*\}$ (fields and antifields of
the classical basis) to new fields and antifields $\{b,b^*,v,v^*\}$ (the
gauge--fixed basis):
\begin{equation}
h=-b^*;\ h^*=b\ \mbox{  and  }\ B=-v^*;\ B^*=v\,.\label{ctrW3gf}
\end{equation}
The extended action in the gauge fixed basis is
\begin{eqnarray}
S&=&-\ft{1}{2}\pl \bpl +b\db c +v\db u -2\kappa \alpha b\db
b(\partial u)u\nonumber\\
&&+X _\mu ^*[\pl c+\ds \plb \plg u-2\kappa (\alpha
+1)b(\partial u)u(\partial X^\mu )]\nonumber\\
&&+b^*[-T+2b\partial c+(\partial b)c+3v(\partial
u)+2(\partial v)u-2\kappa \alpha b(\partial b )(\partial u)u\nonumber\\
&&+v^*[-W+4\kappa Tb\partial u+2\kappa \partial (bT)u+3v\partial
c+(\partial v)c -4\kappa \alpha b\partial v+6\kappa \alpha \partial
(bv\partial uu)]\nonumber\\
&&+c^*\left[(\partial c)c+\kappa \pl \pl (\partial
u)u\right]\nonumber\\
&&+u^*\left[2(\partial c)u-c(\partial
u)\right]\ ,
\end{eqnarray}
where the spin 2 and 3 currents are given by
\begin{equation}
T=\frac{1}{2}(\partial X^\mu )(\partial X^\mu ) \qquad W=\frac{1}{3}d_{\mu
\nu \rho  }(\partial X^\mu )(\partial X^\nu )(\partial X^\rho )\ .
\end{equation}
One may check now that the new antifield independent action, which depends
thus on $\{X^\mu ,b,c,v,u\}$, has no gauge invariances. These are thus
the fields that appear in loops.

It will be useful to summarise some properties of the fields in
the following table~:
\begin{table}[htf]
\label{tbl:fieldsW3}\begin{center}\begin{tabular}{||l|r|r|r||l|r|r|r||}
\hline
 & $gh$& $j$    & $dim-j$&        &$gh$ &$j$ &$dim-j$ \\ \hline
$X$& 0   & 0      & 0      & $X^*$  &  $-1$& 0  & 2 \\
$B$& 0   &$-3$    & 2      & $B^*=v$&  $-1$&$3$ & 0 \\
$h$& 0   &$-2$    & 2      & $h^*=b$&  $-1$&$2$ & 0 \\
$c$& 1   &$-1$    & 0      & $c^*$  &  $-2$&$1$ & 2 \\
$u$& 1   &$-2$    & 0      & $u^*$  &  $-2$&$2$ & 2 \\ \hline
$\partial$ &0  & 1    & 0 &&&& \\
$\bar \partial,\,\nabla$  & 0 &$-1$& 2    &&&& \\
$D$ & 0 &       $-2$     & 2    &&&& \\ \hline
\end{tabular}\end{center}
\caption{We give here the properties of fields and derivative operators.
$gh$ is the ghost number and $j$ is the spin. Further
one can assign an `engineering' dimension to the fields and derivatives,
such that the Lagrangian has dimension 2. We defined $dim(\Phi
^*)=2-dim(\Phi )$ (the number 2 is arbitrary, this freedom is
related to redefinitions proportional to the ghost number). When we
subtract the spin from this dimension,
we find that only a few fields have non zero dimension.}.
\end{table}

\subsection{Calculation of the one--loop anomaly without
antifields} \label{ss:calc1lW3}
We first need
the (invertible) matrix of second derivatives w.r.t. the fields of the
gauge--fixed basis. Using the theorem of section 3.3,
we will only need in these second derivatives the terms
without
fields of negative ghost numbers and at most linear in $c$ and $u$,
the fields of ghost number 1. We will therefore in the entries still use
the names of fields and antifields as in the classical basis.
In the following matrices we first write the
entries corresponding to the bosons $X^\mu $, and then order the fermions
according to ghost number and spin:
$\Phi '^A=\{X^\mu , v=B^*, b=h^*, c,u \}$ we have  $S'_{AB}=\tilde
S'_{AB}+$
terms of antifield number non--zero $+$ terms of pureghost number $>1$.
\begin{eqnarray}
\tilde S'_{AB}&=&\pmatrix{S_{\mu \nu }& q_\mu \cr -q_\nu ^T & \tilde \nabla
\cr}\nonumber\\[0.4cm]
S_{\mu \nu }&=&  \delta
_{\mu \nu }\nabla ^1\partial +d_{\mu \nu \rho }D^2\plg\partial
\nonumber\\[0.4cm]
q_\mu &=&\pmatrix{ 0 & \kappa \partial (D^{-2}u)\pl &0 &0 \cr
}\nonumber\\[0.4cm]
\tilde \nabla &=&\pmatrix{ 0&0& D^{-1}&\nabla ^{-2}\cr
 0&0  & \nabla ^{-1}& \kappa y_h D^{-2}\cr
 D^3& \nabla ^2&0&0\cr
\nabla^3 & \kappa D^4y_h &0&0\cr }
\ , \label{SABgfW3}\end{eqnarray}
where $y_h=T$ is the field equation of the gauge field $h$.
A lot of zeros follow already from the ghost number requirements.

To obtain the regulator we have to choose a
mass matrix $T$ and multiply its inverse with $\tilde S'$.  Looking at the
table~\ref{tbl:fieldsW3} for the spins of the fields, we notice that some
fermionic fields (e.g.  $c$) do not have a partner of opposite spin. This
we need to make a mass term that preserves these rigid
symmetries
(the PV partners of fields have the same properties as the corresponding
fields).  Further, the regulator will not regularise because the
fermion sector of \eqn{SABgfW3} is only linear in the derivatives. This
problem we also met in the previous chapter.
These two problems can be solved by first introducing extra PV fields (this
procedure was already used in \cite{anombv}).
They have no interaction in the massless sector and do not
transform under any gauge transformation. Inspecting the ghost numbers and
spins of the fermions in $\{\Phi '^A\}$, we find that we need extra PV
fields with ghost numbers and spin as in table~\ref{tbl:extraPVW3}.
\begin{table}[htf]
\label{tbl:extraPVW3}\begin{center}\begin{tabular}{||l|r|r|r||}
\hline
 & $gh$& $j$    & $dim-j$\\ \hline
$\bar u$& 1   &$-3$    & 1       \\
$\bar c$& 1   &$-2$    & 1       \\
$\bar b$& $-1$&$1$     & 1       \\
$\bar v$& $-1$&$2$     & 1       \\ \hline
\end{tabular}\end{center}
\caption{The extra non--interacting and gauge invariant PV fields. The
ghost
numbers and spins are chosen in order to be able to construct mass terms
for $c,u,b$ and $v$. The names are chosen such that $gh(\bar x)=gh(x)$
and $j(\bar x)=j(x)-1$. The dimensions follow from the kinetic terms,
although this would still allow less symmetric choices.} \end{table}
Then we can choose the mass matrix $T$ to be
\begin{equation}
T =\pmatrix {\delta _{\mu \nu }& 0 & 0\cr 0 & 0&\frac{1}{M} \unity \cr
0&-\frac{1}{M}\unity &0\cr}\ , \end{equation}
where the second entry refers to the fermions from above, and the third
line to the
four new fermions. The latter are thus ordered as in the table. The kinetic
part of their action can be chosen such that the
enlarged matrix $S'_{AB}$ is \begin{equation}
S'_{AB}=\pmatrix{S_{\mu \nu }& q_\mu &0\cr -q_\nu ^T&\tilde \nabla &0\cr
0&0&-\tilde \partial \cr}\ ;\qquad
\tilde \partial \equiv \pmatrix {0 &0 & 0&\partial \cr 0 & 0 &
\partial & 0\cr 0&\partial &0&0\cr \partial &0&0&0\cr}\ .
\label{kinexPV} \end{equation}
To put everything together, we find as regulator
\begin{eqnarray}
&&{\cal R}={\cal R}_0 + M {\cal R}_1\nonumber\\
&&{\cal R}_0=\pmatrix{
S_{\mu \nu }& q_\mu &0\cr 0 & 0 &0 \cr 0 &0 &0\cr}\ ;\qquad
{\cal R}_1=\pmatrix{0& 0 &0\cr 0 & 0 &\tilde \partial \cr
-q^T_\nu &\tilde \nabla &0\cr}\ .
\end{eqnarray}
The jacobian $J$ is again equal to the transformation matrix $K$
defined in \eqn{defK}.
Dropping again terms of pureghost number 2 or antifield number 1, we find
\begin{eqnarray}
K&=&\pmatrix{K^\mu {}_{\nu }& K^\mu {}_{F}&0\cr
           K^F{}_{\nu }& K^F{}_{F}& 0 \cr
           0 & 0 & 0 \cr}\nonumber\\[0.4cm]
K^\mu {}_{\nu }&=&\delta ^\mu _\nu c\partial +2ud_{\mu \nu \rho }(\partial
X^\rho )\partial \nonumber\\[0.4cm]
K^\mu {}_{F}&=&\pmatrix{0&0&\pl&d_{\mu \rho \sigma }(\partial X^\rho
)(\partial X^\sigma )\cr}\nonumber\\[0.4cm]
K^F{}_{\nu }&=&\pmatrix{-d_{\nu \rho\sigma  }(\partial X^\rho )(\partial
X^\sigma )\partial\cr -(\partial X_\nu )\partial\cr 0\cr 0\cr}
\end{eqnarray}
\begin{equation}
K^F{}_{F}=\pmatrix{
- (c\partial )_{3} & -2\kappa[ y_h (u\partial )_2+u(\partial y_h)]&0&0\cr
-2(u\partial )_{\frac{3}{2}}&-(c\partial )_2&0&0 \cr
0&0&-(c\partial )_{-1}&-2\kappa (1-\alpha )y_h(u\partial )_{-1}\cr
0&0&-2(u\partial )_{-\frac{1}{2}}&-(c\partial )_{-2}\cr }\nonumber\ ,
\end{equation}
where we used the shorthand
\begin{equation}
(c\partial )_{x}=c\partial +x(\partial c)\ .
\end{equation}
The expression of ${\cal R}$ is so far still linear in derivatives for the
fermionic sector. This we solve as in \cite{Diaz} by multiplying in
\eqn{SSMPV} the numerator and the denominator by $(1+{\cal R}_1/M)$. We
obtain
\begin{equation}
(\Delta S)^0=Tr\left[K(1+\frac{1}{M}{\cal R}_1)\frac{1}{\left(
1-\frac{1}{M^2}{\cal R}_0-\frac{1}{M}{\cal R}_1\right) \left(
1+\frac{1}{M}{\cal R}_1\right) }\right] \ , \end{equation}
which leads to the regulator
\begin{equation}
{\cal R}_0 +{\cal R}_1^2+\frac{1}{M}{\cal R}_0{\cal R}_1
=\pmatrix{ S_{\mu \nu }& q_\mu &\frac{1}{M}q_\mu \tilde \partial \cr
-\tilde \partial q^T_\nu & \tilde \partial \tilde \nabla & 0 \cr
 0 &0 &\tilde \nabla\tilde \partial \cr}\ .
\end{equation}
On the other hand
\begin{equation}
K{\cal R}_1 =\pmatrix{0&0&K^\mu {}_{F}\tilde \partial \cr
0&0&K^F{}_{F}\tilde \partial \cr 0&0&0\cr}\ .
\end{equation}
This can not contribute to the trace because grouping the first two
entries together, this regulator is also upper triangular.
Therefore we can omit the term $K{\cal R}_1$,
and then only the first two rows and columns of the
regulator can play a role. This eliminates also the $(1/M)$ terms in the
regulator. So we effectively have to calculate
\begin{equation}
(\Delta S)^0=Tr\left[\pmatrix{K^\mu {}_{\nu }& K^\mu {}_{F}\cr
           K^F{}_{\nu }& K^F{}_{F} \cr}\exp \frac{1}{M^2}\tilde {\cal R}
\right] \ ;\qquad \tilde {\cal R}= \pmatrix{ S_{\mu \nu }& q_\mu \cr
 -\tilde \partial q^T_\nu & \tilde \partial \tilde \nabla\cr} \ .
\end{equation}
The evaluation of this supertrace will be done using
the heat kernel method, was it not that our regulator contains
terms with second derivatives which are not proportional to the unit matrix
in internal space.  We can anticipate on the form of the anomaly which we
have to obtain. The anomaly $\Delta S$ should be an integral of a local
quantity of spin 0 and dimension 2. From the $dim-j$ column of
table~\ref{tbl:fieldsW3}, we then see that the anomaly can be split in a
part without $B$, and a linear part in $B$, which can not contain $h$:
\begin{equation}
(\Delta S)^0=\Delta ^hS +\Delta ^BS\ .
\end{equation}
So we split $\tilde {\cal R}={\cal R}^h+{\cal R}_1$
\begin{eqnarray}
{\cal R}^h&=&diag\left( \delta_{\mu \nu }\partial \nabla ^0,\partial
\nabla
^3,\partial \nabla ^2,\partial \nabla ^{-1},\partial \nabla
^{-2}\right)\nonumber\\
{\cal R}^B&=&\unity \Box +{\cal R}_1\\
{\cal R}_1&=&
\pmatrix{d_{\mu \nu \rho
}D^2\plg\partial &0&\kappa\partial  (D^{-2}u)\pl &0&0\cr
0&0&\kappa \partial D^4y_h & 0&0\cr
\kappa \partial\pl (D^{-2}u)\partial &0&0&0&\kappa \partial y_hD^{-2}\cr
0&0&0&\partial D^{-1}&0\cr}\nonumber \ .
\end{eqnarray}
In the last expression $\Box$ is a flat $\partial \bar \partial $.

First for the calculation at $B=0$ we need for each entry the expression of
\begin{equation}
{\cal A}_j=-i\int dx\,dy\,\delta (x-y) (c\partial )_j G
(x,y;\ft{1}{M^2},\Phi^{(j)})\ ,
\end{equation}
with
\begin{equation}
\Phi ^{(j)}=\left\{ g^{\alpha \beta }=\pmatrix{-h &\ft{1}{2}\cr \ft{1}{2}&
0\cr},
{\cal Y}_+=0, {\cal Y}_- =-j(\partial h), E=-\ft12 j(\partial ^2h)\right\}\
.\label{Phij} \end{equation}
Using the Seeley--DeWitt coefficients\footnote{The
conventions are $ R^\alpha {}_{\beta \gamma \delta }=\partial _\gamma
\Gamma
^\alpha _{\beta \delta } -...$ and $R=R^\gamma {}_{\alpha \beta \gamma
}g^{\alpha \beta }$. Further $W_{\alpha \beta }=\partial _\alpha {\cal Y}
_\beta -\partial _\beta {\cal Y}_\alpha
+[{\cal Y}_\alpha ,{\cal Y}_\beta ]$.
} \begin{eqnarray}
\left.a_0\right|=1\ ; &\qquad & \left.\nabla _\alpha a_0(x,y)\right|=0
\\
\left.a_1\right|= E-\ft16 R=\frac{1-3j}{6}(\partial ^2h)
\ ;&\qquad& \left.\nabla _\alpha
a_1(x,y)\right|=\ft12 \nabla _\alpha (E-\ft16 R) +\ft16 \nabla ^\beta
W_{\alpha \beta }\nonumber
\ , \end{eqnarray}
where $|$ stands for the value at coincident points $x=y$ (after taking
the derivatives).
We will need only $\nabla _+$ for which the
connection is zero:
\begin{equation}
\left.\partial a_1(x,y)\right|=\frac{1-3j}{12}(\partial
^3h)-\frac{j}{12}(\partial ^3h)\ . \end{equation}
This leads to
\begin{equation}
{\cal A}_j=\frac{-i}{24\pi }\int dx\, (6j^2-6j+1)c\,\partial ^3
h\ . \end{equation}
For the overall normalisation of this anomaly, we used $\sqrt{g}=2$, in
accordance with the form of $g_{\alpha \beta }$ which follows from
\eqn{Phij}. In this way, we thus use
the coordinates $x^\alpha =\{  x^+,x^-\}$, and the
integration measure $dx$ in the above integral is then $dx^+\,dx^-$. If one
uses $dx=dx^0\,dx^1$ then the scalars as $R$ and $E$ do not change, but
$\sqrt{g}=\rho ^{-2}$, where $\rho $ is the parameter in the definitions in
footnote~\ref{fn:convrho}. Therefore the overall factor $1/(24\pi) $ in the
above formula, gets replaced by $1/(48\pi \rho ^2)$.
One can either interpret $dx$ as $dx^+\,dx^-$ or as $dx^0\, dx^1$ with
$\rho =1/\sqrt{2}$. The overall normalisation in fact depends just on the
transformation law: if $S^1$ contains $X^*c\partial X$ then the anomaly for
one scalar is
\begin{equation}
{\cal A}_0=\frac{-i}{48\pi }\int dx^0\,dx^1\, c\,\partial (\partial
_0+\partial _1)^2 h\ , \end{equation}
independent of the definition of $\partial $. In all further expressions
for anomalies we again omit $\int dx$ with the normalisation as explained
above.

Denoting the ghost combination in the transformation of the bosons as
\begin{equation}
\tilde c^\mu _\nu =c\delta ^\mu _\nu +2u\ds \plg\ , \label{deftildec}
\end{equation}
we obtain
\begin{equation}
\Delta ^hS = \Delta _{XX} ^hS+\Delta _{FF} ^hS=
\ft{1}{24\pi }(\tilde c^\mu _\mu-100\, c) \partial ^3h
\ ,\label{ANOMB0}
\end{equation}
where $\Delta _{XX}$ is the contribution from the matter entries in the
matrices, and $\Delta _{FF}$ comes from the fermions and gives the factor
$-100$.

For $\Delta ^BS$ we have to evaluate expressions as
\begin{eqnarray} {\cal A}_B&=&tr\left[ K e^{t(\Box+{\cal R}_1)}\right] =
\frac{1}{t}\frac{d}{d\lambda } \left. tr\left[ e^{t(\Box+{\cal R}_1+\lambda
K)}\right] \right|_{\lambda =0}\nonumber\\
&=&
\frac{d}{d\lambda } \left. tr \left[ {\cal R}_1 e^{t(\Box+\lambda
K)}\right] \right|_{\lambda =0}\ ,
\end{eqnarray}
where the last step could be done because ${\cal R}_1$ is linear in $B$,
and we know that the result should be linear in $B$. We have for $K$ and
${\cal R}_1$ the general forms
\begin{equation}
K=k_0 +k_1\partial \ ;\qquad {\cal R}_1=r_0+r_1\partial +r_2\partial ^2\ .
\end{equation}
Then we have to evaluate the heat kernel with
\begin{equation}
\Phi =\left\{ g^{\alpha \beta }=\pmatrix {0&\ft12\cr \ft12&0\cr},{\cal
Y}_+=0,{\cal Y}_-=\lambda k_1, E=\lambda (k_0-\ft12 \partial k_1)\right\}\
. \end{equation}
As the metric is flat, there is no non--trivial contribution from $\Delta
^{1/2}$ and $\sigma (x,y)$ in \eqn{Gexpbn}. The only new coefficient which
we need are the second derivatives at coincident points. For a flat metric,
and being interested only in linear terms in $E$ and ${\cal Y}$, the
coefficients are
\begin{equation}
\left.\nabla _\alpha \nabla _\beta a_0\right|=\ft12 W_{\alpha \beta }\
;\qquad
\left.\nabla _\alpha \nabla _\beta a_1\right|=\ft13 \partial
_\alpha \partial _\beta
E +\ft16 \partial ^\gamma \partial _{(\alpha }W_{\beta )\gamma }+{\cal O}
(\lambda ^2)\ , \end{equation}
where the symmetrisation $(\alpha ,\beta )=\ft12 (\alpha \beta +\beta
\alpha )$. This gives
\begin{equation}
2\pi i{\cal A}_B= k_0\left(r_0-\ft12\partial r_1+\ft13\partial ^2r_2\right)
+(\partial k_1)\left(-\ft12r_0+\ft16\partial r_1-\ft1{12}\partial
^2r_2\right)\ , \end{equation}
which can be used to obtain
\begin{eqnarray}
\Delta ^BS&=&\Delta _{XX}^BS+\Delta _{XF}^BS+\Delta
_{FF}^BS\nonumber\\
i\Delta _{XX}^BS&=&
\frac{1}{12\pi }\tilde c^\mu _\nu
\partial ^3d_{\mu \nu \rho }B \plg \nonumber\\
i\Delta _{XF}^BS&=&-\frac{\kappa }{2\pi } (u\partial B-B\partial
u)(\partial ^3X^\mu )\pl \nonumber\\
i\Delta _{FF}^BS&=&\frac{\kappa }{6\pi }y_h\left\{ (-5+3\alpha
)u(\partial^3 B)+5(\partial ^3u)B\right.\nonumber\\
&&\hspace{1.6cm}\left. +(12-5\alpha )(\partial
u)(\partial ^2B) -12(\partial ^2u)(\partial B)\right\}\ .
 \label{ANOM}
\end{eqnarray}

We have thus obtained the anomaly at antifield number 0. It consists of
three parts. The first can be written in the form
\begin{equation}
i(\Delta S)^0_X= i\Delta _{XX}^hS+i\Delta _{XX}^BS=
\ft{1}{24\pi } \tilde c_{\mu \nu } \partial ^3\tilde h_{\mu \nu }\ ,
\label{cA0X}\end{equation}
where $\tilde c_{\mu \nu }$ was given in \eqn{deftildec}, and
\begin{equation}
\tilde h_{\mu \nu }=h\delta _{\mu \nu }+2d_{\mu \nu \rho }B\partial X^\rho
\ . \end{equation}
It is the total contribution from the matter loops. The other two parts
originate in loops with fermions. They are
\begin{eqnarray}
i(\Delta S)^0_F&=&i\Delta _{XF}S+i\Delta _{FF}^hS=
-\frac{100}{24\pi }\, c\, \partial ^3h
-\frac{\kappa }{2\pi } (u\partial B-B\partial
u)(\partial ^3X^\mu )\pl \nonumber\\
(\Delta S)^0_W&=& \Delta _{FF}^BS\approx 0\ .\label{cA0FW}
\end{eqnarray}
The upper index $0$ indicates that we have so far only the terms of
antifield number 0.

\subsection{Consistency and antifield terms}
{}From the formal argument in \eqn{WZccS} and as we will prove later on,
we know that the anomaly is consistent. At
antifield number zero this implies that $D^0 (\Delta S)^0\approx 0$. We may
check this now, and at the same time obtain the
anomaly at antifield number 1. This will e.g. include the contributions of
$h^*$, which according to \eqn{ctrW3gf} is the antighost.

It will turn out that the three parts mentioned above, $(\Delta S)^0_X$,
$(\Delta S)^0_F$ and $(\Delta S)^0_W$, are separately invariant under $D^0$
\begin{equation}
D^0(\Delta S)^0_X\approx 0\ \qquad
D^0(\Delta S)^0_F \approx 0\ ,   \label{consA}
\end{equation}
while this is obvious for $(\Delta S)^0_W\approx 0$.
To check this, one first obtains that
\begin{eqnarray}
D^0 \tilde c_{\mu \nu }&=&-\tilde c_{\rho (\mu }\partial \tilde
c_{\nu)\rho }+2\kappa
u(\partial u)\left[ 2\pl\plb +(\alpha +1)y_h\delta _{\mu \nu }\right]
\nonumber\\
D^0 \tilde h_{\mu \nu }&=&\left( \delta _{\rho(\mu  }\bar \partial
-\tilde h_{ \rho(\mu }\partial +(\partial \tilde h_{\rho (\mu })
\right)\tilde c_{\nu )\rho }\nonumber\\ &&
+2\kappa \left[2\pl\plb +\delta _{\mu \nu
}y_h\right](B\partial u-u\partial B)-2d_{\mu\nu \rho }y_\rho u\ .
\label{D0tilch}\end{eqnarray}
According to our theorem in section 3.3, this implies that the consistent
anomaly can be split in
\begin{equation}
\Delta S=(\Delta S)_X+(\Delta S)_F+(\Delta S)_W\ ,\label{DelSW3}
\end{equation}
where each term separately is invariant under ${\cal S}$, and starts with
the expressions in \eqn{cA0X} and \eqn{cA0FW}. The theorem
implies that the full expressions are obtainable from the consistency
requirement.

Indeed, from \eqn{consA} one can use \eqn{cSF00} to determine $
(\Delta S)_X^1, (\Delta S)_F^1$ and $(\Delta S)_W^1$: they are obtained by
replacing
the field equations $y_h$, $y_B$ and $y_X$ in the variation under $D^0$ by
$h^*$, $B^*$ and $X^*$. For the first one we obtain:
\begin{eqnarray}
i(\Delta S)_X^1&=&\frac{1}{24\pi }{\tilde c}_{\mu \nu }\partial
^3[-2\kappa
\delta _{\mu \nu }h^*(B\partial u-u\partial B)+2X_\rho ^*d_{\mu \nu \rho
}u]\nonumber\\&+&\frac{1}{24\pi }2\kappa u(\partial u)(1+\alpha )h^*\delta
_{\mu \nu }(\partial ^3{\tilde h}_{\mu \nu})\nonumber\\&+&\frac{1}{24\pi
}h^*[-8\kappa (\partial ^3c)(B\partial u-u\partial B)+8\kappa u(\partial
u)(\partial ^3h)] \label{AX1}
\end{eqnarray}
The first two terms can be absorbed in $(\Delta S)_X^0$ (\eqn{cA0X}) by
adding to ${\tilde c}_{\mu \nu }$ and ${\tilde h}_{\mu \nu }$:
\begin{eqnarray}
\tilde c^{(1)}_{\mu \nu }&=& 2\kappa(1+\alpha ) h^*
u(\partial u)\delta _{\mu \nu }
\nonumber\\
\tilde h^{(1)}_{\mu \nu }&=&-2\kappa \delta
_{\mu \nu }h^*(B\partial u-u\partial B)+2d_{\mu\nu \rho }X^*_\rho u\ ,
\end{eqnarray}
This part originated in the matter--matter entries of the
transformation matrix $K$ and the regulator $S_{\mu \nu }$ including all
antifields. If we consider these entries completely, we get as anomaly
\begin{equation}
i(\Delta S)_m= \ft1{24\pi } c_{\mu \nu } \partial ^3 h_{\mu \nu
}\label{cAm} \end{equation}
where
\begin{eqnarray}
c_{\mu \nu }&=&\tilde c_{\mu \nu }+\tilde c^{(1)}_{\mu \nu }\nonumber\\
h_{\mu \nu }&=&\tilde  h_{\mu \nu }+\tilde h^{(1)}_{\mu \nu }+2\kappa
(1-\alpha )c^*(\partial u)u\delta _{\mu \nu }\ . \end{eqnarray}
Note therefore that computing the matter anomaly by using just these
entries would not give rise to the last term of \eqn{AX1}:
\begin{equation}
i(\Delta S)_{X}-i(\Delta S)_m=\frac{\kappa }{3\pi
}h^*[- (\partial ^3c)(B\partial u-u\partial B)+ u(\partial
u)(\partial ^3h)] +\mbox{ terms of }afn\geq 2\ .
\end{equation}
As $\left(S,(\Delta S)_{X}\right) =0$, it follows that $(\Delta S)_m$
is not a consistent anomaly~! Indeed, the proof of consistency
given in section 6.4 requires that we trace over all the
fields in the theory. One may check that the violation of the consistency
condition for $(\Delta S)_m$ agrees with \eqn{violSA}.
We will see that for $\alpha =0$ this extra term will be cancelled when
adding the fermion contributions.

For the other parts of the anomaly we obtain at antifield number~1:
 \begin{eqnarray}
\frac{6\pi i}{ \kappa}(\Delta S)_W^1&=&h^*\left\{\left(-49(\partial
^3c)+9(\partial ^2c)\partial \right)\left( u\partial B-B\partial u\right)
\right.\nonumber\\
&&\hspace{1cm}+\alpha \left[
u(\partial ^2c)\partial ^2B+10 (\partial u)B(\partial ^3c)+15
(\partial u)(\partial B)(\partial ^2c)\right.\nonumber\\&&\left.
\hspace{13mm} -15u(\partial ^3c)\partial B-6u(\partial ^4c)B\right]
\nonumber\\
&&\left.\hspace{1cm}+49(\partial ^3h)(\partial
u)u-9(\partial ^2h)(\partial ^2u)u+3\alpha
u(\partial ^3\nabla
u)-5\alpha (\partial u)(\partial ^2\nabla
u)\right\}\nonumber\\&&+
B^*\left\{ 9(\partial ^3u)(u\partial B-B\partial u)+(-9+\alpha )(\partial
u)u\partial ^3B+ 10\alpha (\partial ^2u)u\partial ^2B\right\} \nonumber\\
&&+X^*_\mu (\partial X^\mu )\left\{ 5u\partial ^3u+12(\partial ^2u)\partial
u\right\} \nonumber\\[3mm]
\frac{6\pi i}{ \kappa}(\Delta S)^1_F&=&-50 h^*\left\{ (1-\alpha )(\partial
u)u\partial ^3h+(\partial ^3c)(B\partial u-u\partial B)\right\} \nonumber\\
&& -3h^*\left\{  \left( (\partial ^3c)+3 (\partial ^2c)\partial \right)
(u\partial B-B\partial u)
+u(\partial u)\partial ^3h+3u(\partial ^2u)\partial
^2h\right\} \nonumber\\
&&-3B^*\left\{ -3(u\partial B-B\partial u)\partial
^3u+3u(\partial u)\partial ^3B\right\} \nonumber\\
&&-3X_\mu ^*\left\{ -2u(\partial u)\partial ^3X^\mu -(\partial
X^\mu )(\partial (u\partial ^2u))-2u(\partial ^2u)\partial ^2X^\mu \right\}
\end{eqnarray}
Remarkable simplifications occur for the full anomaly:
\begin{eqnarray}
i(\Delta S)^0+i(\Delta S)^1&=& i(\Delta S)_m^0+i(\Delta S)_m^1+i(\Delta
S)_F^0+ i(\Delta S)_W^0\nonumber\\
&&+\frac{\kappa }{6\pi }X^*_\mu \left\{ 6\partial (u(\partial u)\partial
^2X^\mu)+9(\partial ^2u)(\partial u)\partial X^\mu +8u(\partial
^3u)\partial X^\mu \right\}  \nonumber\\
&&
+\frac{\alpha \kappa }{6\pi }h^*\left\{
u(\partial ^2c)\partial ^2B+10 (\partial u)B(\partial ^3c)+15
(\partial u)(\partial B)(\partial ^2c)\right.\nonumber\\&& \hspace{13mm}
 -15u(\partial ^3c)\partial B-6u(\partial ^4c)B\nonumber\\
&&\left.\hspace{13mm}+50(\partial u)u\partial ^3h+3u\partial ^3\nabla
u+5(\partial ^2\nabla
u)\partial u\right\} \nonumber\\&&+\frac{\alpha \kappa }{6\pi
}B^*\left\{ (\partial u)u\partial ^3B+10(\partial ^2u)u\partial
^2B\right\} \ , \end{eqnarray}
where $(\Delta S)_m^0$ and $(\Delta S)_m^1$ are the terms of antifield number 0
and 1 in \eqn{cAm}. Note especially the simplification when using the
parametrisation with $\alpha =0$. The terms with the `antighost' $h^*$ are
included in the `matter anomaly' $(\Delta S)_m$, \eqn{cAm}, and $B^*$
disappears completely.

Let us recapitulate what we have determined. First remark that the
regularisation depends on an arbitrary matrix $T$, and this implies that,
not specifying $T$,  $\Delta S$ is only determined up to $(G,S)$, where $G$
is a local integral. We have chosen
a regularisation (a specific matrix $T$). This determines the value of
$\Delta S$. However, we have calculated only the part of $\Delta S$ at
antifield number 0 (including the weakly vanishing terms). If we would have
calculated up to field equations (which would in principle be sufficient
to establish whether the theory has anomalies), then we would have
determined
$\Delta S$ up to $(G,S)$, where $G$ has only terms with antifield
number~1 or higher.  In our calculation of section~\ref{ss:calc1lW3}, we
determined also the weakly vanishing terms. Therefore the value of $\Delta
S$ has been fixed up to the above arbitrariness with terms $G$ of antifield
number~2 or higher. Indeed, looking at \eqn{cSF00} one can
always shift $(\Delta S)^1\rightarrow (\Delta S)^1+\delta _{KT}G^2$, for
some arbitrary function $G^2$ of antifield number two.

To obtain the complete form of $\Delta S$ up to $(S,G)$ one can continue
the calculations of this subsection to determine the terms of antifield
number 2.

\subsection{Background charges in $W_3$ gravity}
\label{ss:backgrch}
It is well known that the anomalies can be cancelled in chiral
$W_3$ gravity by including background charges \cite{ROMANS}. We will see in
this section how this can be implemented in the BV language.

To cancel the anomalies by local counterterms, we first note that $(\Delta
S)^0_W\approx 0$. Our theorem of section 3 then
implies that there is a
local counterterm, which starts with (we take in this section $\alpha =0$,
but of course
these steps can also be done in parametrisations with $\alpha \neq 0$, as
this involves only a canonical transformation)
\begin{equation}
M_{W1}{}^1=-\frac{\kappa }{\pi } B\left[ \frac{5}{6}u(\partial
^3h^*)+\frac{9}{2}(\partial u)(\partial ^2h^*)+\frac{17}{2}(\partial
^2u)(\partial h^*)+\frac{17}{3}(\partial ^3u)h^*\right]\ , \label{counter1}
\end{equation}
which is the right hand side of the last expression in \eqn{ANOM}, with
$y_h$ replaced by $h^*$, and where we added a total derivative for later
convenience.
The other terms in \eqn{DelSW3} can not be countered by a local integral.

Background charges are terms with $\sqrt{\hbar }$ in the (extended) action.
We can make an expansion \cite{anomw3,hideanom}
\begin{equation}
W=S+\sqrt{\hbar}\,M_{1/2}+\hbar M_1+\ldots \ .
\end{equation}
Therefore the expansion \eqn{MEC} of the master equation \eqn{ME} is now
changed to
\begin{eqnarray}
(S,M_{1/2})&=&0 \label{Mhalf}\\
(S,M_1)&=&i\Delta S-\frac{1}{2}(M_{1/2},M_{1/2})\,. \label{MEsqh}
\end{eqnarray}
Relevant terms $M_{1/2}$ are those which are in the antibracket
cohomology. Indeed, if $M_{1/2}$ which solves \eqn{MEsqh} has a part
$(S,G)$, then
we find also a solution by omitting that part of $M_{1/2}$, and adding to
$M_1$ a term $\ft12\left( M_{1/2},G\right) $.
These terms $M_{1/2}$ are thus again determined by their part at antifield
number zero. In chiral $W_3$ gravity one considers
\begin{equation} M_{1/2}{}^0=(2\pi )^{-1/2}\left[ a_\mu
h(\partial ^2X^\mu )+e_{\mu \nu }B\pl (\partial ^2 X^\nu )\right] \ ,
\end{equation}
where the numbers $a_\mu $ and $e_{\mu \nu }$ are the background charges,
and the numerical factor is for normalisation in accordance with previous
literature. We first consider \eqn{Mhalf}.
Using our theorem, $D^0M_{1/2}{}^0$ should be weakly zero
in order to find a solution. This gives the
following conditions on the background charges:
\begin{eqnarray}
e_{(\mu \nu )}-\ds a_\rho&=&0\nonumber\\
\ds (e_{\rho \sigma }-e_{\sigma \rho })+2e_{(\mu }{}^\rho d_{\nu )
\sigma \rho }&=& b_\sigma \delta _{\mu \nu }\ , \label{condbM12}
\end{eqnarray}
where $b_\sigma $ is determined to be $2\kappa a_\sigma $ by the
consistency of the symmetric part
$(\mu \nu \sigma )$ of the last equation with the first one and
\eqn{dsym}. If these conditions are satisfied,
we know that we can construct
the complete $M_{1/2}$ perturbatively in antifield number. The solution is
\begin{eqnarray}
M_{1/2}&=&
M_{1/2}{}^0+M_{1/2}{}^1+M_{1/2}{}^2
\nonumber\\
M_{1/2}{}^1&=&(2\pi )^{-1/2}\left[  -a_\mu X^*_\mu (\partial
c)+e_{\mu
\nu }X^*_\mu u(\partial ^2X^\nu )+e_{\nu \mu }(\partial X^*_\mu) u(\partial
X^\nu)\right.\nonumber\\
& &\hspace{25mm}\left.+\kappa a_\mu h^*(D^{-2}u)(\partial ^2X^\mu
)\right] \nonumber\\
M_{1/2}{}^2&=&(2\pi )^{-1/2}2\kappa a_\mu\left[ (\partial
X^*_\mu) h^* - c^*(\partial ^2X^\mu )\right]u (\partial u)\ .
\end{eqnarray}
We calculated here the terms of antifield number 2 (and checked that there
are no higher ones), but note that these are not necessary for the analysis
below.

Let us discuss the solutions of \eqn{condbM12}, together with the solutions
of \eqn{dsym}. The
general solutions of \eqn{dsym} are related to specific realisations of
real Clifford algebras ${\cal C}(D,0)$ of positive signature. We have then
$n=1+D+r$, where $r$ is the dimension of the Clifford algebra realisation.
We obtain solutions of \eqn{dsym} for the following values :
$(D=0,r=0)$, $(D=1,r$ arbitrary), $(D=2,r=2)$, $(D=3,r=4)$ (these are the
$SU(3)$ $d$--symbols), $(D=5,r=8)$,
$(D=9,r=16)$. The latter four are the so--called `magical cases'. Then the
Clifford algebra representation is irreducible, and the $d$--symbols are
traceless as it is the case for $(D=1,r=0)$. All the solutions can be
given in the following way.
$\mu$ takes the values $1,a$ or $i$, where $a$ runs over $D$ values and $i$
over $r$ values. The non--zero coefficients are (for $\kappa =1$)
$d_{111}=1$, $d_{1ab}=-\delta _{ab}$, $d_{1ij}=\frac{1}{2}\delta _{ij}$,
and $d_{aij}=\frac{\sqrt{3}}{2}(\gamma _a)_{ij}$. For $D=1$ the gamma
matrix is $(\gamma _a)_{ij}=\delta _{ij}$ (reducible). In that case the
form of the solution can be simplified by a rotation between the index 1,
and $a$, which takes only one value, see below: \eqn{solD1}. All
representations
of all real Clifford algebras ${\cal C}(D,0)$ appear as solutions of a
generalisation of \eqn{dsym} and classify the homogeneous special
K\"ahler and quaternionic spaces \cite{alhom}.

In this set of solutions
we can shown that there is only a solution for \eqn{condbM12}
in the case $D=1$ and $r$ arbitary. In \cite{ROMANS} the generic
(i.e. $D=1$, $r$ arbitrary) solution was already found,
and no solution was found for the
magical cases, but this was also not excluded. In \cite{Mohamm}, it was
shown that the first twomagical realisations did not survive quantisation.
Later, it was shown that also the other two cases lead to anomalous
theories \cite{anomw3,jose}.
Indeed, in all the other cases (i.e.
the four magical and the $D=r=0$) we have that the d--symbols are
traceless. By taking traces of the equations in \eqn{condbM12}, we find
that the only (trivial) solutions are $a_\mu =e_{\mu \nu }=0$, i.e. no
background charges. This means that the four magical cases lead to
anomalous theories.

There is thus exactly one solution for each value of $n$, the range of the
index $\mu $. For these models, the solution of \eqn{dsym} can be simply
written as
\begin{equation}
d_{111}=-\sqrt{\kappa }\ ;\qquad d_{1ij}=\sqrt{\kappa }\,\delta _{ij}\ ,
\label{solD1} \end{equation}
where $i=2, \ldots, n$. The solution to \eqn{condbM12} is
\begin{eqnarray}
&&e_{00}=-\sqrt{\kappa }\,a_0\ ;\qquad
e_{ij}=\sqrt{\kappa }\,a_0\delta _{ij}
\nonumber\\
&&e_{0i}=2\sqrt{\kappa }\,a_i\ ;\qquad  e_{i0}=0\ , \label{solD1e}
\end{eqnarray}
where $a_\mu $ is arbitrary.

The final relation for absence of anomalies up to one loop is that we have
to find an $M_{1}$ such that the last equation of \eqn{MEsqh} is satisfied.
Again we only have to verify this at zero antifield number, and up to field
equations of $S^0$, due to our theorem. We calculate
$Q\equiv i\Delta
S-\frac{1}{2}(M_{1/2},M_{1/2})$ at antifield number zero. It contains
terms proportional to $ c\partial ^3h$, which can not be removed by a
local counterterm. So in order to have no anomaly, the
multiplicative factor of this term has to vanish.
This implies the relation
\begin{equation}
c_{mat}\equiv n-12a_\mu a_\mu =100
\end{equation}
For the other terms in $Q$, one needs a counterterm
\begin{equation}
M_{B1}{}^0=\frac{1}{6\pi }e_{\mu \nu }a_\mu B\partial ^3 X\ ,
\end{equation}
and one imposes the relations
\begin{eqnarray}
&&2e_{\mu \nu }a_\mu -6a_\mu e_{\nu \mu }+d_{\nu \mu \mu }=0\nonumber\\
&&e_{\mu \rho }e_{\rho \nu }=\kappa Y\delta _{\mu \nu }\nonumber\\
&&-e_{\mu \rho }e_{\nu \rho }+\ft13 d_{\mu \rho \sigma }d_{\nu \rho \sigma
}+\ft23d_{\mu \nu \rho }e_{\sigma \rho }a_\sigma =\kappa \delta _{\mu \nu
}Z\nonumber\\
&&-4 \kappa a_\mu a_\nu +e_{\rho \mu }e_{\rho \nu }+2a_\rho e_{\rho \sigma
}d_{\sigma \mu \nu }=(3Z+4Y-2)\kappa \delta _{\mu \nu }\ , \end{eqnarray}
where $Y$ and $Z$ are arbitrary numbers, to be determined by consistency
requirements.
This set of equations on the background charges are exactly
the same as in \cite{ROMANS}.
The solutions \eqn{solD1} and \eqn{solD1e} give now
\begin{eqnarray}
&&a_0{}^2=-\frac{49}{8}\ ;\qquad  a_i{}^2= \frac{-53+2n}{24}\nonumber\\
&&Y=a_0{}^2\ ;\qquad Z=\frac{1}3 \left(2-a_0{}^2\right)=\frac{87}{8}\ .
\end{eqnarray}
Then we obtain that
\begin{eqnarray}
&&2\pi \left( (M^0_{1/2},M^1_{1/2})-i\Delta
S+(M_{B1}^0,S^1)\right) =-\ft13 e_{\nu \mu }a_\nu (\partial
^2u)y_\mu\\
&&+\kappa B\left( 2Z (\partial ^3u) y_h +3Z(\partial ^2u)\partial
y_h+ (3Z-2
+2Y)(\partial u)\partial ^2y_h+(Z+Y-1 )u\partial ^3y_h\right) \ .
\nonumber \end{eqnarray}
This determines then the counterterm $M_{B1}$ at antifield number 1.
Inserting the values for $Y$ and $Z$ gives
\begin{eqnarray}
M_{B1}{}^1&=&\frac{\kappa }{16\pi } B\left[ 30u(\partial
^3h^*)+147(\partial u)(\partial ^2h^*)+261(\partial
^2u)(\partial h^*)+174(\partial ^3u)h^*\right]\nonumber\\&&+
\frac{1}{6\pi }X^*_\mu e_{\nu \mu }a_\nu \partial ^2u\ .
\label{counter2} \end{eqnarray}
Combining this term with \eqn{counter1}, we get for the total
counterterm at antifield number one
\begin{eqnarray}
M_1{}^1&=&
\frac{25\kappa }{48\pi } B\left[ 2u(\partial
^3h^*)+9(\partial u)(\partial ^2h^*)+15(\partial
^2u)(\partial h^*)+10(\partial ^3u)h^*\right] \nonumber\\
&&+\frac{1}{6\pi }X^*_\mu e_{\nu \mu }a_\nu \partial ^2u\ .
\label{counter}
\end{eqnarray}
The form of this counterterm is also obtained in the last reference of
\cite{partanomw3} and also in
\cite{SSvNMiami,Bergshoeff}. As mentioned before, this does not only
determine the quantum corrections to the action, but also the corrections
to the BRST transformations, by looking to the linear terms in antifields
in the gauge--fixed basis.

The value of $\kappa $ has been irrelevant here. In fact, one can remove
$\kappa $ rescaling $d _{\mu \nu \rho }$, $e_{\mu \nu }$, $B^*$ and
$u^*$ with $\sqrt{\kappa }$ and $B$ and $u$ by $(1/\sqrt{\kappa })$. For
the usual normalisations in operator product expansions of the $W$--algebra
one takes
\begin{equation}
\kappa =\frac{1}{6Z}=\frac{8}{22+5c_{mat}}=\frac{4}{261}\ .
\end{equation}

\subsection{Higher loop anomalies in the BV formalism}
The discussion so far concerned the one loop theory. We have seen that, in
order to cancel the anomaly, one needs to add a counterterm to the
classical action, and this counterterm was antifield dependent.
The next step is of course to go to the anomaly computation at two loops.
At order $\hbar ^2$, the master equation reads~:
\begin{equation}
\cA _2=\Delta M_1+\frac{i}{2}(M_1,M_1)+i(M_2,S)\ .\label{2loop}
\end{equation}
To make computations at
order $\hbar ^2$ and to investigate the anomaly equation, one must
regularise at two loop. Pauli-Villars is then not sufficient anymore and
one must use another regularisation scheme.

It is very important to realise that the master equation
must be understood inside the path integral. This can be seen from
the anomalous Zinn--Justin equation \eqn{anZJ}. In fact, the right hand
side of this equation must be written as
\begin{equation}
<{\cal A}>=\lim _{\Lambda \rightarrow \infty}<{\cal A}^L>_\Lambda =
\lim _{\Lambda \rightarrow \infty}<\hbar {\cal A}^\Lambda _1+\hbar
^2{\cal A}_2^\Lambda +...>_\Lambda \ .
\end{equation}
By this we mean the following. When one regularises the theory, one
introduces a cutoff $\Lambda $, e.g. for (one loop) PV regularisation,
this is simply the mass $M$. This cutoff appears in the regularised action.
The $\Lambda $ after the brackets in this expression means that one
considers the expectation value with this cutoff dependent action.
On the other hand, the anomaly one computes
is also dependent on the cutoff, as we have seen before in explicit
examples.

For the two loop anomaly, this means that
there can be two contributions. The first is coming from
computing the first two terms in the r.h.s of \eqn{2loop}.
Of course, if no
counterterm $M_1$ was needed to cancel the one loop anomaly, these do not
contribute. The second contribution is coming from inserting the one loop
anomaly ${\cal A}_1^\Lambda $ in the path integral. In order to have an
anomaly free theory, the
sum of these two contributions should be absorbed in a local counterterm
$M_2$.

The two loop anomalies in chiral $W_3$ gravity were first computed in
\cite{partanomw3}, directly from the effective action. In
the context of the antifield formalism, the two loop master
equation was recently studied, first in \cite{jordiw3} and later in
\cite{jorwal}, using non--local regularisation.
In the latter, it was shown that the well known 2--loop anomaly
follows, with the correct coefficient, directly from inserting the one loop
anomaly ${\cal A}_1^\Lambda $ in the path integral. However, one
must pay special attention to the regularisation procedure. As we already
said, the one loop anomaly depends on the cutoff.
One can then isolate the finite part of the one--loop anomaly by
sending the cutoff to infinity. To compute the finite part of the two
loop anomaly, one would expect to insert only the finite part of the
one--loop anomaly. However, as was shown in \cite{jorwal}, this is not
true. One has
to insert the {\it complete} (including the terms that explicitly depend
on the cutoff) expression for the one--loop anomaly, i.e. ${\cal
A}_1^\Lambda $ in the path integral, before sending the cutoff to infinity.
A careful analysis shows that
also these terms can
contribute to the finite part of the two loop anomaly, and, moreover, it
leads to the correct result. It is therefore important not to bring in the
limit into the path integral.

An analogous procedure was also applied
to quantum mechanical systems, see \cite{gervjev}.

\section{Consistency of the regulated anomaly}
\label{app:consista}
\setcounter{equation}{0}
The PV regularisation should give a consistent anomaly, as all
manipulations can be done at the level of the path integral. Here we want
to check this explicitly from the final expression for the anomaly after
integrating out the PV fields. Then we will consider the expression which
is obtained after integrating out only parts of the fields. We will see
that in that case the resulting expression does not satisfy this
consistency equation.

The expression for the anomaly depends on an invertible matrix $T_{AB}$:
\begin{equation}
\Delta S = str\left[ J \frac{1}{\unity - {\cal R}/M^2}\right]\
,\label{DelSreg} \end{equation}
where
\begin{eqnarray}
K^A{}_{B}=\dl^A  S\dr_B\ ;&\qquad& \underline{S}_{AB}=\dl_A S\dr_B
\nonumber\\
J=K+\ft12 T^{-1}({\cal S}T)\ ;&\qquad&{\cal R}=T^{-1}\underline{S}\ ,
\label{defKSJO}\end{eqnarray}
and for matrices $M^A{}_B$ or $M_{AB}$ we define the nilpotent operation
\begin{equation}
\left( {\cal S}M\right) _{AB}= \left( M_{AB},S\right) (-)^B\ .
\end{equation}
Matrices can be of bosonic or fermionic type. The Grassmann parity
$(-)^M$ of a matrix is the statistic of $M^A{}_B(-)^{A+B}$. So of the above
matrices, $J$, $K$ and $({\cal S}T)$ are fermionic, the other are bosonic.
The definition of supertraces and supertransposes depend on the position of
the indices.
\begin{eqnarray}
\left( R^T\right) _{BA}=(-)^{A+B+AB+R(A+B)}R_{AB}\ ;&\qquad &
\left( T^T\right) ^{BA}=(-)^{AB+T(A+B)}T_{AB} \nonumber\\
\left( K^T\right)_B{}^A=(-)^{B(A+1)+K(A+B)}K^A{}_{B}\ ;&\qquad &
\left( L^T\right)^A{}_B=(-)^{B(A+1)+L(A+B)}L_B{}^A\nonumber\\
str\ K = (-)^{A(K+1)} K^A{}_{A}\ ;&\qquad&
str\ L = (-)^{A(L+1)} L_A{}^A
\ . \label{Trules}
\end{eqnarray}
This leads to the rules
\begin{eqnarray}
&&\left( M^T\right) ^T =M\ ;\qquad (MN)^T =(-)^{MN} N^T M^T\ ;\qquad
\left( M^T\right) ^{-1}=(-)^M\left( M^{-1}\right) ^T\nonumber\\
&&str\ (MN)= (-)^{MN}str\ (NM)\ ;\qquad str\ (M^T)= str\ (M)\nonumber\\
&&{\cal S}(M\,N)=M({\cal S}N)+(-)^N ({\cal S}M)N\ ;\qquad
{\cal S}\,(str\,M)=str\ ({\cal S}M)\,.
\label{propstr}\end{eqnarray}
The second derivatives of the master equation lead to
\begin{eqnarray}
{\cal S}\underline{S} &=& -K^T \underline{S}-\underline{S}\,K\nonumber\\
{\cal S}K&=&  \overline{S}\,\underline{S}-K\,K \ ,
\end{eqnarray}
where,
\begin{equation}
\overline{S}=\dl^A  S\dr^B\ ;\qquad
(K^T)_A{}^B=-\dl_A  S\dr^B\ ,
\end{equation}
the first being a graded antisymmetric matrix, and the second is in
accordence with \eqn{defKSJO} and the previous rules of supertransposes.
We rewrite the expression of the anomaly as
\begin{equation}
\Delta S=M^2\ str\left[ T\,J\, P^{-1}\right] \ ,
\end{equation}
where
\begin{equation}
P=M^2\,T-\underline{S}\ .
\end{equation}
We also easily derive the following properties
\begin{eqnarray}
{\cal S}(T\,J)&=&{\cal S}(T\,K)=T\overline{S}\underline{S}-TKK-({\cal
S}T)K\nonumber\\
{\cal S}P&=&M^2(TJ+J^TT)-(PK+K^TP) \label{varyP}\ .
\end{eqnarray}
This leads to
\begin{eqnarray}
{\cal S}\Delta S&=&M^2\
str\left[\left(T\overline{S}\underline{S}-TKK-({\cal
S}T)K\right)P^{-1}\right.\nonumber\\&&\hspace{5mm}\left.-TJP^{-1}\left(M^2
( T
J + J ^ T T ) - ( P K + K^T P ) \right) P ^ { - 1 } \right] \end{eqnarray}
In the first term of the first line we write $\underline{S}=M^2T-P$. The
trace of both these terms is zero due to \eqn{propstr} and the
(a)-symmetry properties of the matrices given above. For the same reason
the second term of the second
line vanishes. The first term of
the second line is a square of a fermionic matrix which vanishes under the
trace. The remaining terms again combine into matrices which are traceless
by using their symmetry and by \eqn{propstr}.
This means we have proven that PV-regularisation guarantees
consistent one--loop anomalies,
\begin{equation}
{\cal S}\Delta S=0\ .
\end{equation}
In \cite{jordi} a formula has been given for the non--local counterterm
for any $\Delta S$ defined by \eqn{DelSreg}:
\begin{equation}
\Delta S=-\frac{1}{2}{\cal S}\left( str \ln \frac{{\cal R}}{M^2-{\cal R}
}\right) \ . \end{equation}
This can be proven also from the above formulas.
\vspace{1cm}

Consider the part of the anomaly originating from the path integral over
some subset of fields, e.g. the matter fields in $W_3$.
The question arises whether this gives already a consistent anomaly.
To regulate this anomaly we only have to introduce
PV--partners for this subset of fields. In the space of all
fields, and taking the basis with first the fields which are integrated, we
write the $T$--matrix as \begin{eqnarray}
T=\left(\begin{array}{cc}
\tilde T & 0\\
0 & 0
\end{array}\right)\ .
\end{eqnarray}
Because only this subsector is integrated, we have to project out the
other sectors (mixed and external) in the full matrix of second
derivatives $\underline{S}$ before inverting it to define a
propagator. This can be done by defining the projection operator $\Pi $ as
\begin{eqnarray}
\Pi =\left(\begin{array}{cc}
\unity & 0\\
0 & 0
\end{array}\right)\ .
\end{eqnarray}
The inverse propagator is then $Z=(\Pi P\Pi)$. Further we understand by
inverses, as e.g. $T^{-1}$, the inverse in the subspace. So we
have \begin{equation}
TT^{-1}=T^{-1}T=\Pi \ .
\end{equation}
The matter anomaly takes the form
\begin{equation}
(\Delta S)_m=M^2\ str[TJZ^{-1}]\ ,
\end{equation}
and \eqn{varyP} remains valid. In the variation of this anomaly, we now
often encounter
\begin{equation}
P\,Z^{-1}=\Pi + Y \qquad \mbox{where}\qquad Y=-(1-\Pi )\underline{S}Z^{-1}\
. \end{equation}
Using again the properties of supertraces and transposes, we obtain
\begin{equation}
{\cal S}(\Delta S)_m=M^2\ str\left[-T\overline{S}Y-TK(\unity -\Pi
)KZ^{-1}-Y^TKZ^{-1}(TJ+J^TT)\right] \ .\label{violSA}
\end{equation}
This thus shows that one does in general not obtain a consistent anomaly
when integrating in the path integral only over part of the fields.  One
can see that for ordinary chiral gravity the structure of the extended
action
implies that each term of \eqn{violSA} vanishes.  For chiral $W_3$ gravity
however, some terms remain.  A similar result for Ward identities was
obtained in \cite{FBWardid}.

\section{Remarks on the classical and quantum cohomology}
Let us first repeat again how one defines the physical states of the
classical theory. One has a nilpotent operator $\cS=(\cdot,S)$, such that
the spectrum is defined as the elements of the antibracket cohomology of
$\cS$. Suppose now that we have two elements of the cohomology $F$ and $G$,
satisfying
\begin{equation}
\cS F= 0 \qquad \cS G=0 \ ,
\end{equation}
and they are not $\cS$ exact. Then, we can find two new invariants namely
the product and the antibracket of $F$ and $G$~:
\begin{equation}
\cS [FG]=0 \qquad \cS (F,G)=0\ .
\end{equation}
This is because the operator $\cS$ works as a derivation on the product as
well as on the antibracket. It could however be that the product or the
antibracket are $\cS$ exact, so that they drop out of the cohomology. If
not, one generates new elements in the cohomology. Let us still mention
that, if $F$ or $G$ is $\cS$ exact then also the product and the
antibracket is $\cS$ exact, as one can easisly check.

This procedure defines a ring structure of observables, $\cR$. For local
functions
there is no antibracket cohomology at negative ghost number. So the lowest
ghosts number of an element in the cohomology is zero. By taking the
product of two elements one stays at ghost number zero. Then the product
defines a ring at ghost number zero, $\cR _0$. By taking the antibracket of
two
elements of zero ghost number one obtains a (possible) observable at ghost
number one. If one consider also local functionals, there can also be
cohomology at negative ghost number. The lowest are the functionals at
ghost number minus one. By taking the antibracket between two such local
functionals, one again obtains a local functional at ghost number minus
one, because the antibracket increases the ghost number with one. So this
procedure defines yet another ring, which we can call $\cR _{-1}$.

In the quantum theory, we also have a nilpotent operator, which involves
the $\Delta $ operator. It is
\begin{equation}
\cS _q=(W,\cdot )+\frac{\hbar}{i}\Delta \ ,
\end{equation}
and satisfies $\cS _q^2=0$ if the quantum master equation is satisfied.
This operator enables us to define physical
states at the quantum level, namely elements of the quantum cohomology. If
$F=F_0+\hbar F_1+\hbar ^2 F_2+ ...$ is an invariant under $\cS _q$, it must
satisfy
\begin{eqnarray}
(F_0,S)&=&0\nonumber\\
i\Delta F_0-(F_0,M_1)&=&(F_1,S) \ ,\label{quobs}
\end{eqnarray}
and further conditions at higher order in $\hbar$. The above equation
already tells us that a classical observable, satisfying $(F_0,S)=0$, can
not necessarily be extended to a quantum observable, since there is an
extra condition on $F_0$ coming from the quantum theory. It says that the
left hand side of the second equation of \eqn{quobs} must be $\cS$ exact.

Suppose now we have found two elements $F$ and $G$ in the quantum
cohomology.
Analogous to the classical theory, we can try to construct new observables,
but now they must be invariant under $\cS _q$. Since the $\Delta $ does not
act as a derivative on the product, this product will in general not be
invariant. This can be seen from the formulas
\begin{eqnarray}
\cS _q[FG]&=&\cS _q[F]G+(-)^FF\cS _qG-(-)^Fi\hbar(F,G)\nonumber\\
\cS _q[(F,G)]&=&(\cS _qF,G)+(-)^{F+1}(F,\cS _qG)\ .
\end{eqnarray}
That means that, in general, the product is not invariant, but is
proportional to the antibracket of $F$ with $G$. The latter means that also
the
bracket can not be an element from the quantum cohomology since it is $\cS
_q$ exact. The construction of making new observables out of $F$ and $G$
can only work when $(F,G)=0$. If this condition is satified, the product
$FG$ is quantum invariant (of course it can still be exact). This can occur
when $F$ and $G$ have no antifield dependence. So the ring structures
discussed in the classical case can not be maintained in the quantum
theory. Only when $(F,G)=0$ we can build up a ring under ordinary
multiplication.

All this is rather formal. It would be very interesting to investigate
these structures in explicit examples. At the classical level, a starting
point was give in the second paper of \cite{Brandt} for the bosonic string.
For the
quantum cohomology in the BV framework, almost nothing is known. However, a
connection with the ground ring structure given in \cite{grring} in the
BRST framework certainly must exist.
We leave this for future research.

\chapter{BV and topological field theory}
\section{Introduction}
Topological field theories (TFT) \cite{TFT,TFTrep} have attracted a lot
of attention recently. They are interesting both from a physical and
mathematical point of view. For physics, the interest is twofold. On
the one hand, they form "subtheories" of the general class of $N=2$ (or
also $N=4$) supersymmetric
invariant theories. By this we mean that any $N=2$ \footnote{For local
$N=2$
in four dimensions, one also needs the existence of $R$ symmetries, see
\cite{AnsFre}} theory contains
a topological subsector in which correlation functions can be computed
exactly. This is because in TFT the semiclassical limit is exact, so the
only contribution to correlation functions comes from the classical regime
and from instanton corrections. So these models can give new information
about non--perturbative effects (instantons) of $N=2$ theories. For recent
applications in non--perturbative gauge theories, see \cite{SeiWi}. The
procedure to go
from the $N=2$ theory to the TFT and backwards is called {\it twisting}, as
was explained in \cite{TFT,EY}.
On the other hand, as we will see, TFT have a very large gauge symmetry
group.
It is possible that gauge theories, like e.g. string theory, are in a
broken
phase of TFT. One could then study which properties of the string are still
described by the TFT, after the symmetry is broken. As an example, one can
think about the discrete states in the spectrum of the non-critical $c=1$
string \cite{c=1str}. Unfortunately, we do not yet have a good mechanism to
describe this symmetry breaking.

From the mathematical point of view, TFT provide a framework to compute
topological invariants of certain spaces. For four manifolds, these are the
Donaldson invariants, which can be obtained by computing the spectrum of
topological Yang-Mills theory on a general four manifold, see the first
reference in \cite{TFT}. In
fact, this method has led to spectacular new results in mathematics.
E.g., the long
standing open problem of computing how many rational curves there are on
the
quintic three-fold was solved in \cite{ratcurv} using techniques from TFT.

A general definition is that a TFT is characterised by the fact
that
its partition function is independent of the metric, which is considered to
be external and thus not included in the set of dynamical fields of the
theory~:
\begin{equation}
\frac{\delta Z}{\delta g^{\alpha \beta }}=\frac{\delta }{\delta g^{\alpha
\beta  }}\int {\cal D}\phi e^{\frac{i}{\hbar}S(\phi ,g_{\alpha \beta })}=0\
.\label{MIPI}
\end{equation}
As mentioned above, TFT have local gauge symmetries. In order to define a
path integral, one first must gauge fix. The resulting gauge fixed theory
then possesses a BRST operator, under which the action is invariant. The
condition of metric independence is then
satisfied owing to the Ward identities, provided the
energy momentum tensor
\begin{equation}
T_{\alpha \beta }=\frac{2}{\sqrt |g|}\frac{\delta S}{\delta
g^{\alpha \beta }}
\end{equation}
is BRST exact.
Soon after their discovery by Witten,
topological field theories were shown \cite{bvtft} to
be generally of the form
\begin{equation}
S=S_0+\delta _QV\ ,    \label{usual}
\end{equation}
where $S_0$ is either zero or a topological invariant (i.e. independent of
the metric) and BRST invariant. $\delta _QV$ is then the gauge fixing
term that corresponds to the gauge symmetry of $S_0$. Using the formal
arguments of \cite{topomeasure} based on Fujikawa variables to prove the
metric independence of the measure, one then has that
\begin{equation}
\frac{\delta Z}{\delta g^{\alpha \beta }}=\frac{i}{\hbar}\int {\cal D}\phi
\frac{\delta
}{\delta
g^{\alpha \beta  }} [\delta _QV]e^{\frac{i}{\hbar}S(\phi ,g_{\alpha
\beta })}\ . \end{equation}
Usually, differentiating with respect to the metric and taking the BRST
variation are freely commuted, which then leads to the desired result.
We will show in the next section, that the assumed commutation is not
allowed in general.

In order to investigate several steps of this process in more detail, we will
use the BV formalism. Although the BV scheme was used in \cite{bvtft} to
treat specific examples, the full power of this scheme was not
exploited. This was done in \cite{FSbv}, and we will discuss it also here.
Below we will define an energy momentum tensor with the
property $\cS _q T^q_{\alpha \beta} = 0$, by carefully specifying the
metric dependence of the antibracket and the $\Delta$-operator.
Hence, this $T^q_{\alpha \beta}$ is quantum BRST invariant.
For the theory to be topological, its energy momentum tensor
has to satisfy
\beq
             T^q_{\alpha \beta} = \cS _qX_{\alpha \beta },
\eeq
which makes $T^q_{\alpha \beta}$ cohomologically equivalent
to zero. As both tensors appearing in this equation can have an expansion
in $\hbar$,
this is a tower of equations, one for every order in $\hbar$. At the
classical level, we are looking for an $X_{\alpha \beta}^0$ such that
$T_{\alpha \beta} = (X^0_{\alpha \beta}, S)$, where $S$ is the classical
extended action. We will show that non trivial
conditions appear for higher orders in $\hbar$. Even when no
quantum counterterms are
needed to maintain the Ward identity, the order $\hbar$
equation is non trivial.

\section{The energy-momentum tensor in BV}

As explained in the introduction, the metric plays an important role in
TFT. Therefore, we have to be precise on its occurences in all our
expressions. This will depend on the chosen convention. The two obvious
possibilities for a set of conventions are the following~:\\
{\bf{1.}} All integrations are with the volume element $dx\sqrabsg $. The
functional derivative is then defined as
\beq
   \frac{\delta \phi^A}{\delta \phi^B} = \frac{1}{\sqrabsg_B} \delta_{AB}\ ,
\label{FD}
\eeq
and the same for the antifields. $\delta_{AB}$ contains both space-time
$\delta$-functions (without $\sqrabsg$) and Kronecker deltas ($1$ or
zero) for the discrete indices . $g$ is
$\det g_{\alpha \beta}$, and its subscript $B$ denotes that we evaluate it
in the space-time index contained in $B$.  Using this, the antibracket and
box operator are defined by
\begin{eqnarray}
 ( A, B ) & = & \sum_i \int dx \sqrabsg_X \left( \frac{\rd A}{\delta
\phi^X} \frac{\ld B}{\delta \phi^*_X} - \frac{\rd A}{\delta \phi^*_X}
\frac{\ld B}{\delta \phi^X} \right) \nonumber \\
  \Delta A & = & \sum_i \int dx \sqrabsg_X (-1)^{X}
\frac{\ld}{\delta \phi ^X}\frac{\ld}{\delta \phi^*_X} A\ ,
\end{eqnarray}
For once, we made the summation that is hidden in the De Witt summation
more explicit. $X$ contains the discrete indices $i$ and the space-time
index $x$.
We then have that $(\phi ,\phi
^*)=\frac{1}{\sqrabsg}$. In this convention the extended action takes the
form
\begin{equation}
S=\int dx\sqrabsg
[{\cal L}_0+\phi ^*_iR^i_a( \phi )c^a+\phi ^*_i\phi ^*_j...]\ .
\end{equation}
Demanding that
the total lagrangian is a scalar
amounts to taking the antifield of a scalar to be a
scalar, the antifield of a covariant vector to be a contravariant vector,
etc. .
For differentiating with respect to the metric, we use the
following rule~:
\begin{equation}
\frac{\delta g^{\alpha \beta }(x)}{\delta g^{\rho \gamma }(y)}=\frac{1}{2}
(\delta ^\alpha _\rho \delta ^\beta _\gamma +\delta ^\alpha _\gamma \delta
^\beta _\rho )\delta (x-y)\ ,
\end{equation}
where the $\delta $--function does not contain any metric, i.e.
$\int dx \delta (x-y)f(x)=f(y)$. This we do in order to agree with the
familiar recipe to calculate the energy-momentum tensor.
Let us now define the differential operator
\beq
     D_{\alpha \beta} = \frac{2}{\sqrabsg} \frac{\delta}{\delta g^{\alpha
\beta}} + g_{\alpha \beta} \sum_i \phi^*_X \frac{\ld}{\delta \phi^*_X}\ .
\eeq
Then it follows that
\begin{eqnarray}
D_{\alpha\beta} (A,B)&=& (D_{\alpha\beta} A, B) + (A , D_{\alpha\beta}
B)\nonumber\\
D_{\alpha\beta} \Delta  A&=&\Delta D_{\alpha\beta} A\ .\label{propD}
\end{eqnarray}
The antifield dependent terms in the second term of the right hand side are
really necessary for these properties, due to the chosen
notation.
If we now define the energy
momentum tensor as
\beq
T_{\alpha \beta} =  D_{\alpha\beta}  S,\label{EMTBV}
\eeq
then it follows immediately that
\beq
D_{\alpha\beta} (S,S) = 0 \Leftrightarrow ( T_{\alpha \beta} , S ) =0.
\eeq
Again, we remark that the second term in $D_{\alpha \beta }$ is necessary
to make the energy momentum tensor BRST invariant, as a consequence of
our conventions. By adding to this expression for
$T_{\alpha \beta}$ terms of the form $(X_{\alpha \beta},S)$, one can
obtain
cohomologically equivalent expressions. For example, by subtracting the
term $(\frac{1}{2} g_{\alpha \beta} \sum_i \phi^*_X \phi^X, S)$,
$T_{\alpha
\beta}$ takes a form that is more symmetric in the fields and antifields.
Analogously, for the quantum theory, if we define
\beq
     T^q_{\alpha \beta} =D_{\alpha\beta} W,
\eeq
then it follows by letting $D_{\alpha\beta}$ act on the quantum master
equation that
\beq
  \cS _q T^q_{\alpha \beta} = (T^q_{\alpha \beta} , W ) - i\hbar \Delta
T^q_{\alpha \beta} = 0\ .
\eeq
Now we turn to a second set of conventions.\\
{\bf{2.}} We integrate
with the volume element $dx$ without metric, and define the functional
derivative (\ref{FD}) without $\sqrabsg$. Also the antibracket is defined
without explicit metric dependence,
and so we have that $(\phi ,\phi ^*)'=1$. With this
bracket the extended action takes the form
\begin{equation}
S'=\int dx[\sqrabsg{\cal L}_0+
\phi ^*_iR^i_a(\phi )c^a+\frac{1}{\sqrabsg}\phi ^*\phi ^*...]\ .
\end{equation}
The energy momentum tensor is then defined as
\begin{equation}
T_{\alpha \beta }=\frac{2}{\sqrabsg}\frac{\delta S'}{\delta g^{\alpha
\beta }}\ ,
\end{equation}
and is gauge invariant because, in this convention, the derivative w.r.t.
the metric acts like an ordinary derivative on the antibracket, i.e.
\begin{eqnarray}
\frac{\delta }{\delta g^{\alpha\beta}} (A,B)&=& (\frac{\delta }
{\delta g^{\alpha\beta}} A, B) + (A
, \frac{\delta }{\delta g^{\alpha\beta}} B)\nonumber\\
\frac{\delta }{\delta g^{\alpha\beta}} \Delta  A&=&
\Delta \frac{\delta }{\delta g^{\alpha\beta}} A\ .
\end{eqnarray}
These variables have the advantage not to work with the operator
$D_{\alpha \beta }$. However general covariance is
not explicit and requires a
good book--keeping of the $\sqrabsg$ 's in the extended action and in other
computations. Therefore, we will not use
this convention. Let us finally mention that
the relation between the two sets of conventions is a transformation
that scales the antifields with the metric, i.e. $\phi ^*\rightarrow
\sqrabsg \phi ^*$. Both conventions show however, that one cannot
simply commute the derivative w.r.t. the metric (or $D_{\alpha \beta }$)
with the BRST charge,
as the BRST operator is simply the antibracket, i.e.
$\delta _Q\,A(\phi)=(A,S)$.

From now on, we will work in the first convention. All our arguments can
be repeated in the second convention. In the remaining part of this section
we will show that our definition of the energy momentum tensor
is invariant under (infinitesimal) canonical transformations with
generating fermion $f(\phi ,\phi ^{'*})$, up to a term that is BRST exact.
The expression in the primed coordinates for any function(al) given in the
unprimed coordinates can be obtained by direct substitution of the
transformation rules. Owing to the infinitesimal nature of the
transformation, we can expand in a Taylor series to linear order and we
find
\beq
   X'(\phi ',\phi^{*'})
       =  X (\phi^{A'} , \phi^{*'}_A ) - (X,f)\ .
\eeq
Especially, the
classical action and the energy-momentum tensor transform as follows:
\begin{eqnarray}
      S^{'} & = & S - (S,f) \nonumber \\
      T^{'}_{\alpha \beta} & = & T_{\alpha \beta} - (T_{\alpha \beta}, f)
      \ . \label{Tcantr}
\end{eqnarray}
Here, $T_{\alpha \beta}$ is the energy-momentum tensor that is obtained
following the recipe given above starting from the extended action $S$.
Analogously, we can apply the recipe to the transformed action $S^{'}$,
which leads to an energy-momentum tensor $\tilde T_{\alpha \beta}$. Using
\eqn{Tcantr} and \eqn{propD}, it is easy to show that
\begin{eqnarray}
  \tilde T_{\alpha \beta} &=&  D_{\alpha\beta} S' \nonumber \\
  & = & T^{'}_{\alpha \beta} + (D_{\alpha\beta} f , S'),
\end{eqnarray}
as for infinitesimal transformations terms of order $f^2$ can be neglected.
We will use below that if $T_{\alpha \beta} = (X_{\alpha \beta}, S)$, then
$\tilde T_{\alpha \beta} = (\tilde X_{\alpha \beta}, S')$ as canonical
transformations do not change the antibracket cohomology. For
infinitesimal transformations we have that
\beq
\label{nice}
\tilde X_{\alpha \beta} = X_{\alpha \beta} - (X_{\alpha \beta},f)
+D_{\alpha \beta }f\ .
\eeq
This argument can easily be repeated for the quantum theory. This last
formula will be used in section (4).

\section{Topological field theories in BV}
After carefully introducing the energy momentum tensor, we define a
topological field theory by the condition
\begin{equation}
      T^q_{\alpha \beta} = \sigma X_{\alpha \beta}\ .
      \label{TME}
\end{equation}
First we remark that this condition is gauge independent (canonical
invariant), due to the presence of the antifields, as was explained in the
previous section. If the energy momentum satisfies the above equation, we
have
\begin{equation}
D_{\alpha \beta }Z=0 \qquad {\mbox {or}} \qquad \frac{\delta }{\delta
g^{\alpha \beta }}Z=0\ , \end{equation}
depending on which set of conventions one chooses. If one fixes the gauge
by dropping the antifields after a suitable canonical transformation, one
recovers \eqn{MIPI}. We also assumed that one can construct a metric
independent measure.

In general, both $W$ and $X_{\alpha \beta}$ have an expansion in terms of
$\hbar $~:
\begin{eqnarray}
W&=&S+\hbar M_1+\hbar^2M_2+\ldots \ .\nonumber \\
X_{\alpha \beta}&=&X^0_{\alpha \beta} + \hbar X^1_{\alpha \beta} +
\ldots \ .
\end{eqnarray}
Thus we see that (\ref{TME}) leads to a tower of equations, one for each
order in $\hbar$. The first two orders are
\begin{equation}
      T_{\alpha \beta} = (X^0_{\alpha \beta}, S)\ ,  \label{ctme}
\end{equation}
at the $\hbar^0$ level and
\begin{equation}
\frac{2}{\sqrabsg} \frac{\delta M_1}{\delta g^{\alpha \beta }}
+ g_{\alpha \beta} \sum_i \phi ^*_X \frac{\dl M_1}{\delta \phi^*_X}
=(X^0_{\alpha \beta },M_1)+(X^1_{\alpha \beta },S)-i\Delta
X^0_{\alpha \beta }\ ,\label{OLTME}
\end{equation}
at the one loop level.
Once $M_1$ is known from the one loop master equation,
one has to solve (\ref{OLTME}) for $X^1$. This is an important
equation that must be satisfied at the one loop level.
To solve these equations one needs a regularisation
scheme, such that one can calculate (divergent) expressions like $\Delta S$
and $\Delta X^0_{\alpha \beta}$. If no (local) solution can be
found for $X_{\alpha \beta}$ then the proof of the topological nature
of the theory, based on the Ward identity, breaks down\footnote{In
\cite{TFTrep}, and references therein, it was shown that the one loop
renormalisation procedure
does not break the topological nature of the theory. However, the finite
counterterm $M_1$, to cancel possible anomalies, and the $X^1$ term have
not been discussed.}. Notice that
even when no counterterm
$M_1$ is needed, one still has to solve the one loop equation if $\Delta
X^0_{\alpha \beta} \neq 0$.

Let us come back to the classical part of the discussion.
As is mentioned in the introduction, and as we will see in our example, the
gauge--fixed
action turns out to be BRST exact in the antibracket sense (up to a metric
independent term):
\begin{equation}
S=S_0+(X,S) ,     \label{SGF}
\end{equation}
where $S_0$ is a topological invariant. However, we want to stress
that this is not the fundamental equation to characterise TFT.
From this, it does not follow that (\ref{ctme}) is satisfied. Rather,
\beq
   T_{\alpha \beta} = ( \frac{2}{\sqrabsg} \frac{\delta X}{\delta g^{\alpha
\beta}} + g_{\alpha \beta} \sum_i \phi^*_X \frac{\dl X}{\delta \phi^*_X} ,
S ) + (X,T_{\alpha \beta})\ .
\eeq
In order for the theory to be topological, the second term should be BRST
exact.

\section{Example 1~: Topological Yang--Mills theory}
Topological Yang-Mills theory was first constructed by Witten in
\cite{TFT}.
Although he obtained the model via twisting, it can also be obtained via
gauge fixing a topological invariant \cite{bvtft,FSbv}, as we will show
below.

We start from a four manifold ${\cal M}$ endowed with a metric
$g_{\alpha \beta }$. On
this manifold we define the Yang--Mills connection $A_\mu =A_\mu ^aT_a$,
where $T_a$ are the generators of a Lie group $G$. We use
the same conventions as for ordinary Yang Mills theory (see chapter 3 and
6). The first step is to choose a classical action. We can take it to be
the integral over the 4 dimensional manifold ${\cal M}$ of the second Chern
class of a principle $G$--bundle $P\rightarrow {\cal M}$~:
\begin{equation}
S_0=\int _{\cal M}d^4x \sqrabsg \tr F_{\mu \nu }\tilde {F}^{\mu \nu }\ .
\label{TYM}
\end{equation}
It is known from topology that this action is a integer number, also
called winding number.

The dual of an antisymmetric tensor $G_{\mu \nu }$ is
defined by \begin{equation}
\tilde {G}_{\mu \nu }=\frac{1}{2}[\epsilon ]_{\mu \nu \sigma \tau
}G_{\alpha \beta }g^{\alpha \sigma }g^{\beta \tau }\ .
\end{equation}
The Levi-Civita
tensor is defined by $[\epsilon ]_{\mu \nu \sigma \tau
}=\sqrabsg \epsilon
_{\mu \nu \sigma \tau }$, where $\epsilon _{\mu \nu \sigma \tau }$ is the
permutation symbol and $g=\det g_{\alpha \beta }$.
By raising the indices, we also have that
$[\epsilon ]^{\mu \nu \sigma \tau }=\pm\frac{\sqrabsg}{g}\epsilon
_{\mu \nu \sigma \tau }$.
It is then clear that the classical is a topological invariant
in the sense that it is independent of the metric.

The classical action
is invariant under continous deformations of the gauge fields that do not
change the winding number~:
\begin{equation}
\delta A_\mu =\epsilon _\mu \ .  \label{SS}
\end{equation}
Following the approach of e.g. \cite{bvtft} we will
now gauge fix this shift symmetry by
introducing ghosts $\psi _\mu $. Then we immediately
obtain the BV extended action
\begin{equation}
S=S_0+A^{*\mu}\psi _\mu \ . \label{S1}
\end{equation}
Remember that an overall $\sqrabsg$ is always understood in the volume
element of the space-time integration.
The usual approach is to include the
Yang--Mills gauge symmetry $\delta
A_\mu =D_\mu \epsilon $ as an extra symmetry, which then leads to a
reducible set of gauge
transformation as $D_\mu \epsilon $ is clearly a specific choice for
$\epsilon _\mu $. The description with a reducible set of gauge
transformations can be obtained from the description using only the shift
gauge symmetry, by adding a trivial system $(c,\phi )$ (ghost $c$ and ghost
for ghost $\phi $) to the action
\begin{equation}
S=S_0+A^{*\mu}\psi _\mu+c^*\phi\ ,
\end{equation}
and performing a canonical transformation, generated by the fermion
\begin{equation}
f=-\psi '^{*\mu }D_{\mu }c-\phi ^{'*}cc \ .
\end{equation}
We then obtain that (after dropping the primes)
\begin{eqnarray}
S&=&S_0+A^{*\mu }(\psi _\mu +D_\mu c)+\psi ^{*\mu }(\psi _\mu c+c\psi _\mu
-D_\mu \phi )\nonumber\\&&+ c^*(\phi +cc)-\phi ^*[c,\phi ] \ .
\end{eqnarray}
Notice that the correct reducibility transformations have appeared in the
action, i.e. $A^\mu $ transforms under the shifts as well under
Yang--Mills.
Of course, this extra symmetry with ghost $\phi$, has to be gauge fixed
too. This is done in the literature by introducing a Lagrange
multiplier and antighost (sometimes called $\eta $ and $\bar \phi$).
As the BV scheme allows us
to enlarge the field content with trivial systems and perform canonical
transformations at any moment, we are free to choose to include them
or not. However, as was explained in \cite{eqcoh}, including the
Yang-Mills symmetry enables one to define the equivariant cohomology.
This is an important concept, since the antibracket cohomology in the
space of all fields and their derivatives is empty, and so, there are no
physical states. The equivariant cohomology is defined as the antibracket
cohomology computed in the space of gauge (Yang-Mills) invariant
functions. This precisely leads to the correct spectrum, i.e. the Donaldson
invariants. However, in this chapter we will not compute the cohomology and
therefore, we choose the description with only the shift symmetry.

Let us now gauge--fix the shift symmetry (\ref{SS}) in order to obtain the
topological
field theory that is related to the moduli space of self dual YM instantons
\cite{TFT}. We take the usual gauge fixing conditions
\begin{eqnarray}
\label{GF}
F^+_{\mu \nu }&=&0\nonumber\\
\partial _\mu A^\mu &=&0\ ,
\end{eqnarray}
where $G_{\mu \nu }^{\pm}=\frac{1}{2}(G_{\mu \nu }\pm \tilde {G}_{\mu \nu
})$. These projectors are orthogonal to each other, so that we have
for general $X$ and $Y$ that $X^+_{\mu \nu }Y^{-\mu \nu }=0$. The
above gauge choice does not fix all the gauge freedom. The
dimension of the space solutions of (\ref{GF}) depends on the winding
number in \eqn{TYM}. This space is called the moduli space of instantons.
However, this gauge choice is admissible in the sense that the gauge fixed
action will have well defined propagators. As in the usual BRST quantisation
procedure, one has to
introduce auxiliary fields in order to construct a gauge
fermion. To fix the gauge, we start from the non--minimal action
\begin{equation}
S^1_{nm}=S+\frac{1}{2}\chi _0^{*2}+\frac{1}{2}b^{*2}\ ,
\label{S2}
\end{equation}
where $\chi _0^{*2}\equiv \chi _{0,\mu \nu }^*\chi _{0,\rho \sigma }^*
g^{\mu \rho }g^{\nu \sigma }$. The first step in the gauge fixing
procedure is to perform the canonical transformation generated by
the gauge fermion
\begin{equation}
\Psi _1=\chi _0^{\mu \nu }F^+_{\mu
\nu }+b\partial _\mu A^\mu\ .
\label{CT1}
\end{equation}
We introduced here an
antisymmetric field $\chi _0^{\mu \nu }$ and its antifield. This field has
six components, which we use to impose three gauge conditions. This
means that three components of $\chi _0$ are not part of the gauge
fixing procedure and are free. Indeed,
our action after the canonical transformation (\ref{CT1}) has the gauge
symmetry $\chi _0^{\mu \nu }\rightarrow \chi _0^{\mu \nu
}+\epsilon
_0^{-\mu \nu }$.
So, we have to constrain this field, e.g. by considering only self dual
fields $\chi _0^{\mu \nu }= \chi _0^{+\mu \nu }$. We can do
this by gauge
fixing this symmetry with the condition $\chi
_0^{-\mu \nu }=0$.
This can be done by adding an extra non--minimal sector to the action, i.e.
$S_{nm}^2=\chi _1^{*\mu \nu} \lambda _{1,\mu \nu }$. After that, we perform
the canonical transformation with gauge fermion $\Psi _2=\chi
_{1\mu \nu }\chi _0^{-\mu \nu }$. But then we have again
introduced too many fields,
and this leads to a new symmetry $\chi _{1\mu \nu }\rightarrow
\chi _{1\mu \nu }+\epsilon^+_{1\mu \nu }$ which we have to
gauge fix. One easily sees that this procedure repeats itself ad infinitum.
We could, in principle, also solve this problem by only
introducing $\chi _0^{+\mu \nu }$ as a field. Then we have
to integrate over the space of self dual fields.
To construct the measure on this space, we have to solve the constraint
$\chi =\chi ^+$.
Since this in general can
be complicated (as e.g. in the topological $\sigma
$--model) we will keep the $\chi _{\mu \nu }$ as the
fundamental
fields. The path integral is with the measure ${\cal D}\chi _0^{\mu
\nu }$ and we do not split this into the measures in the spaces of
self and anti--self dual fields. The price we have to pay is an infinite
tower of auxiliary fields. These we denote by
$(\chi _n^{\mu \nu},\lambda _n^{\mu \nu
})$\footnote{One remark has to be made here concerning the place of the
indices. We choose the indices of $\chi _n$ and $\lambda _n$ to be upper
resp. lower indices when $n$ is even resp. odd. Their antifields have
the opposite property, as usual.}
with statistics $\epsilon (\lambda _n)=n,\epsilon (\chi _n)=n+1$ (modulo
$2$) and ghost
numbers $gh(\lambda _n)$ equal to zero for $n$ even and one for $n$ odd.
Similarly, $gh(\chi _n)$ equals $-1$ for $n$ even and zero for $n$ odd.

To obtain the gauge fixed action, we start from the total non--minimal
action
\begin{equation}
S_{nm}=S_{nm}^1+\sum _{n=1}^{\infty}\chi ^*_{n,\mu \nu }\lambda _n^{\mu \nu
}\ ,\label{nmact}
\end{equation}
and take as gauge fixing fermion
\begin{equation}
\Psi =\Psi _1+\sum _{n=1}^{\infty}\chi _n^{\alpha \beta }\chi _{n-1,\alpha
\beta }^{(-)^{n}}, \label{IJK}
\end{equation}
where $\chi ^{(-)^n}$ denotes the self dual part of
$\chi $
if $n$ is even and the anti self dual part if $n$ is odd.
After doing the gauge fixing we end up with
the following non--minimal solution of the classical master equation
\footnote{Note that from $(\chi ,\chi ^*)= \frac{1}{\sqrabsg}$, it follows
that
$(\chi ^{\pm},\chi ^{*\pm})=\frac{1}{\sqrabsg} P^{\pm}$ and $(\chi ^+,\chi
^{*-})=0$, where
$P^{\pm}$ are the projectors onto the (anti)-self dual sectors.}~:
\begin{eqnarray}
S&=&S_0+\frac{1}{2}(\partial _\mu A^\mu
+b^*)^2+\frac{1}{2}(F^++\chi^-_1 +\chi
_0^*)^2\nonumber\\&&+b\partial _\mu \psi ^\mu +\chi _0^{+\alpha \beta
}D_{[\alpha }\psi _{\beta ]}+A_\mu ^*\psi ^\mu \nonumber \\
& &
+\sum _{n=1}^{\infty}(\chi ^*_{n\alpha \beta }+\chi
_{n+1,\alpha \beta }^{(-)^{(n+1)}}+\chi _{n-1,\alpha \beta
}^{(-)^{n}})\lambda _n^{\alpha \beta }\ .  \label{EA}
\end{eqnarray}
Notice that we now have terms quadratic in the antifields. This means that
the BRST operator defined by $Q\phi ^A=(\phi ^A,S)|_{\phi ^*=0}$ is only
nilpotent using field equations. Indeed, $Q^2b=\frac{1}{2x}\partial _\mu
\psi ^\mu \approx 0$, using the field equation of the field $b$.

We can write this extended action as $S=S_0 +(X,S)$, with $X$ given by
\beq
   X = \frac{1}{2} b (\partial_{\mu} A^{\mu} + b^*) - \psi_{\mu}^*
\psi^{\mu} + \frac{1}{2} \chi^{\mu \nu}_0 (F_{\mu \nu}^+ + \chi_{1 \mu
\nu}^- + \chi_{0\mu \nu}^* ) - \sum_{n=1}^{\infty} \lambda_n \lambda^*_n\ .
\eeq
We will now calculate the energy-momentum tensor, as defined in
section 2. As for notation, when a term is followed by $(\alpha
\leftrightarrow \beta)$, this means that this term, and only this one, has
to be copied but with the indices $\alpha$ and $\beta$ interchanged.
We find~:
\begin{eqnarray}
T_{\alpha \beta }&=&
(\partial _\mu A^\mu +b^*)(\partial _\alpha
A_\beta +\partial _\beta A_\alpha )
\nonumber\\&&
-\frac{1}{2}(\partial _\mu
A^\mu )^2g_{\alpha \beta }+\frac{1}{2}(b^{*})^2g_{\alpha \beta
}-\frac{1}{4}g_{\alpha \beta }F^2+\frac{1}{2}g_{\alpha \beta }{\tilde
F}_{\mu \nu} \chi^{*\mu \nu} _0+\frac{1}{2}g_{\alpha \beta }(\chi _0
^*)^2\nonumber\\&&+\frac{1}{2}(F_{\alpha \nu }+\chi ^*_{0\alpha \nu
})(F_\beta {}^\nu +{\chi _0^*}_\beta{}^{\nu })+(\alpha \leftrightarrow
\beta )  \nonumber\\&&
+b(\partial _\alpha \psi _\beta +\partial _\beta \psi _\alpha
)-g_{\alpha \beta }b\partial _\mu \psi ^\mu
\nonumber\\&&-\frac{1}{2}g_{\alpha \beta }\chi^{\mu \nu}_0  D_{[\mu}
\psi_{\nu ]} -{\chi _0}^\rho
{}_\alpha {\widetilde {D_{[\rho }\psi _{\beta ]}}}-{\chi _0}^\rho {}_\beta
{\widetilde {D_{[\rho }\psi _{\alpha
]}}}\nonumber\\&&-\frac{1}{2}g_{\alpha \beta }\chi _0^{*\mu \nu} {\tilde
\chi }_{1\mu \nu}  + \frac{1}{2} (\chi_{1\alpha}{}^{\mu} +
\chi^*_{0\alpha}{}^{\mu} ) (\chi _{1\beta \mu} + \chi^*_{0 \beta \mu}) +
(\alpha \leftrightarrow \beta)-\frac{1}{4}\chi _1^2\nonumber\\
&&- g_{\alpha \beta}
\sum _{n=1}^{\infty}[(\chi _{n+1,\mu \nu }^{(-)^{(n+1)}}
+\chi _{n-1,\mu \nu }^{(-)^{n}})\lambda _n^{\mu \nu }+\frac{1}{2}({\tilde
\chi }_{n+1,\mu \nu }-{\tilde \chi }_{n-1,\mu \nu })\lambda _n^{\mu \nu
}]\nonumber\\&&+\sum_{n=1}^{\infty}({\tilde \chi }_{n+1,\mu \alpha }-
{\tilde \chi }_{n-1,\mu \alpha }){\lambda _n}^\mu {}_\beta +
(\alpha \leftrightarrow \beta)\ .
\end{eqnarray}
We now determine $X_{\alpha \beta }^0$ such that
\begin{equation}
T_{\alpha \beta }=(X_{\alpha \beta }^0,S)\ , \label{CTME}
\end{equation}
which is the classical part of (\ref{TME}). Finding a solution of this
equation is a problem of antibracket cohomology. We could try to
construct the solution by an expansion in antifieldnumber,
but this still turns out to be quite tedious. Instead, we will
take a different
strategy, using canonical transformations.
We know already that (\ref{EA}) is
canonically equivalent, with generating fermion $\Psi $, to \eqn{nmact}.
Therefore, we can calculate the energy-momentum tensor in this set
(before $\Psi $ was done) of
coordinates, verify that it is cohomologically trivial and transform the
result using (\ref{nice}).
For \eqn{nmact}, we find
\begin{eqnarray}
X^{0}_{\alpha \beta }=\frac{1}{2}g_{\alpha \beta
}b^*b+\frac{1}{2}g_{\alpha
\beta }\chi _{0\mu \nu }^*\chi _0^{\mu \nu }+\chi ^*_{0\mu \alpha }{\chi
_0}^\mu {}_\beta\ .
\end{eqnarray}
Then it follows that in the new variables the solution is given by
\begin{eqnarray}
X^0_{\alpha \beta }&=&b(\partial _\alpha A_\beta +\partial _\beta A_\alpha
)-\frac{1}{2}g_{\alpha \beta }b\partial _\mu A^\mu +\frac{1}{2}g_{\alpha
\beta }b^*b\nonumber\\
&&+\chi _{0\mu \alpha }F^{-\mu }{}_\beta
+(\alpha \leftrightarrow
\beta )-\frac{1}{2}g_{\alpha \beta }\chi _0^{\mu \nu }F_{\mu \nu}^-
\nonumber\\
&&+\frac{1}{2}g_{\alpha \beta }\chi ^*_{0\mu \nu }\chi _0^{\mu \nu }+\chi
^*_{0\mu \alpha }{\chi _0}^\mu {}_\beta +(\alpha \leftrightarrow \beta
)\nonumber\\
&& + \frac{1}{2} g_{\alpha \beta} \chi_{1\mu \nu}^- \chi_0^{\mu \nu}
+ \chi^-_{1 \mu \alpha} \chi_0^{\mu} {}_{\beta} + (\alpha \leftrightarrow
\beta) \\
&&-g_{\alpha \beta }\sum _{n=1}\chi _n( \chi _{n-1}^{(-)^n} + \frac{1}{2}
\tilde \chi_{n-1} ) + \sum_{n=1} \tilde \chi_{n \alpha \mu} \chi_{n-1
\beta }{}^{\mu} + (\alpha \leftrightarrow \beta )\nonumber \ .
\end{eqnarray}
One can indeed check that this expression satisfies (\ref{CTME}).
Notice that it contains $b^*b$ and $\chi^*_0 \chi_0$ terms.
Therefore, it is expected that the one loop equation (\ref{OLTME}) becomes
non-trivial.

\section{Example 2~: Topological Landau--Ginzburg models}
Analogous to topological Yang--Mills (YM) theory, one can obtain
topological
Landau Ginzburg (LG) models via twisting $N=2$ LG theories. This was done
in \cite{Vafa}.
Since topological YM can also be obtained from gauge fixing
a topological invariant, we expect the same for the LG models. Indeed, it
was shown in \cite{tlg} how to obtain topological LG models by gauge
fixing zero action using the BRST formalism.
We will present a part of that construction here in the context
of BV theory.

We start by defining the classical fields in the theory. These are scalar
fields which we denote by $\bX ^i$ and $H^i$. They are defined on a
Riemann surface of genus $g$. The classical action, which is the integral
of the Lagrangian over the Riemann surface, is taken to be zero~:
\begin{equation}
S^0(\bX,H)=0\ .
\end{equation}
This action has of course two gauge symmetries, namely arbitrary shift
symmetries on both fields. For these symmetries, we introduce ghosts $\bx
^i$ and $\chi ^i$. The extended action then is
\begin{equation}
S=\bX ^{i*}\bx ^i-iH^{i*}\chi ^i\ ,
\end{equation}
where the factor $-i$ is for convention, as we will see later.
To do the gauge fixing we first have to add non-minimal sectors to the
extended action
\begin{equation}
S_{nm}=\bX ^{i*}\bx ^i-iH^{i*}\chi ^i+\frac{1}{2}\bp ^{i*}\bF ^i
+\frac{i}{2}\rho ^{i*}X^i\ .
\end{equation}
$\bp ^i$ and $\rho ^i$ play the role of antighosts with ghost number
minus one, while $\bF ^i$ and $X^i$ are Lagrange multipliers with ghost
number zero. Again we have put some numerical factors for convenience.
We now perform
the canonical transformation with generating fermion
\begin{equation}
f=-\partial _-\rho ^i[-2i\partial _+\bX ^i+4\kappa i\partial _+H^j\partial
_i\partial _jW(X)]-\bp ^i[2\partial _-\partial _+H^i+4\kappa \partial
_{{\bar i}}{\bar W}(\bX)\ .
\end{equation}
It is the gauge fixing that defines the content of the theory. The gauge
fixing conditions in topological YM theory led to the study of the moduli
space of self dual instantons. Here we have two differential equations
\begin{eqnarray}
\partial _-\partial _+H^i&=&-2\kappa \partial _{{\bar i}}{\bar
W}\nonumber\\
\partial _+\bX ^i&=&2\kappa \partial _+H^j\partial _i\partial _jW\ ,
\end{eqnarray}
where $x^{\pm}$ are the coordinates on the Riemann surface and $\kappa $ is
a coupling constant.
We also introduced here a potential $W(X)$,
which is polynomial in the $X^i$. The derivatives w.r.t. $X^i$ are denoted
by $\partial _iW$. Its conjugated ${\bar W}(\bX)$ is
the same as $W$, but with $X$ and ${\bX}$ interchanged. The choice of the
potential defines the model. Indeed, by choosing a different potential one
obtaines different solutions of the above differential equations, and so,
one studies a different moduli space. It is however less clear what the
geometrical interpretation is of these equations.

After the canonical transformation, the action is (dropping the primes)~:
\begin{eqnarray}
S&=&-\partial _+\bX ^i\partial _+X^i+\partial _+H^i\partial _-\bH
^i-2\kappa \bF ^i \partial _{{\bar i}}{\bar W}+2\kappa \partial
_+H^i\partial _-X^j\partial _i\partial _jW\nonumber\\
&&-2i\partial _+\bx ^i\partial _-\rho ^i-2i\partial _-\bp ^i\partial _+
\chi ^i+4\kappa \partial _+\chi ^i\partial _-\rho ^j\partial _i\partial
_jW-4\kappa \bp ^i\bx ^j \partial _{{\bar i}}\partial _{{\bar j}}{\bar
W}\nonumber\\
&&+\bX ^{i*}\bx ^i-iH^{i*}\chi ^i+\frac{1}{2}\bp ^{i*}\bF ^i
+\frac{i}{2}\rho ^{i*}X^i\ .
\end{eqnarray}
As an intermediate step and as a check, one can compute the energy momentum
tensor using \eqn{EMTBV}, and see if it is antibracket exact. One finds
\begin{eqnarray}
T_{\pm \pm}&=&-\partial _{\pm}\bX ^i\partial _{\pm}X^i+\partial
_{\pm}H^i\partial _{\pm}\bF ^i+2\kappa \partial _{\pm}H ^i\partial
_{\pm}X^j\partial _i\partial _jW\nonumber\\
&&-2i\partial _{\pm}\bx ^i\partial _{\pm}\rho ^i-2i\partial _{\pm}\bp
^i\partial _{\pm}\chi ^i+ 4\kappa \partial
_{\pm}\chi ^i\partial _{\pm}\rho ^j\partial _i\partial _jW \nonumber\\
&=&(2i\partial _{\pm}\bX ^i\partial _{\pm}\rho ^i-4\kappa i\partial
_{\pm}H^i\partial _{\pm}\rho ^j\partial _i\partial _jW+2\partial
_{\pm}H^i\partial _{\pm}\bp ^i,S)\nonumber\\
T_{+-}&=&2\kappa \bF ^i\partial _{{\bar i}}{\bar W}+4\kappa \bp ^i\bx
^j\partial _{{\bar i}}\partial _{{\bar j}}{\bar W}\nonumber\\
&=&(4\kappa \bp ^i\partial _{{\bar i}}{\bar W},S)\ ,\label{EMTLG}
\end{eqnarray}
so the metric independence is proven at the classical level.

This theory is not yet LG theory. To make the connection with
\cite{Vafa,tlg} we make the non-local change of variables~:
\begin{eqnarray}
\psi ^i=\partial _+\chi ^i   &\qquad & \chi ^{i*}=-\partial _+\psi
^{i*}\nonumber\\
\xi ^i=\partial _-\rho ^i &\qquad & \rho ^{i^*}=-\partial _-\xi
^{i*}\nonumber\\
F^i=\partial _-\partial _+H^i &\qquad & H^{i*}=\partial _-\partial
_+F^{i*}\ .\label{FR}
\end{eqnarray}
All the fields in the old basis were scalar fields. Doing this field
redefinition one defines the forms $\psi ^idx^+$ and $\xi ^idx^-$ of degree
(1,0) and (0,1) (under holomorphic coordinate transformations) and a (1,1)
form ${\bf F}^i=F^idx^+\wedge dx^-$.
Remark that the Jacobian of this transformation is one, at least formally,
since the contributions from the fermions cancel against the bosons.

After this field redefinition, the extended action is
\begin{eqnarray}
S&=&-\partial _+\bX ^i\partial _-X^i-F^i\bF ^i-2\kappa \bF ^i\partial
_{{\bar i}}{\bar W}-2\kappa F^i\partial _iW\nonumber\\
&&+2i\bx ^i\partial _+\xi ^i+2i\bp ^i\partial _-\psi ^i+4\kappa \psi ^i\xi
^j\partial _i\partial _jW-4\kappa \bp ^i\bx ^j\partial _{{\bar i}}
\partial _{{\bar j}}{\bar W}\nonumber\\
&&+\bX ^{i*}\bx ^i-iF^{i*}\partial _-\psi ^i+\frac{1}{2}\bp ^{i*}\bF
^i+\frac{i}{2}\xi ^{i*}\partial _-X^i\ .
\end{eqnarray}
One can indeed check that this satisfies the master equation $(S,S)=0$ in
the new basis of fields and antifields.

This is the familiar action of topological Landau--Ginzburg theory. The
fields $F^i$ and $\bF ^i$ are auxiliary fields. They are needed to make the
BRST rules
nilpotent off shell. If one integrates them out, one obtains the condition
$F^i=-2\kappa \partial _{{\bar i}}{\bar W}$. This equation is precisely
the first of our gauge conditions, written in the new basis. Eliminating
the auxiliary fields gives the action of \cite{Vafa}.
However, the BRST rules of the fields obtained here are not the same as in
\cite{Vafa}~! This crucial difference is explained in \cite{tlg}, which we
will summarise now.

First notice that the energy momentum tensor changes after
the field redefinition. The expression for $T_{\mu \nu }$ in the new
basis is not obtained
by taking \eqn{EMTLG} and substituting the fields in the old basis by the
fields in the new basis. This leads to a non-local expression for the
energy momentum tensor. This can not be correct since the action in the new
variables is still local, so its derivative w.r.t. the metric is also
local. The reason is that the field redefinitions \eqn{FR} change the form
degree and this changes the coupling to the metric. Take for instance the
term in the (old) action $\partial _+H^i\partial _-\bF ^i$. When
covariantising and writing the metric explicit this becomes $\sqrt g
g^{\mu \nu }\partial _\mu  H^i\partial _\nu  \bF ^i$. It is clear
that this gives a contribution to the energy momentum tensor as given in
\eqn{EMTLG}. However, in the new basis, we write this part of the action as
an integral of the two form $\bF ^i{\bf F}^i$, which has no metric
dependence and so, there is no contribution to $T_{\mu \nu }$. Following
this procedure, one finds for the energy momentum tensor
\begin{eqnarray}
     T_{++}=&- \partial_+ \bX ^{i} \partial_+ X^i + 2i \psi^i
     \partial_+\bp ^{i}& \nonumber \\
     T_{--}=&- \partial_- \bX ^{i} \partial_- X^i + 2i \xi^i
     \partial_- \bx ^{i}&=( 2i \xi^i \partial_- \bX ^{i},S)\nonumber\\
     T_{+-}=&4\kappa \bp ^{i} \bx ^{j}
     \partial_{{\bar i}} \partial_{{\bar j}}{\bar W}+2 \kappa  \bF ^{i}
\partial_{{\bar i}}{\bar W}&=(4\kappa \bp ^i\partial _{{\bar i}}{\bar W},S)
\ .
\end{eqnarray}
It is striking that in the new basis the (++) component is not BRST exact~!
After the change of variables we seem to have lost the metric independence
of the theory. There is however a way out, as was shown in \cite{tlg}. One
can introduce an anti-BRST operator such that the (++) component becomes
anti-BRST exact. Then the topological nature of the theory is proven using
the Ward identities for the anti-BRST operator. Moreover the BRST operator
constructed in \cite{Vafa} is nothing but the sum of the anti-BRST and BRST
operators of our formalism. It would be a good exercise to translate the
construction with the anti-BRST operator of \cite{tlg} into the BV
language. This can be done using the tools of \cite{antiBV,DJTH}.

\chapter{The geometry of the antifield formalism}
\section{Motivation}
As we have explained in chapter 2, there is some analogy between the
Hamiltonian and Lagrangian formalism. The conjugated momenta in the
Hamiltonian formalism are replaced by the antifields (with opposite
statistics) in the Lagrangian formalism. Also the Poisson brackets, which
we will call "even" get replaced by antibrackets, which are "odd" due to
the
change of statistics. One can do canonical transformations that preserve
Poisson brackets or antibrackets. The advantage of the Lagrangian method is
however that it is a covariant formalism while the Hamiltonian method is not
covariant.

In the Hamiltonian language, there is an underlying geometrical structure
that determines the dynamics. Time evolution is generated by taking the
Poisson bracket with the Hamiltonian. Poisson brackets are defined by
introducing a symplectic two form. Locally one can always take Darboux
coordinates such that the Poisson brackets take the usual form. The
underlying geometry is that of ("even") symplectic geometry.

The aim of this last chapter is to discuss the geometrical structure behind
the Lagrangian method, in the context of the BV formalism. Here, gauge
transformations are generated by taking the antibracket with the extended
action. We want to know the analogue of the symplectic two form of the
Hamiltonian formalism. The geometry is then that of "odd" symplectic
geometry.

The geometrical meaning of the BV formalism was first discussed by
Witten, as we mentioned at the end of chapter 2. More recently
the BV formalism has been set up on a curved supermanifold of fields and
anti-fields with an odd symplectic structure \cite{Schwarz1}. It has been
applied to study quantisation of string field theory \cite{strfthBV,HaZw}.
The application went far beyond the original motivation of the BRST
quantisation. As such an example we would also like to mention the work by
Verlinde \cite{Verlinde}.
In whatever circumstance it is used,  the ultimate goal of the BV formalism
is to determine the odd symplectic structure of the supermanifold and
solve the master equation. Therefore it is important to understand the
geometry of the fermionic symplectic structure.

We will start by recalling some basic facts about even symplectic
geometry. Then we will show how this geometry induces an odd symplectic
structure. The BV formalism will then be written down in a covariant way,
not using Darboux coordinates. Finally, we will discuss some applications,
mainly based on \cite{Shogo}.

\section{The "even" geometry}
Let us here very briefly repeat the basic ingredients of symplectic
geometry. For a reference, see for instance chapter one in \cite{Wood}.
Let us start with a
(bosonic) $D$-dimensional symplectic manifold ${\cal M}$ with
coordinates
$x^i,i=1,...,D$. This means there exists a symplectic 2-form on ${\cal
M}$~:
\begin{equation}
\omega =\omega _{ij}dx^i\wedge dx^j\ , \label{sf}
\end{equation}
which is nondegenerate and closed~:
\begin{equation}
d\omega =0\ ,\label{closed}
\end{equation}
where $d$ is the exterior derivative on ${\cal M}$. In components this
equation reads
\begin{equation}
\partial _i\omega _{jk}+\partial _j\omega _{ki}+\partial _k\omega _{ij}=0\
.\label{clcomp}
\end{equation}
The nondegeneracy ($det(\omega _{ij})\neq 0$) implies that
${\cal M}$ is even dimensional, since an odd dimensional antisymmetric
matrix has zero determinant. Therefore we write $D=2d$.
It also implies the existence of an inverse
\begin{equation}
\omega ^{ik}\omega _{kj}=\delta ^i{}_j\ ,
\end{equation}
whith $\omega _{ij}=-\omega _{ji}$ and $\omega ^{ij}=-\omega ^{ji}$.

Now consider the space of functions on ${\cal M}$, which
we denote
by ${\cal F}({\cal M})$. On this space we can define a Poisson bracket
\begin{equation}
\{f(x),g(x)\}=\dr_i f\omega ^{ij}\dl _jg\ ,
\end{equation}
for all $f,g\in {\cal F}({\cal M})$. This Poisson bracket satisfies
the Jacobi identy due to the closure of $\omega $.

Canonical transformations are a special kind of general coordinate
transformations. They are defined as those transformations that leave
the symplectic structure invariant. Under general coordinate
transformations
\begin{equation}
x^i\rightarrow {\tilde x}^i(x)=x^i+\epsilon ^i(x)\ ,
\end{equation}
the symplectic structure changes according to
\begin{equation}
{\tilde \omega }^{ij}({\tilde x}(x))=\{{\tilde x}^i(x),{\tilde x}^j(x)\}=
{\tilde x}^i(x)\dr _k\omega ^{kl}(x)\dl _l{\tilde x}^j(x)\ .
\end{equation}
For a canonical transformation one must have
\begin{equation}
{\tilde \omega }^{ij}({\tilde x}(x))=\omega ^{ij}(x)\ .
\end{equation}
As a consequence, canonical transformation leave the Poisson brackets
invariant.

To define integration on ${\cal M}$, we have to define a
measure. This can be done by using the symplectic 2 form~:
\begin{equation}
\omega ^d\equiv \omega \wedge ... \wedge \omega ={\sqrt \omega }d^Dx\ ,
\label{imev} \end{equation}
where the product has $d$ factors and the $\omega $ under the square root
is the determinant
of $\omega _{ij}$. It is clear that this measure is invariant under
canonical transformations.

One can repeat this construction for a supermanifold with $M$ bosonic and
$N$ fermionic coordinates. The non-degeneracy of the symplectic two form
then requires that $M$ is even and $N$ is arbitrary. The Darboux theorem
then states that, locally, there are coordinates such that half of
the
bosonic coordinates are conjugated (the momenta) to the other half, and the
fermions are conjugated to itself. This is the main difference with odd
symplectic structures, as we will see. In the latter case, the bosonic
coordinates are conjugated to the fermionic coordinates, and one must have
that $M=N$, with $M$ odd or even.

\section{The "odd" geometry}
As argued at the end of the previous section, we start with a
$2D$ manifold parametrised by real coordinates $y^i = (x^1,x^2,\cdots,x^D,
\xi_1,\xi_2,\cdots,\xi_D) $ with $x$'s and $\xi$'s bosonic and fermionic
respectively. An odd
symplectic structure is given by a non-degenerate 2-form
\begin{equation}
\omega = dy^j \wedge dy^i \omega_{ij}   \ ,\label{OSS}
\end{equation}
which is closed
\begin{equation}
d\omega = 0\ .
\end{equation}
These equations read in components (a derivative without arrow will always
be a left derivative)
\begin{equation}
(-)^{ik}  \partial_i \omega_{jk} + (-)^{ji} \partial_j \omega_{ki} +
(-)^{kj}\partial_k \omega_{ij} = 0 \label{closcomp}
\end{equation}
\begin{equation}
\omega_{ij}  = -(-)^{ij}\omega_{ji}\ .\label{symprop}
\end{equation}
The statistics of the coefficients of the symplectic form is
$\varepsilon (\omega_{ij}) = i + j + 1$. We define the
anti-bracket by
\begin{equation}
(A,B) = A\overleftarrow \partial_i \omega^{ij} \partial_j B\
,\label{ABcov}
\end{equation}
in which $\omega^{ij}$ is the inverse matrix of $\omega_{ij}$ such that
\begin{equation}
\omega_{ij} \omega^{jk} = \omega^{kj}\omega_{ji} = \delta^k_i\ .
\end{equation}
Note that the right-derivative $\overleftarrow \partial_i$ is related with
the left-one by
\begin{equation}
A\overleftarrow \partial_i = (-)^{i(\varepsilon (A) + 1)} \partial_i A \ .
\end{equation}
In terms of $\omega^{ij}$, \eqn{closcomp} and \eqn{symprop} become
respectively
\begin{equation}
(-)^{(i+1)(k+1)} \omega^{il}  \partial_l \omega^{jk} +
(-)^{(j+1)(i+1)} \omega^{jl}\partial_l \omega^{ki} +
(-)^{(k+1)(j+1)} \omega^{kl}\partial_l \omega^{ij}  = 0\nonumber
\end{equation}
\begin{equation}
\omega^{ij}  = -(-)^{(i+1)(j+1)} \omega^{ji} \ .
\end{equation}
Because of this, the anti-bracket \eqn{ABcov} satisfies the Jacobi
identity. Also canonical transformations can be done. They preserve the odd
symplectic structure and so, the antibrackets. We have discussed this in
previous chapters.

Remarkably, with every even symplectic form $\omega ^e$ on a manifold
with coordinates $x^i$, one can associate
an odd symplectic form $\omega ^o$ \cite{KuNer}. One first introduces
fermionic coordinates $\xi _i$ and defines
\begin{equation}
\omega ^o=\omega ^e_{ij}dx^i\wedge d\xi ^j+\omega ^e_{ij,k}\xi ^kdx
^i\wedge dx ^j\ ,\label{eveninodd}
\end{equation}
where $\omega ^e_{ij,k}$ is the derivative of $\omega ^e_{ij}$ w.r.t.
$x^k$. We will use this formula in the next sections.

Now we have to define integration on our supermanifold. Let us first
mention that integration on superspace with an even symplectic structure
can still be defined as in \eqn{imev} with the determinant replaced by the
berezinian. In the case of odd symplectic structures, \eqn{imev} does
not make sense anymore, since $\omega \wedge \omega =0$. This is one of the
essential differences between the even and odd geometry. To define
integration theory on an odd symplectic manifold, we have to provide our
space with an additional structure, namely the volume form.

Following \cite{Schwarz1}, we introduce a volume element by
\begin{equation}
d\mu (y)=\rho (y)\Pi  _{i=1}^{2D}dy^i\ ,
\end{equation}
where $\rho (y)$ is a density. The delta operator on a function $A$ is
then defined as the divergence of the (Hamiltonian) vector field $V_A$
associated with $A$~:
\begin{eqnarray}
V_A&=&V_A^i\partial _i=\omega ^{ij}(\partial _jA)\partial _i\nonumber\\
\Delta _\rho A&\equiv &div_\rho V_A=\frac{1}{\rho
}(-)^i\partial _i(\rho \omega ^{ij}\partial _jA)\ .\label{defdel}
\end{eqnarray}
Locally, one can go to coordinates such that $\omega =dx^a\wedge d\xi _a$.
These coordinates are called Darboux coordinates. In these coordinates,
one can choose the density function to be one, i.e.
$\rho =1$. Then
one recovers the standard $\Delta $ operator used in previous chapters.
It is then also clear that there is no analogue of this operator in the
Hamiltonian formalism, since Hamiltonian vector fields are divergenceless.
One could however interpret it as some odd Laplacian.

The $\Delta _\rho $ operator satisfies the properties~:
\begin{eqnarray}
\Delta _\rho (A,B)&=&(\Delta _\rho A,B)+(-)^{A+1}(A,\Delta _\rho
B)\nonumber\\
\Delta _\rho [AB]&=& [\Delta _\rho A]B+(-)^AA\Delta _\rho B+(-)^A(A,B)\ .
\end{eqnarray}
However, the nilpotency condition $\Delta _\rho^2=0$ is only satisfied if
$\rho $ obeys the equation
\begin{equation}
\Delta _\rho[\frac{1}{\rho }(-)^i\partial _i(\rho \omega ^{ij})]=0\
.\label{nilpcond}
\end{equation}
Finally the master equation in the covariant BV formalism is given by
\cite{Schwarz1}
\begin{eqnarray}
\Delta _\rho e^S=0 \Leftrightarrow \Delta _\rho +\frac{1}{2}(S,S)=0\ .
\label{covME}
\end{eqnarray}

\section{Even and odd K\"ahler structures}
In this section we will require that our manifold is K\"ahler. Again we
start
with a bosonic manifold ${\cal M}$ and then generalise to supermanifolds.
This section is based on \cite{KuNer,Shogo}.

So, we assume there
exist an almost complex structure $J^i{}_j(x)$ on ${\cal M}$ with bosonic
coordinates $x^i, i=1,...,D=2d$ satisfying
\begin{equation}
J^i{}_kJ^k{}_j=-\delta ^i{}_j\ .
\end{equation}
The complex structure $J$ is said to be compatible with the symplectic
structure $\omega $ if
\begin{equation}
J^k{}_i\omega _{kl}J^l{}_j=\omega _{ij}\ .\label{ks}
\end{equation}
Using these two objects we can define a metric
\begin{equation}
g_{ij}=\omega _{ik}J^k{}_j\ .
\end{equation}
This is a symmetric nondegenerate tensor because of \eqn{ks}. One also has
that
\begin{equation}
J^k{}_ig_{kl}J^l{}_j=g _{ij} \qquad g^{ij}=\omega ^{ik}J^j{}_k\ .
\end{equation}
The triple $({\cal M},g,J)$ is said to be an almost K\"ahler manifold. When
the almost complex structure is integrable (i.e. the Nijenhuis tensor
vanishes), the manifold is said to be K\"ahler.
The integration measure \eqn{imev} then also reduces to the standard
volume element $\sqrt gd^Dx$.
For a K\"ahler manifold one can find holomorphic coordinates $\{z^a,
{\bar z}^{\underline a}\}$ such that the complex structure, the metric
and the symplectic form are canonical, i.e.
\begin{equation}
J^a{}_b=i\delta ^a{}_b \qquad J^{\underline a}{}_{\underline b}=-i\delta
^{\underline a}{}_{\underline b}\ ,
\end{equation}
and
\begin{eqnarray}
\omega_{ab}=\omega_{\underline a \underline b} = 0, &\qquad&
g_{ab}=g_{\underline a \underline b} = 0, \nonumber\\
\omega_{a \underline b}=i g_{a \underline b}, &\qquad&
\omega_{\underline a b} = -i g_{\underline a b},
\end{eqnarray}
together with
\begin{eqnarray}
\omega_{a \underline b}= -\omega_{\underline b a},
&\qquad& \omega_{a \underline b}^* = \omega_{\underline a
b}, \nonumber\\
g_{a \underline b} = g_{\underline b a},
&\qquad& g_{a \underline b}^* = g_{\underline a b}\ ,\label{gversom}
\end{eqnarray}
where the $*$ means complex conjugation.
By means of these equations, the symplectic form \eqn{sf} takes the form
\begin{equation}
\omega = 2i d \overline z^{\underline b}\wedge dz^a \
g_{a \underline b},
\end{equation}
and is called the K\"ahler 2-form.
Then \eqn{closed}, or equivalently \eqn{clcomp}, is solved by
\begin{equation}
\gamma_{a \underline b} = \partial_a \partial_{\underline b} K,
\label{Kmetric}
\end{equation}
in which $K$ is called the K\"ahler potential.

Now we will repeat this for the odd symplectic structure. The coordinates
on our supermanifold are again denoted by $y^i$ as in the previous
section. Now $i=1,...,2D=4d$, because there are as many bosons as fermions.
On this supermanifold we can define a complex structure
\begin{equation}
J^k_j = \pmatrix{ 0  & \unity & 0 & 0  \cr
         -\unity & 0 & 0 & 0  \cr
         0  & 0 & 0 & \unity  \cr
         0  & 0 & -\unity & 0 \cr}\ ,
\end{equation}
with the $d \times d$ unit matrix $\unity$. The matrix
$\gamma_{ij} $ may be defined by
\begin{equation}
\gamma_{ij} = \omega_{ik}J^k_{\ j}  \ .\label{oddmetric}
\end{equation}
We shall now impose the K\"ahler condition on $\omega_{ij}$
\begin{equation}
\omega_{kl}J^k_{\ i}J^l_{\ j}  = \omega_{ij}  \ ,\label{Kahlcon}
\end{equation}
or equivalently
\begin{equation}
\gamma_{kl}J^k_{\ i}J^l_{\ j}  = \gamma_{ij}\ .
\end{equation}
Since  $J^k_{\ j}$ is a bosonic matrix, the metric $\gamma_{ij}$ is fermionic,
$ \varepsilon (\gamma_{ij}) = i + j + 1$. From the symmetry properties of
the odd symplectic structure in its two indices, we have
\begin{eqnarray}
\gamma_{ij} &=& (-)^{ij} \gamma_{ji}\nonumber\\
\gamma_{ij}\gamma^{jk} &=& \gamma^{kj}\gamma_{ji} = \delta^k_i\nonumber\\
\gamma^{ij} &=& (-)^{(i+1)(j+1)}\gamma^{ji} \ .
\end{eqnarray}
We also find the relation
\begin{equation}
\gamma_{ij}\omega^{jk}\gamma_{kl} = -\omega_{il} \ .
\end{equation}
The affine connection may be defined by postulating
\begin{equation}
D_k \gamma_{ij} \equiv \partial_k \gamma_{ij} - \Gamma^l_{ki}\gamma_{lj}
- (-)^{ij}\Gamma^l_{kj}\gamma_{li} = 0 \ .
\end{equation}
By solving this we obtain
\begin{equation}
\Gamma^k_{ij} = {1 \over 2}[\partial_i \gamma_{jl} + (-)^{ij}\partial_j
\gamma_{il} - \gamma_{ij}\overleftarrow \partial_l ]\gamma^{lk}\ ,
\end{equation}
which is bosonic, $\varepsilon (\Gamma^k_{ij}) = i+j+k$.

We shall go to the complex coordinate basis
$ y^i \rightarrow ({\bf z}^a, \overline {\bf z}^{\underline a}) $
with
\begin{equation}
{\bf z}^a  = (z^\alpha,  \zeta^\alpha)  \qquad
\overline {\bf z}^{\underline a}  = (\overline z^{\underline \alpha},
\overline \zeta^{\underline \alpha})  \qquad
\alpha  = 1,2,\cdots,d \ ,
\end{equation}
defined by
\begin{equation}
z^{\alpha} = x^\alpha + i x^{d+\alpha} \qquad
\zeta^\alpha = \xi^\alpha + i \xi^{d+\alpha}\ ,
\end{equation}
and complex conjugation.
Then the K\"ahler condition \eqn{Kahlcon} reduces to
\begin{eqnarray}
\omega_{ab} = \omega_{\underline a \underline b} = 0 &\qquad&
\gamma_{ab}  = \gamma_{\underline a \underline b} = 0 \nonumber\\
\omega_{a \underline b}  = i \gamma_{a \underline b} &\qquad&
\omega_{\underline a b} = -i \gamma_{\underline a b}\ ,
\end{eqnarray}
together with \footnote{Complex conjugation of fermion products is
chosen as $(\zeta \eta)^\ast = \overline \zeta \overline \eta$.}
\begin{eqnarray}
\omega_{a \underline b}  = -(-)^{ab}\omega_{\underline b a} &\qquad&
\omega_{a \underline b}^\ast = \omega_{\underline a b} \nonumber\\
\gamma_{a \underline b}  = (-)^{ab}\gamma_{\underline b a} &\qquad&
\gamma_{a \underline b}^\ast = \gamma_{\underline a b}\ .
\end{eqnarray}
By means of these equations the odd symplectic form takes the K\"ahler 2-
form
\begin{equation}
\omega = 2i d \overline {\bf z}^{\underline b}\wedge d {\bf z}^a \
\gamma_{a \underline b}\ .
\end{equation}
Then the closureness condition $d\omega =0$ is solved by
\begin{equation}
\gamma_{a \underline b} = \partial_a \partial_{\underline b} K\ ,
\end{equation}
in which $K$ is a fermionic K\"ahler potential.
Thus the supermanifold acquires a fermionic K\"ahler geometry as the
consequence of \eqn{closcomp} and \eqn{Kahlcon}.
The affine connection is simplified as
\begin{equation}
\Gamma^c_{ab}  = \partial_a \gamma_{b \underline d}\gamma^
{\underline d c} \qquad
\Gamma^{\underline c}_{\underline a \underline b}  = \partial_{\underline a}
\gamma_{\underline b d}\gamma^{d \underline c} \ ,
\end{equation}
and all the other components are vanishing.
The covariant derivative of a holomorphic vector is defined by
\begin{equation}
D_a A_b \equiv \partial_a A_b - \Gamma^c_{ab}A_c  \qquad
D_a A^b \equiv \partial_a A^b + (-)^{a A}A^c\Gamma^b_{ca}\ .
\end{equation}

\section{The isometry}
The fermionic K\"ahler manifold admits an isometry. It is realised by a set
of
Killing vectors $V^{Ai}(y), A = 1,2,\cdots,N$, with  $\varepsilon (V^{Ai}) = i$
in the real coordinates. They satisfy the Lie algebra of a group $G$
\begin{equation}
V^{Ai}\partial_i V^{Bj} - V^{Bi}\partial_i V^{Aj} =
f^{ABC}V^{Cj}\ , \label{LieKV}
\end{equation}
with (real) structure constants $f^{ABC}$. The metric $\gamma_{ij}$
and the odd symplectic 2-form obey the Killing conditions
\begin{eqnarray}
{\cal L}_{V^A}\gamma_{ij} &\equiv &V^{Ak}\partial_k \gamma_{ij} +
\partial_i
V^{Ak}\gamma_{kj}+ (-)^{ij}\partial_j V^{Ak}\gamma_{ki} = 0\nonumber\\
{\cal L}_{V^A}\omega_{ij} &\equiv &V^{Ak}\partial_k \omega_{ij} +
\partial_i
V^{Ak}\omega_{kj}- (-)^{ij}\partial_j V^{Ak}\omega_{ki} = 0\ ,
\end{eqnarray}
or equivalently
\begin{eqnarray}
{\cal L}_{V^A}\gamma^{ij} & \equiv &V^{Ak}\partial_k \gamma^{ij} -
\gamma^{ik}\partial_k V^{Aj} - (-)^{(i+1)(j+1)}\gamma^{jk}\partial_k V^{Ai} =
0\ ,\nonumber\\
{\cal L}_{V^A}\omega^{ij} & \equiv &V^{Ak}\partial_k \omega^{ij} -
\omega^{ik}\partial_k V^{Aj} + (-)^{(i+1)(j+1)}\omega^{jk}\partial_k V^{Ai}
= 0\ .
\end{eqnarray}
From consistency of these Killing conditions and the definition of the
metric \eqn{oddmetric} we find the constraints on $V^{Ai}$
\begin{equation}
\partial_k V^{Al} J^k_{\ i}J^j_{\ l}   = -\partial_i V^{Aj}\ .
\end{equation}
In the complex coordinates this equation implies that the Killing vectors
$V^{Ai}$ are holomorphic:
\begin{equation}
V^{Ai} = ({\bf R}^{Aa} ({\bf z}), \overline {\bf R}^{A \underline a}
(\overline {\bf z}))\ .\label{holKV}
\end{equation}
Due to this property the Killing condition reduces to the form
\begin{equation}
\partial_c ({\bf R}^{Aa}\gamma_{a \underline b})\ + \  (-)^{cb}
\partial_{\underline b} (\overline {\bf R}^{A \underline a}\gamma_
{\underline a c})
= 0   \quad \quad
 ({\rm Killing \  equation})\ .
\end{equation}
It then follows that the Killing vectors ${\bf R}^{Aa}$ and $\overline
{\bf R}^{A \underline a}$ are given by a set of real potentials  $\Sigma^A$
such that
\begin{equation}
{\bf R}^{Aa} \gamma_{a \underline b}  = i \partial_{\underline b} \Sigma^A
\quad \quad \overline {\bf R}^{A\underline a} \gamma_{\underline a b}  = -i
\partial_b \Sigma^A\ .\label{defKP}
\end{equation}
$\Sigma^A$ are fermionic and are called the Killing potentials. It is worth
noting
that the isometry transformations given by the Killing vectors \eqn{holKV}
can be nicely rewritten in terms of the antibracket
\begin{eqnarray}
\delta {\bf z}^a &\equiv& \epsilon^A {\bf R}^{Aa}  = \{{\bf z}^a,
\epsilon^A \Sigma^A \}\nonumber\\
\delta \overline {\bf z}^{\underline a} &\equiv& \epsilon^A \overline {\bf
R}^ {A\underline a} =
\{\overline {\bf z}^{\underline a}, \epsilon^A \Sigma^A \}\ .
\end{eqnarray}
Here $\epsilon^A ,  A = 1,2,
\cdots,N$
are constant (real) parameters of the transformations. One can
show that by these transformations the Killing and K\"ahler potentials
respectively transform as
\begin{equation}
\delta \Sigma^B  = \epsilon^A f^{ABC}\Sigma^C \ ,\label{trKP}
\end{equation}
and
\begin{equation}
\delta K = \epsilon^A F^A ({\bf z}) + \epsilon^A \overline F^A (\overline
{\bf z})\ ,\label{trK}
\end{equation}
with some holomorphic  functions $F^A({\bf z})$ and their complex conjugates.
Equation \eqn{trKP} is equivalent to the
anti-bracket relation
\begin{equation}
\{\Sigma^A, \Sigma^B \} = f^{ABC}\Sigma^C\ .
\end{equation}
The solution of \eqn{trKP} can also be put in the form
\begin{equation}
\Sigma^A = i f^{ABC}{\bf R}^{Bb}\gamma_{b\underline c}\overline {\bf R}^
{C\underline c}\ ,\label{compKP}
\end{equation}
by multiplying \eqn{defKP} by $f^{ABC}$
and using  $f^{ABC}f^{ABD} = 2\delta^{CD}$. This way of calculating the Killing
potentials $\Sigma^A$
is more practical than using \eqn{trKP}, if the metric
$\gamma_{a\underline b}$ is given. It was already known for the bosonic
case \cite{KPHull,KPShogo}.

\section{The metric of the fermionic irreducible hermitian symmetric space}
The holomorphic Killing vectors ${\bf R}^{Aa}$ and $\overline {\bf R}^
{A\underline a}$ in \eqn{holKV} independently satisfy the Lie-algebra
\eqn{LieKV}, i.e.
\begin{equation}
{\bf R}^{Ai}\partial_i {\bf R}^{Bj} - {\bf R}^{Bi}\partial_i {\bf R}^{Aj} =
f^{ABC}{\bf R}^{Cj},
\end{equation}
and the complex conjugate. These equations can be solved by
\begin{equation}
{\bf R}^{Aa} = (R^{A\alpha}(z), S^{A\alpha}(z,\zeta) ), \quad \quad \quad
\alpha = 1,2,\cdots,d\ ,
\end{equation}
with
\begin{equation}
S^{A\alpha} = \zeta^\beta {\partial \over \partial z^\beta } R^{A\alpha}\ ,
\end{equation}
and $\varepsilon (R^{A\alpha})= 0=\varepsilon (S^{A\alpha}) - 1$, in
which $R^{A\alpha}$ satisfy the Lie-algebra
\begin{equation}
R^{A\alpha}{\partial \over \partial z^\alpha} R^{B\beta} - R^{B\alpha}
{\partial
 \over \partial z^\alpha} R^{A\beta}
= f^{ABC}R^{C\beta}\ .
\end{equation}
The Killing vectors $R^{A \alpha}$ define a bosonic K\"ahler manifold.
For a class of bosonic K\"ahler manifolds, called the irreducible hermitian
symmetric spaces, they
can be explicitly constructed by extending the strategy developed in
\cite{IHSS}.
They are related to the metric of these manifolds by
\begin{equation}
R^{A\alpha}\overline R^{A\underline \beta} = g^{\alpha \underline \beta}\ ,
\end{equation}
with
\begin{equation}
R^{A\alpha}R^{A\beta} = 0\ ,\label{RR=0}
\end{equation}
and complex conjugation.
The fermionic  metric $\gamma^{ij}$ can be given in terms of these bosonic
Killing  vectors :
\begin{equation}
\gamma^{a \underline b}  =
\pmatrix{ \gamma^{z\underline z} & \gamma^{z \underline \zeta} \cr
         \gamma^{\zeta \underline z} & \gamma_{\zeta \underline \zeta} }
=\pmatrix{ 0  &  R^{A\alpha}\overline R^{A\underline \beta} \cr
       R^{A\alpha}\overline R^{A\underline \beta} &
       R^{A\alpha}\overline S^{A\underline \beta} +
       S^{A\alpha}\overline R^{A\underline \beta} }\label{metric}
\end{equation}
and
\begin{equation}
\gamma^{ab}  = \gamma^{\underline a \underline b} = 0\ ,
\end{equation}
in which $\gamma^{z\underline z} , \gamma^{z \underline \zeta},
\gamma^{\zeta \underline z}$ and $\gamma^{\zeta \underline \zeta}$ are
$d\times d$  matrices.
The symplectic form that follows from this metric
indeed satisfies the closure
property and the Killing condition. To prove it one uses the
the Lie-algebra \eqn{LieKV} and the formulae
\begin{equation}
f^{ABC}R^{B\beta} R^{C\gamma} = 0, \qquad {\rm c.c.}\ .
\end{equation}
The last formulae are consequences of the condition \eqn{RR=0}.
For the proof, see \cite{KPShogo}.
The metric \eqn{metric} can be inverted merely by knowing the inverse of
$R^{A\alpha}
\overline R^{A\underline \beta}$, denoted by $g_{\alpha \underline
\beta}$~:
\begin{eqnarray}
\gamma_{\underline b a} & =&
\pmatrix{ \gamma_{\underline z z} & \gamma_{ \underline z \zeta} \cr
         \gamma_{\underline \zeta z} & \gamma_{\underline \zeta \zeta} }
=\pmatrix{-g_{\underline \beta \gamma}g_{\alpha \underline \delta}
[R^{A\gamma}\overline S^{A\underline \delta} +
S^{A\gamma}\overline R^{A\underline \delta}]  &
g_{\underline \beta \alpha} \cr
     g_{\underline \beta \alpha} & 0  }\nonumber\\
 \gamma_{ba} & = &\gamma_{\underline b \underline a} = 0 \ .
\end{eqnarray}
Thus we have explicitly constructed the odd symplectic structure \eqn{OSS}.
It is precicely the one one would obtain by using \eqn{eveninodd}. We call
the manifold with a symplectic structure given by this 2-form a fermionic
irreducible
hermitian symmetric space.
For this class of K\"ahler manifolds the Killing potentials can be
explicitly calculated by \eqn{compKP}.

As an example we show the fermionic CP$^1$ space. It is parametrised by the
supercoordinates $(z, \zeta)$ and their complex conjugates. The $SU(2)$
transformations of the coordinates are given by the Killing vectors:
\begin{eqnarray}
\delta z & = &\epsilon^A R^{Az} = i[\epsilon^- + \epsilon^0 z -{1\over 2}
\epsilon^+z^2]  \nonumber\\
\delta \zeta &= &\epsilon^A \zeta {\partial \over \partial z^\alpha} R^{Az}
= i[\epsilon^0\zeta -
\epsilon^+ z\zeta] \ ,\label{KillV}
\end{eqnarray}
and the complex conjugates \footnote{We have chosen the structure
constants to be $f^{+-0}=-i$. Then the scalar product of the adjoint vectors
is given by
$a^Ab^A = a^0b^0 + a^+b^- + a^-b^+$.}.
We calculate the metric of the fermionic CP$^1$ space from \eqn{metric}
\begin{equation}
\gamma^{a\underline b} =
\pmatrix{ \gamma^{z\underline z} & \gamma^{z \underline \zeta} \cr
         \gamma^{\zeta \underline z} & \gamma^{\zeta \underline \zeta } \cr}
= \pmatrix{ 0  &  (1 + {1 \over 2}z\overline z)^2    \cr
        (1 + {1 \over 2}z\overline z)^2  &
        (1 + {1 \over 2}z\overline z)(z\overline \zeta + \overline z\zeta)
        \cr}  .   \label{5.6}
\end{equation}
Its inverse metric is given by
\begin{equation}
\gamma_{b \underline a} =
\pmatrix{ \gamma_{\underline z  z} & \gamma_{ \underline z \zeta} \cr
         \gamma_{ \underline \zeta z} & \gamma_{\underline \zeta \zeta} \cr}
= \pmatrix{ -{z\overline \zeta + \overline z\zeta \over (1 + {1 \over 2}z
  \overline z)^3 } &  {1 \over (1 + {1 \over 2}z\overline z)^2}   \cr
   {1 \over  (1 + {1 \over 2}z\overline z)^2}  &   0     \cr}\ .
    \label{5.7}
\end{equation}
Plugging this metric together with the Killing vectors \eqn{KillV} in
\eqn{compKP}, we obtain the Killing potentials
\begin{eqnarray}
\Sigma^+ & =& {\overline \zeta \over 1 + {1 \over 2}z\overline z } -
{\overline z (z\overline \zeta + \overline z\zeta) \over
2(1 + {1 \over 2}z\overline z)^2 } \nonumber\\
\Sigma^- & =& {\zeta \over 1 + {1 \over 2}z\overline z } -
{z (z\overline \zeta + \overline z\zeta) \over
2(1 + {1 \over 2}z\overline z)^2 }\nonumber\\
\Sigma^0 & = &{z\overline \zeta + \overline z \zeta \over (1 + {1 \over 2}z
\overline z)^2 }\ .\label{5.8}
\end{eqnarray}
It is worth checking that \eqn{defKP} is indeed satisfied by these
quantities.
The fermionic K\"ahler potential follows simply by integrating
\eqn{Kmetric}~:
\begin{equation}
K = {z\overline \zeta + \overline z\zeta \over 1 + {1 \over 2}z\overline z}.
\label{5.9}
\end{equation}
It is a consistency check of our calculations to see that both potentials
given by \eqn{5.8} and \eqn{5.9} satisfy the properties \eqn{trKP} and
\eqn{trK} respectively.

\section{The master equation}
Now we discuss the BV formalism on the fermionic CP$^1$ space. The closed
symplectic form $\omega$ is known explicitly from the metric \eqn{5.7} by
\eqn{gversom}
With this symplectic form we define the second order differential operator
according to \eqn{defdel}.
Then the function $\rho$ is fixed by the nilpotency condition
\eqn{nilpcond}. We find the unique $U(1)$-invariant solution
\begin{equation}
\rho = p + q{i \zeta \overline \zeta  \over (1 + {1 \over 2}z\overline z)^2 },
\end{equation}
with arbitrary constants $p(\not= 0)$ and $q$. To check this it is useful to
note that the metric \eqn{metric} in general satisfies
\begin{equation}
(-)^a\partial_a \gamma^{a \underline b} = 0 , \quad \quad \quad {\rm c.c.}.
\end{equation}
We may be interested in solving the master equation \eqn{covME} with these
$\omega^{ij}$ and $\rho$. The solution is given by
\begin{equation}
S = S_0  -({q \over 2p} + re^{-S_0}){i\zeta \overline \zeta \over (1 +
{1 \over 2}
z \overline z)^2}\ ,      \label{6.1}
\end{equation}
in which $S_0$ is an arbitrary function of $z$ and $\overline z$, and $r$ is
an integration constant.
We have assumed that $z$ and $\zeta$ have no space-time dependence. They can
be interpreted as coupling parameters for physical variables. Then $Z(= e^S)$
looks like the partition function of matrix models or 2-dim. topological
conformal field theories, being a function of the coupling space. The BV
formalism in the coupling space has been discussed by Verlinde
\cite{Verlinde}.

One can search for the solutions \eqn{6.1} satisfying the classical
equation $(S,S)=0$,
which implies that $S$ is BRST invariant. Assuming reality of $S$ we find
that
\begin{equation}
S = S_0\ ,   \label{6.3}
\end{equation}
or
\begin{equation}
S = a + b{i \zeta \overline \zeta  \over (1 + {1  \over 2}z\overline z)^2
}\ , \label{6.4}
\end{equation}
with some arbitrary constants $a$ and $b$. In the language of the BRST
quantisation, the first solution can be taken as a classical limit of the
full solution \eqn{6.1}. Namely its BRST transformation is trivial.
The second solution is invariant by the $SU(2)$ transformations.

The master equation \eqn{covME} may be solved also by allowing $z$
and $\zeta$ to be
space-time dependent. We will here only study the classical
master equation $(S,S)=0$,
which is still of great interest. Remarkably there is a solution for the
general K\"ahler group manifold discussed above, although it would not be the
unique one. It is given by
the action of a 2-dim. field theory
\begin{equation}
S = \int d^2 x G\partial_- \Sigma^0.  \label{6.5}
\end{equation}
Here $G$ is an arbitrary $U(1)$-invariant function of bosons $z^\alpha
(x^+,x^-)$ , chiral fermions $\zeta^\alpha (x^+,x^-)$ and their complex
coordinates. $\Sigma^0$
is the $U(1)$-component of the Killing potentials given by
\eqn{compKP}.
We can verify that this action satisfies the classical mater equation
by calculating
\begin{eqnarray}
(S,S)& = &i(-)^a\partial_a S \gamma^{a\underline b}\partial_{\underline
b}S - i(-)^a\partial_{\underline a} S \gamma^{\underline a b}\partial_b
S\nonumber\\
  & =& 2\int d^2x [-i \partial_a G \gamma^{a \underline
b}\partial_{\underline
 b}\Sigma^0 + i \partial_{\underline a} G \gamma^{\underline a b}\partial_b
 \Sigma^0 ]\partial_- G \partial_- \Sigma^0 \nonumber\\
 & = &-2\int d^2x [{\bf R}^{0a}\partial_a G + \overline {\bf R}^{0
\underline a}
 \partial_{\underline a} G] \partial_- G\partial_- \Sigma^0
= 0\ ,
\end{eqnarray}
using \eqn{defKP} and $U(1)$-invariance of $G$.
In the CP$^1$ case the solution \eqn{6.5} can be written in the general
form
\begin{equation}
S = \int d^2 x [(z \overline \zeta + \overline z \zeta)f
               + i(z \overline \zeta - \overline z \zeta)g]
               \partial_- [{z \overline \zeta + \overline z \zeta \over
               (1 + {1\over 2}z \overline z)^2} ]\ ,\label{actCP1}
\end{equation}
in which $f$ and $g$ are arbitrary real functions of $z\overline z$.
Finally, the BRST
transformations of the fields are given by $\{z, S\}$ and $\{\zeta, S\}$.
One can check explicitly that the action \eqn{actCP1} is invariant under
these BRST transformations.



\end{document}